\documentclass[11pt,a4paper]{article}
 \pdfoutput=1
\usepackage{amsmath,amssymb}
\usepackage{graphicx}
\usepackage{dcolumn}
\usepackage{bm,color}
\usepackage{hyperref}
\usepackage{accents}
\usepackage{amssymb,float}
\usepackage{amsmath}
\usepackage{multirow}
\usepackage{siunitx}
\usepackage{tabularx}
\usepackage{booktabs}
\usepackage{authblk}
\usepackage{url}
\usepackage{geometry}
\usepackage{enumitem}
\usepackage{caption}
\DeclareRobustCommand{\ion}[2]{\textup{#1\,\textsc{\lowercase{#2}}}}
\newcommand*\element[1][]{%
  \def\aa@element@tr{#1}%
  \aa@element
}

\DeclareSIUnit\parsec{pc} 
\hypersetup{
    colorlinks=true,
    linkcolor=red,
    filecolor=magenta,      
    citecolor=blue
}

\newcommand{\udt}[3]{#1^{#2}_{\phantom{#2}#3}}

\newcommand{\dut}[3]{#1_{#2}^{\phantom{#2}#3}}

\newcommand{\lc}[1]{\accentset{\circ}{#1}}

\title{\texorpdfstring{$f(T)$}{f(T)} cosmology in the regime of quasar observations} 

\author[1]{Rodrigo Sandoval-Orozco,\thanks{\href{rodrigo.sandoval@correo.nucleares.unam.mx}{rodrigo.sandoval@correo.nucleares.unam.mx}}}
\author[1]{Celia Escamilla-Rivera,\thanks{\href{celia.escamilla@nucleares.unam.mx}{celia.escamilla@nucleares.unam.mx}}}
\author[2,3]{Rebecca Briffa, \thanks{\href{rebecca.briffa.16@um.edu.mt}{rebecca.briffa.16@um.edu.mt}}} 
\author[2,3]{\\ Jackson Levi Said, \thanks{\href{jackson.said@um.edu.mt}{jackson.said@um.edu.mt}}}

\affil[1]{Instituto de Ciencias Nucleares, Universidad Nacional Aut\'{o}noma de M\'{e}xico, Circuito Exterior C.U., A.P. 70-543, M\'exico D.F. 04510, M\'{e}xico}
\affil[2]{Institute of Space Sciences and Astronomy, University of Malta, Malta, MSD 2080}
\affil[3]{Department of Physics, University of Malta, Malta}

\begin{document}

\maketitle
\abstract{
The open problems
related to cosmological tensions in current times have opened new paths to study new probes to constrain cosmological parameters in standard and extended cosmologies, in particular, to determine at a local level the value of the Hubble constant $H_0$, through independent techniques. However, while 
standard Cosmological Constant Cold Dark Matter ($\Lambda$CDM) model has been well constrained and parts of extended cosmology have been intensively studied,
the physics behind them aspects restrains our possibilities of selecting the best cosmological model that can show a significant difference from the first model.
Therefore, to explore a possible deviation from a such model that can explain the current discrepancy on the $H_0$ value, in this work we consider adding the current local observables, e.g. Supernovae Type Ia (SNIa), $H(z)$ measurements, and Baryon Acoustic Observations (BAO) combined with two new calibrated Quasars (QSO) datasets using ultraviolet, x-ray and optical plane techniques. While these can be identified as part of the high-redshift standard candle objects, the main characteristics of these are based on fluxes distributions calibrated up to $z \sim 7 $. We consider five $H_0$ prior scenarios to develop these calibrations.
Furthermore, we found that our estimations provide 
the possibility to relax the $H_0$ tension at 2$\sigma$ using a QSO ultraviolet sample in combination with late measurements showing higher values of $H_0$.
Our results can be an initial start for more serious treatments in the quasars physics from ultraviolet, x-ray, and optical plane techniques behind the local observations as cosmological probes to relax the cosmological tensions problems.
}

\section{\label{sec:intro}Introduction}

The $\Lambda$ Cold Dark Matter ($\Lambda$CDM) model \cite{peebles1984} 
has been the most successful model up to the last few years
cosmological backbone. Its theoretical simplicity and phenomenological characteristic allow it to be constrained successfully with early \cite{planck2018} and late times \cite{riess1998,perlmutter1999} observations. Moreover, its theoretical structure in a spatially flat 
geometric, and homogeneous and isotropic description relies on the existence of cold dark matter \cite{Primack:2015kpa,Lisanti:2016jxe}, which should allow the structure formation and a Cosmological Constant ($\Lambda$) associated with dark energy \cite{caroll-2001}, which should drive the late time cosmic acceleration. 

However, the fundamental nature and the properties of the dark sector are still unknown. From exploring a possible particle (or particles) to explain dark matter effects, up to explaining the fine-tuning issue on the Cosmological Constant $\Lambda$, have made the $\Lambda$CDM model a scenario that needs to be tested with several observational surveys.

Recent studies related to the flat geometry assumed in this model have brought the possibility to consider a spatially non-flat Universe \cite{Efstathiou:2020wem,DiValentino:2019qzk,Bargiacchi:2021hdp}. While this standard cosmological model can accept non-flat schemes, its simple theoretical characteristic is now bent. Additionally, the constraint analyses using Cosmic Microwave Background (CMB) data can raise significant statistical interpretations on the geometry, which would imply relevant consequences in the understanding of cosmic evolution. Furthermore, the effectiveness of this model has come into question with the appearance of statistical tensions between early and late time surveys. The so-called $S_8$ and $H_0$ tensions \cite{Abdalla:2022yfr} have opened a wide door to propose new gravity theories and cosmological models to alleviate all the mentioned issues, along with a good agreement with the observational methodology employed to constraint the cosmological parameters of interest. 

From the theoretical perspective, there have been proposals for alternative theories of gravity from fundamental setups \cite{Sotiriou:2008rp,Clifton:2011jh}. Their main characteristics are based on extra terms in the Einstein-Hilbert action associated with the curvature and its Levi-Civita connection. Moreover, another setup in the direction of extended theories of gravity considers teleparallel torsion rather than curvature as a mechanism to communicate with the gravitational field \cite{Aldrovandi:2013wha,Bahamonde:2021gfp}. 
In the latter case, Teleparallel gravity (TG) has been studied since it produces a scalar $T$ which is equal to the Ricci scalar. Under this relation, we can have the action in a linear form of the $T$, which is dynamically equivalent to General Relativity (GR). This is the so-called Teleparallel Equivalent of General Relativity (TEGR). Furthermore, TEGR can be generalised to a form of $f(T)$ gravity \cite{Ferraro:2006jd,Ferraro:2008ey,Bengochea:2008gz,Linder:2010py,Chen:2010va,Bahamonde:2019zea, RezaeiAkbarieh:2018ijw,Farrugia:2016pjh,Cai:2015emx}, which has been well-constrained at astrophysical \cite{Finch:2018gkh} and cosmological level \cite{Nunes:2018evm}.

From the observational analysis perspective, statistically significant deviations from the standard $\Lambda$CDM have already been analysed using SNIa, $H(z)$ measurements, 
and Baryon Acoustic Observations (BAO) \cite{Briffa:2021nxg}. However, there is a strong restriction on the priors considered for $H_0$ giving lower values of the matter density parameter $\Omega_m$. 
At high redshifts, $f(T)$ cosmologies have been constrained through cosmography by considering non-flat and flat geometries \cite{Shabani:2023xfn} with a gamma-ray bursts (GRB) observables, using quasars (QSO) objects \cite{Sabiee:2022iyo} detected through high-quality UV and X-ray fluxes up to $z\sim 5.1$ \cite{Lusso:2019akb}. Also, some studies use quasars as standard rulers \cite{Sabiee:2022iyo} by its angular size–luminosity using very-long-baseline interferometry \cite{1985AJ.....90.1599P}.
However, these efforts show analogous results compared with the standard $\Lambda$CDM predictions and are well consistent with the observational data from the Hubble diagrams of each observable. A relevant point to discuss in this analysis is the method used behind the use of quasars as standard candles. This requirement sets us to a fixed information on $H_0$ priors in each scenario.

Furthermore, several $f(T)$ cosmological models have been tested at background level giving consistent results a high value of the Hubble constant \cite{Briffa:2021nxg}. Recently, there have been studies that includes the ${\rm Pantheon}^{+}$ catalog \cite{Briffa:2023ern,Briffa:2020qli,Cai:2019bdh,Ren:2022aeo}. At early times, some works incorporate Big Bang Nucleosynthesis (BNN) constraints \cite{Benetti:2020hxp}, or CMB constraints \cite{Nunes:2018evm,Kumar:2022nvf,Nunes:2018xbm}. At astrophysical scales, constraints from primordial black holes (BH) in these models seems consistent with independent cosmologies \cite{Papanikolaou:2022hkg}.

As a step further, in this paper, we discuss the treatment of two different quasar samples: the xA sample \cite{negrete2018highly} and the nUVX sample \cite{Lusso:2020pdb}. On one hand, the first quasar sample is based on the 4D Eigenvector 1 empirical formalism (4DE1) formalism to locate $\sim 250$ extreme accretors, or Active Galactic Nuclei (AGNs) galaxies with the highest accretion rates. These objects can be considered standard candles using the Eddington luminosity and could bring extra information to the SNIa trend \cite{Dultzin2020-xA}. 

On the other hand, the nUVX sample is based on the empirical relation between the UV and X-ray emission of $\sim 2500$ quasars fitted through different redshifts $z$ to obtain a relation that could allow us to determine their distance in a model-independent manner. 
As we can notice, the two QSO samples have different formalisms, therefore we will analyse them separately for the selected cosmological models along with standard baseline data sets. 
For example, for the standardization of these sources as cosmological candles in a joint analysis, e.g. SNIa + QSO, we need to consider the observed non-linear relation between the ultraviolet and the X-ray luminosity in QSOs \cite{2010A&A...512A..34L}. This method extends the empirical distance ladder used for the SNIa to have access to higher redshifts.

According to the latter characteristics, we will describe key points on how the methodology employed for quasars can be improved to constraint and found \textit{true} deviations from standard cosmologies.

We discuss each step comprehensively to show how these observables could play a significant role in the statistical credibility of the constraints.  Furthermore, we consider adding the current local observables (we denote these as \textit{baseline} sample) and two new calibrated QSO datasets using ultraviolet, x-ray, and optical plane techniques up to $z \sim 7 $. We extend the analysis by considering five $H_0$ prior scenarios to develop the calibrations of the QSO. Our methodology
presents new constraints using a baseline constructed with SN + $H(z)$ \& BAO, and two newly calibrated QSO datasets along the latter. Both baselines (SN + $H(z)$ + BAO, SN + $H(z)$ + QSO) are employed to analyse the $f(T)$ cosmologies at higher redshifts. The impact of considering objects as QSO will be fundamental to study if there is any possible deviation from the standard $\Lambda$CDM model, and if this deviation can relax the current statistical tension on $H_0$.

This paper is divided as follows:
In Sec.~\ref{sec:introfT} we summarise the TG background theory and the most promising $f(T)$ cosmologies available in the literature. All of these models are described through their normalised $E(z)$ Friedmann evolution equation. Furthermore, we are going to consider a standard $w$CDM model in addition to the four $f(T)$ cosmologies to proceed with comparisons between them.
In Sec.~\ref{sec:obs_data} we present the methodology employed for the baseline datasets. We divided this discussion into local and quasar measurements, including two samples through ultraviolet, x-ray, and optical plane techniques.
Our results on new constraints are developed in Sec.~\ref{sec:results-baseline}.
Finally, the discussion is presented in Sec.~\ref{sec:conc}.

\section{Teleparallel Cosmology and its $f(T)$ models} 
\label{sec:introfT}

We can characterise TG as the interchange between the curvature Levi-Civita connection $\udt{\lc{\Gamma}}{\sigma}{\mu\nu}$ with the teleparallel connection $\udt{\Gamma}{\sigma}{\mu\nu}$ \cite{Hayashi:1979qx,Aldrovandi:2013wha}. Notice that we consider over-circle quantities as the objects determined using the Levi-Civita connection. Under this definition, all curvature-inspired geometric bodies can vanish through the calculations using this connection \cite{Krssak:2018ywd,Bahamonde:2021gfp,Aldrovandi:2013wha}. 
In this scheme, TG can be expressed through the tetrad $\udt{e}{A}{\mu}$, its inverses $\dut{E}{A}{\mu}$), and a spin connection define by $\udt{\omega}{A}{B\mu}$, where we identify Latin indices as coordinates on the tangent space and Greek indices denotes coordinates on the manifold. Furthermore,
the tetrad can be constructed using the standard metric  $g_{\mu\nu} = \udt{e}{A}{\mu}\udt{e}{B}{\nu}\eta_{AB}$, where $\eta_{AB} = \dut{E}{A}{\mu}\dut{E}{B}{\nu}g_{\mu\nu}$.
Notice that as with the metric, the tetrad fulfills orthogonality conditions  $\udt{e}{A}{\mu}\dut{E}{B}{\mu}=\delta^A_B$, and $\udt{e}{A}{\mu}\dut{E}{A}{\nu}=\delta^{\nu}_{\mu}$.

To have local Lorentz transformation invariance in the field equations, the spin connection $\udt{\omega}{A}{B\mu}$ must be flat. The connection between the tetrad and this spin connection can be defined through \cite{Weitzenbock1923}
\begin{equation}
    \udt{\Gamma}{\sigma}{\nu\mu} := \dut{E}{A}{\sigma}\left(\partial_{\mu}\udt{e}{A}{\nu} + \udt{\omega}{A}{B\mu}\udt{e}{B}{\nu}\right)\,,
\end{equation}
where the torsion tensor can be written from the teleparallel connection and its antisymmetric operator as $\udt{T}{\sigma}{\mu\nu} \equiv 2\udt{\Gamma}{\sigma}{[\nu\mu]}$ \cite{Hayashi:1979qx}, where its contraction can be expressed as \cite{Krssak:2018ywd,Bahamonde:2021gfp}
\begin{equation}
    T=\frac{1}{4}\udt{T}{\alpha}{\mu\nu}\dut{T}{\alpha}{\mu\nu} + \frac{1}{2}\udt{T}{\alpha}{\mu\nu}\udt{T}{\nu\mu}{\alpha} - \udt{T}{\alpha}{\mu\alpha}\udt{T}{\beta\mu}{\beta}\,,
\end{equation}
and the teleparallel equivalent (TEGR) action is denoted by a linear Lagrangian form of the torsion scalar considering $ R=\lc{R} + T - B = 0,$ \cite{Bahamonde:2015zma,Farrugia:2016qqe}
with $R\equiv0$ denoting a curvature-less teleparallel connection and $\lc{R} \neq 0$. The boundary term denoted by $B$ is a total divergence term. 

If we parametrise an arbitrary $\tilde{f}(T) = -T +f(T)$ gravity we can write a modified version of a TEGR action of the form
\cite{Linder:2010py,Chen:2010va,RezaeiAkbarieh:2018ijw} through the action
\begin{equation}
\label{f_T_ext_Lagran}
    \mathcal{S}_{f(T)}^{} =  \frac{1}{2\kappa^2}\int \mathrm{d}^4 x\; e\left[-T + f(T)\right] + \int \mathrm{d}^4 x\; e\mathcal{L}_{\text{m}}\,,
\end{equation}
with $e=\det\left(\udt{e}{a}{\mu}\right)=\sqrt{-g}$ as the tetrad determinant,
$\kappa^2=8\pi G$, and $\mathcal{L}_{\text{m}}$ denotes the matter Lagrangian contribution. 
Cases where $f(T) \rightarrow 0$ can bring healthy scenarios, while the standard $\Lambda$CDM model can be recovered when this function is equal to a constant, e.g. $\Lambda$. 

To analyse a flat homogeneous and isotropic cosmology we should consider the tetrad \cite{Krssak:2015oua,Tamanini:2012hg}
\begin{equation}
    \udt{e}{A}{\mu} = \text{diag}\left(1,\,a(t),\,a(t),\,a(t)\right)\,,
\end{equation}
where $a(t)$ is the scale factor in cosmic time $t$. Notice that a standard flat Friedmann--Lema\^{i}tre--Robertson--Walker (FLRW) metric is recovered using the relation between the metric and the tetrad so that the line element in cartesian coordinates is \cite{misner1973gravitation}
\begin{equation}\label{FLRW_metric}
     \mathrm{d}s^2 = \mathrm{d}t^2 - a^2(t) \left(\mathrm{d}x^2+\mathrm{d}y^2+\mathrm{d}z^2\right)\,,
\end{equation}
where $T = -6 H^2$ and $B = -6\left(3H^2 + \dot{H}\right)$. Finally, the $f(T)$ gravity Friedmann equations are given by \cite{Bahamonde:2021gfp}
\begin{align}
    H^2 + \frac{T}{3}f_T - \frac{f}{6} &= \frac{\kappa^2}{3}\rho\,,\label{eq:Friedmann_1}\\
    \dot{H}\left(1 - f_T - 2Tf_{TT}\right) &= -\frac{\kappa^2}{2} \left(\rho + p \right)\label{eq:Friedmann_2}\,,
\end{align}
where we can define the Hubble parameter as $H=\dot{a}/a$, and the over-dots denotes derivatives with respect to $t$, and  
the energy density and pressure of the total matter contribution are by $\rho$ and $p$, respectively.

Since now we have the Friedmann equation in this TEGR scheme, we can consider $f(T)$ models suitable to cosmological late time constraints to test them with our observables. In this work, we consider the following models:

\begin{itemize}
\item $\Lambda$CDM model.
This is the simplest model that we will use as a test to compare with TG-inspired models. By defining the normalised Hubble parameter $E(z)\equiv H(z)/H_0$, the Friedmann equation for this case is of the form
\begin{equation}
\label{eq:wcdm}
  E^2(z) = \Omega_0 (1+z)^3 + (1-\Omega_m),
\end{equation}
where $\Omega_m$ denotes the fractional density of matter and the subindex $0$ denotes parameters evaluated at current times. 

\item Power Law Model -- \texorpdfstring{$f_1(T)$}{} Model.
This model was first studied in \cite{Bengochea:2008gz} due to its ability to reproduce the late-time cosmic acceleration behaviour in the $f(T)$ scheme. We can describe it through
\begin{equation}
\label{eq:f1}
    f_1 (T) = \left(-T\right)^{b_1}\,,
\end{equation}
where  $b_1$ is a constant. We can write the Friedmann equation for this model as
\begin{equation}
    E^2(z) = \Omega_{m} \left(1+z\right)^3 + \Omega_{r}\left(1+z\right)^4 + \left(1 - \Omega_{m} - \Omega_{r}\right) E^{2b_1}(z)\,,
\end{equation}
which recovers $\Lambda$CDM model for $b_1 = 0$. 
For $b_1 = 1$, the extra component in the Friedmann equation gives a re-scaled gravitational constant term in the density parameters, i.e. the GR limit. Also, we can obtain an upper bound such that $b_1 < 1$ for an accelerating Universe.

\item Linder Model.

This model was proposed to produce late-time accelerated expansion through \cite{Linder:2010py} 
\begin{equation}\label{eq:f2}
   f_2 (T) =  T_0 \left(1 - \text{Exp}\left[-b_2\sqrt{T/T_0}\right]\right)\,,
\end{equation}
where $b_2$ is a constant and $T_0 = T\vert_{t=t_0} = -6H_0^2$.
The corresponding Friedmann equation for this model can be written as
\begin{equation}
    E^2\left(z\right) = \Omega_{m} \left(1+z\right)^3 + \Omega_{r}\left(1+z\right)^4 + \frac{1 - \Omega_{m} - \Omega_{r}}{(b_2 + 1)e^{-b_2} - 1} \left[\left(1 + b_2 E(z)\right) \text{Exp}\left[-b_2 E(z)\right] - 1\right]\,,
\end{equation}
which reduces to $\Lambda$CDM as $b_2 \rightarrow +\infty$. 

\item Variant Linder Model -- \texorpdfstring{$f_3(T)$}{} Model.
A variant version of the latter model can be described by \cite{Nesseris:2013jea}
\begin{equation}\label{eq:f3}
    f_3 (T) = T_0\left(1 - \text{Exp}\left[-b_3 T/T_0\right]\right)\,,
\end{equation}
where $b_3$ is constant. The Friedmann equation for this model can be written as
\begin{equation}
    E^2\left(z\right) = \Omega_{m} \left(1+z\right)^3 + \Omega_{r}\left(1+z\right)^4 + \frac{1 - \Omega_{m} - \Omega_{r}}{(1 + 2b_3 )e^{-b_3} - 1} \left[\left(1 + 2b_3 E^2 (z)\right)\text{Exp}\left[-b_3 E^2 (z)\right] - 1\right]\,,
\end{equation}
which goes to $\Lambda$CDM as $b_3 \rightarrow +\infty$ similar to $f_2$CDM. 

\item Logarithmic Model -- \texorpdfstring{$f_4(T)$}{} Model.
Proposed in \cite{Bamba:2010wb}, this model is described by
\begin{equation}\label{eq:f4}
    f_4 (T) =  T_0 \sqrt{\frac{T}{b_4 T_0}} \log\left[\frac{b_4 T_0}{T}\right]\,,
\end{equation}
where $b_4$ is a constant. The Friedmann equation for this form is 
\begin{equation}
    E^2\left(z\right) = \Omega_{m} \left(1+z\right)^3 + \Omega_{r}\left(1+z\right)^4 + \left(1 - \Omega_{m} - \Omega_{r} \right) E(z)\,.
\end{equation}
Notice that this expression does not feature $b_4$, therefore the background behaviour of this model is intriguing because it cannot feature confirmation of any bias with the standard $\Lambda$CDM.
\end{itemize}


\section{\label{sec:obs_data}Observational data treatment}

In this analysis, we consider the four 
$f(T)$ models described, which have been analysed in previous works \cite{Briffa:2021nxg} and give well-consistent constraints using local Universe measurements. Each $f(T)$ cosmological model will be tested using the constraining parameters method through MCMC (Monte Carlo
Markov Chain) analysis using the publicly available \footnote{\href{https://emcee.readthedocs.io/en/stable/}{emcee.readthedocs.io}} for our cosmology and the baseline (or base for further reference) and the extract of constraints using \texttt{GetDist}\footnote{\href{https://getdist.readthedocs.io/en/latest/}{getdist.readthedocs.io}}. The baseline contains the $H(z)$ measurements, Supernovae Type Ia (SNIa) data set and BAO measurements.
As a step forward, we will focus on using two kinds of quasars datasets using ultraviolet, x-ray, and optical plane techniques.

For our analyses, we used different priors on $H_0$ to analyse the behaviour of different models and data sets. These are reported in Table \ref{tab:priors}. The following priors are used: the estimation of the Hubble constant by SH0ES team $H_0 = 73.3 \pm 1.04$ km s$^{-1}$ Mpc$^{-1}$ using SNIa and Cepheid calibrations \cite{Riess:2021jrx}. We will call this R21 in our analysis. The GAIA Early Data Release 3 using Cepheid stars to calibrate $H_0 = 74.03 \pm 1.42$ km s$^{-1}$ Mpc$^{-1}$ \cite{Abdalla:2022yfr}. The calibration of the constant using the Tip of the Red Giant Branch (TRGB) as a standard candle with $H_0 = 69.8 \pm 0.8$ km s$^{-1}$ Mpc$^{-1}$ \cite{Freedman:2019jwv} denoted by F20. We should mention that in \cite{Anderson:2023aga} it was reported a higher value of $H_0 =71.8\pm 1.5$ km s$^{-1}$ Mpc$^{-1}$, however we keep our current study with the F20.

The indirect measurement of $H_0 = 67.36 \pm 0.54$ km s$^{-1}$ Mpc$^{-1}$ by the Planck Collaboration \cite{Planck:2018vyg} using TT+TE+EE+lowE+Lensing denoted by P18. Finally, the indirect estimation of the Hubble constant using an independent probe with Atacama Cosmology Telescope (ACT) $H_0 = 67.9\pm 1.5$ km s$^{-1}$ Mpc$^{-1}$ \cite{ACT:2020gnv}.


\subsection{Local measurements}

\begin{itemize}
\item $H(z)$ measurements. 
Data in which the Hubble function is calculated through analyses of different galactic spectra in a redshift range from $z = 0$ to $z \sim 2$, used for detecting very small redshift differences for two galaxies in a cluster that formed at the same time \cite{Moresco:2016mzx}. 
These estimations can be used to calculate $\Delta z/\Delta t$ which allows us to have an estimate of $H(z)$. 
\textit{In this case we tested a newly calculated covariance matrix for the data}\footnote{\href{https://gitlab.com/mmoresco/CCcovariance}{gitlab.com/mmoresco/CCcovariance}}. 
The corresponding $\chi^2_{H(z)}$ is given by: 
\begin{equation}
    \chi^2_{H(z)} = \Delta H(z_i,\Theta)^T C^{-1}_{H(z)} \Delta H(z_i,\Theta), 
\end{equation}
where $\Delta H(z_i,\Theta) = H(z_i,\Theta) - H_{obs}(z_i)$ and $ C^{-1}_{H(z)}$ is the covariance matrix generated.  
\item Pantheon SNIa dataset. 
We use the 1048 data points provided by the \textit{Pantheon} collaboration \cite{Scolnic:2017caz} that measure the apparent distance for several SNIa events in a redshift range $0.01 < z < 2.3$. The dataset for the Pantheon sample provides SN magnitudes corrected for the stretch and colour effects along with the maximum brightness, the mass of the host galaxy, and sky position bias, so to obtain a cosmological useful quantity we need to calculate the distance modulus $\mu = m - M$, where $M$ is the absolute magnitude that is considered as free parameter in our analyses \cite{Alestas:2020mvb}. Thus, the $\chi^2_{\text{SN}}$  for the Pantheon sample is given by
\begin{equation}
    \chi^2_{\text{SN}} = \Delta\mu(z_i,\Theta)^TC^{-1}_{\text{SN}} \Delta \mu(z_i,\Theta) + \ln\left(\frac{S}{2\pi}\right) - \frac{k^2(\Theta)}{S},
\end{equation}
where $C^{-1}_{\text{SN}}$ is the total covariance matrix for the data, $S$ is the sum of all components of the inverse of the matrix and $k(\Theta) = \Delta\mu(z_i,\Theta)^TC^{-1}_{\text{SN}} $, using $\Delta\mu(z_i,\Theta) = \mu(z_i,\Theta) - \mu_{\text{obs}}(z_i)$. 

In this case, the distance modulus $\mu(z)$ can be calculated as: 
\begin{equation}
    \label{eq:distance_modulus}
    \mu(z_i,\Theta) = 5\log[D_L(z_i,\Theta)] + 25, 
\end{equation}
and where $D_L(z_i,\Theta)$ is the luminosity distance given as: 
\begin{equation}
    D_L(z_i,\Theta) = c(1+z_i)\int_0^{z_i} \frac{dz'}{H(z',\Theta)},
\end{equation}
where $c$ is the speed of light and $H(z_i,\Theta)$ is the Hubble parameter. So, to calculate the apparent magnitude $m$ for the SNIa we will use:

\begin{equation}
    m = \mu(z_i, \Theta) + M
\end{equation}
where the distance modulus $\mu(z_i,M)$ is computed as Eq. \eqref{eq:distance_modulus}. We will be referring to Pantheon SNIa simply as SNIa from now on. 
\item Baryon acoustic oscillation (BAO).
For this observable, we will consider the following independent surveys: 
\begin{enumerate}[label=(\alph*)]
    \item The result obtained from six-degree Field Galaxy Survey measurement (6dFGS) at $z = 0.106$ \cite{2011MNRAS.416.3017B}.
    \item The result from Sloan Digital Sky Survey (SDSS) Main Galaxy Sample Measurement (MGS) from Data Release 7 (DR7) at $z =0.15$ \cite{Ross:2014qpa}. 
    \item The data points from Baryon Oscillation from Spectroscopic Survey (BOSS) DR12 at $z = 0.38, 0.51, 0.61, 1.52$ \cite{Alam:2016hwk}. 
    \item The data points from the BAO measurements from SDSS DR14 at $z=0.978,1.23$, $1.526,1.944$ \cite{Zhao:2018gvb}. 
\end{enumerate}
To use these datasets we need to compute different cosmological quantities such as the averaged distance
\begin{equation}
    D_V(z) = \left[\frac{cz}{H(z)} \frac{D_L^2(z)}{(1+z)^2} \right]^{1/3},
\end{equation}
the comoving sound horizon at the baryon drag epoch
\begin{equation}
    r_s(z_d) = \int_{z_d}^\infty \frac{c_s(z')}{H(z')}dz', 
\end{equation} 
and the fixed value for a fiducial cosmology through $r_{s,\text{fid}}$. In the case the SDSS DR7 data point we consider $r_{s,\text{fid}} = 148.69$ Mpc \cite{Ross:2014qpa}. For the other data surveys described we use the value $r_{s,\text{fid}} = 147.78$ Mpc \cite{Alam:2016hwk,Zhao:2018gvb}.

Also, the sound horizon can be obtained by
\begin{equation}
    c_s(z) = \frac{c}{\sqrt{3\left[1 + \frac{3\Omega_{b}}{4\Omega_\gamma} \frac{1}{1+z} \right]}}.
\end{equation}
Therefore, we require additional calculations like the redshift of the baryon drag epoch $z_d$ \cite{Ade:2015xua}: 
\begin{equation}
    z_d = \frac{1291(\Omega_m h^2)^{0.251}}{1+0.659(\Omega_m h^2)^{0.828}}\left[1+b_1(\Omega_b h^2)^{b_2} \right],
\end{equation}
defined as the time at which the baryons are released from the Compton drag given by a weighted integral over the Thomson scattering. Under this assumption we can perform a numerical fit to the recombination results \cite{Hu:1995en} obtaining
\begin{equation}
    b_1 = 0.313(\Omega_m h^2)^{-0.419} \left[1 +0.607(\Omega_m h^2)^{0.674} \right], \quad\text{and}\quad  b_2 = 0.238(\Omega_m h^2)^{0.223}. 
\end{equation}
$D_m(z)$ is simply a quantity related to luminosity distance by $D_m(z) = D_L(z) (1+z)^{-1}$. The calculations need also the dimensionless Hubble constant $h = H_0/100$ km/s/Mpc, the baryon density parameter $\Omega_b$, and its photon counterpart $\Omega_\gamma$ and for the purpose of this work, the quantities $\Omega_b h^2 = 0.0224$, $\Omega_\gamma h^2 = 2.469 \times 10^{-5}$ are fixed in agreement with \cite{Planck:2018vyg}. So, for the fit of the BAO datasets the corresponding $\chi^2_{\text{BAO}}(\Theta)$ will be defined as: 
\begin{equation}
    \chi^2_{\text{BAO}} = \Delta \Xi(z_i,\Theta)^{T} C_{\text{BAO}}^{-1}\Delta \Xi(z_i,\Theta),
\end{equation}
where $\Delta \Xi(z_i,\Theta) = \Xi_{\text{obs}}(z_i,\Theta) - \Xi(z_i)$ and $C_{\text{BAO}}$ is the covariance matrix for the considered observations, including the SDSS DR12 and the SDSS DR14 that have correlation among the observations and therefore we need a correlation matrix \cite{Zhao:2018gvb,Alam:2016hwk}.    
However, notice that BAO measures $D_V/r_{\text{drag}}$, therefore to obtain a lower value of $r_{\text{drag}}$ we require a higher value of $H_0$. This aspect implies that we should expect a different early physics to reduce this value \cite{Evslin:2017qdn}.
\end{itemize}


\subsection{Quasars measurements}
\label{sec:qso_sec}

Quasars are one of the most luminous energy sources in the Universe, therefore their use at cosmological scales can make the most on studies at higher redshifts, e.g. up to $z \sim 7$ \cite{Lusso:2020pdb}. This is a key aspect to exploring possible different models that can be indistinguishable at low $z$. 
Although, there are some examples of its cosmological use such as the reverberation mapping technique \cite{WatsonReverberation2011} or the relationship between variability in X-ray amplitude and Black Hole mass scatter of these objects \cite{LaFranca2014}. However, this analysis 
remains at very high $z$ and the samples are only applicable to a certain redshift range due to their strong dependency on the technique used to determine the data points. In this line of thought, there is a lack of a clear definition of the quasar to order the diversity of AGN objects \cite{MartinezAldama2018}.  

We should mention that several QSO catalogs have been used to constraints standard cosmological models under certain theoretical assumptions, e.g. in \cite{Dainotti:2022bzg} was presented a feasibility Gamma-ray Bursts (GRB) with QSO cosmology analysis for future landscapes. Also, in \cite{Bargiacchi:2023jse} QSO catalogs were combined with GRB observations achieving small uncertainties on some cosmological parameters. Moreover, in \cite{Lenart:2022nip} was combined the Risaliti-Lusso relation for QSO with SNIa finding that in the case of non calibrated QSOs it can be corrected the redshift evolution as function of cosmology. Furthermore, it was been showed that through Risaliti-Lusso relation the $\Omega_m$ value seems larger when only it is considered QSO baselines \cite{Yang:2019vgk,Khadka:2020whe,Khadka:2020vlh,Khadka:2020tlm,Colgain:2022nlb}, and some systematics in the QSOs measurements should be expected in future new cosmological probes \cite{Zajacek:2023qjm}.

\begin{figure}[H]
    \centering
    \includegraphics[scale=0.7]{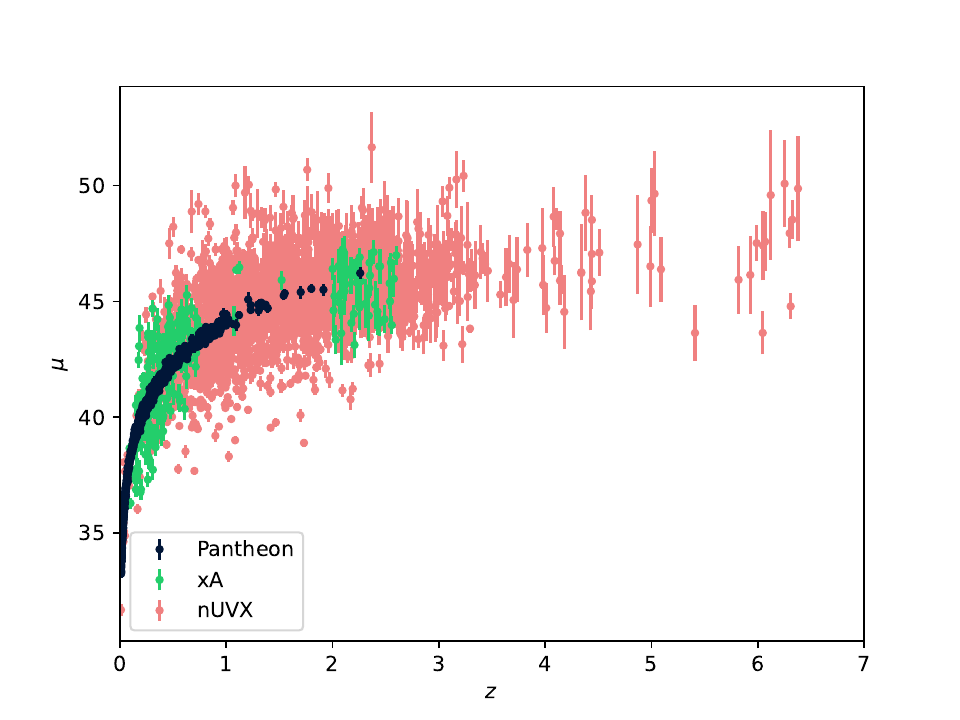}
    \caption{Hubble diagram for the QSO samples described in Sec.~\ref{sec:qso_sec}. The dark blue dots represent the Pantheon data using an $M = -19.3$. The green color points denote the xA sample and the coral color denotes the observed results for the nUVX sample. The $x$-axis represents the redshift $z$ and $y$-axe the distance modulus $\mu(z)$.}
    \label{fig:samples}
\end{figure}

In particular, and to tackle the latter issues, in this work, we use two different quasar samples:
\begin{itemize}
\item \textbf{QSO non-linear UV/X-ray sample} 
\cite{Lusso:2020pdb}. For this sample, we denote non-linear UV and X-ray samples as nUVX. 
This data sample includes 2421 selected objects from the Sloan Digital Sky Survey Data Release 14 (SDSS-DR14) \cite{SDSS-DR14} with other several surveys to make a selection of objects that cover a range from $0 < z \leq 7.54$. The detailed procedures to create the sample are well described in \cite{Lusso:2020pdb}, and references therein. Moreover, here we will only describe the cosmological essentials to use the sample.
The method to threat quasars as cosmological candles is based on the relation between the flux in Ultraviolet Light at 2500 \AA\ and the X-ray flux at 2 keV of objects ($F_{\text{UV}}$-$F_{\text{X}}$). Although the nature of this relation is not fully explained yet in the literature, we can obtain the distance modulus as $\mu = 5\log(d_L) + 25$, and the luminosity distance is written as 

\begin{equation}
    \log(d_L) = \frac{[\log F_{\text{X}} - \gamma F_{\text{UV}}]}{2(\gamma -1)} + \beta',
\end{equation}
where both fluxes are observable quantities and $\gamma$ is the slope of the relation between the fluxes. $\beta'$ is related to the interception of those slopes and we will consider it as a parameter to be fitted. The fluxes relation shows no correlation with redshift and has a $\gamma = 0.702$ and $\delta = 0.21$ fixed for the complete redshift range \cite{Lusso:2020pdb}. 
In this direction, we can calculate the distance modulus as 
\begin{equation}
    \mu = \frac{5}{2(\gamma -1)} (\log F_{\text{X}} - \gamma F_{\text{UV}}) + 5\beta', 
\end{equation}

So, for this sample is used the following $\chi^2_{\text{F}}$: 
\begin{equation}
    \chi^2_{\text{F}} = - \frac12 \sum_i \left(\frac{[\mu_i - \mu(\Theta)]^2}{s_i^2} - \ln s_i^2 \right), 
\end{equation}
where $s_i = dy_i+ \gamma^2 dx_i + \exp(2 \ln \delta)$ and it takes into account the uncertainties for both UV ($x_i$) and X-ray fluxes ($y_i$). $\mu(\Theta)$ is the modeled theoretical distance modulus that can be obtained using the constructed cosmological distance. The data set for this sample contains the fluxes in UV, X, and both their uncertainties for all the selected objects to reconstruct the $\chi^2$. 

\item \textbf{QSO xA sample} 
\cite{Dultzin2020-xA}. 
    This data sample is a selection from $\sim 250$ objects from the SDSS DR-14 \cite{SDSS-DR14} survey up to $z \sim 2.5$. The reason why this sample is smaller is because the selection of these objects requires certain elements in the quasar's spectrum to appear and this selection is not possible for every redshift \cite{Marziani:2018front}. However, to treat them we use the technique so-called Eigenvector 1 of quasars \cite{Dultzin2020-xA}, which is based on the detection of common spectra characteristics. To employ it we require the intensity ratio between the \ion{Fe}{II} blend at $\lambda$4570 and H$\beta$ called $R_{\text{\ion{Fe}{II}}}$, and the full-width half maximum (FWHM) of H$\beta$ to create a relationship between these two parameters to determine quasar types for redshifts up to $z \sim 0.8$. In scenarios where the previous elements do not appear on the spectrum, the lines $\ion{C}{III]}$$\lambda$1909, \ion{Al}{III}$\lambda$1860 and \ion{Si}{III}$\lambda$1892 are useful for $z \sim 2$ because they behave as substitutes for the $R_{\text{\ion{Fe}{II}}}$ \cite{Marziani:2014mnras}. This creates the \emph{optical plane} in which our analysis goes to the so-called xA group; the quasars that have a measured $R_{\text{\ion{Fe}{II}}} > 1 $, and FWHM(H$\beta$) $< 4000$ km/s. 
The xA quasar type has characteristics that make them candidates for standard candles \cite{Dultzin2020-xA}: 
\begin{enumerate}[label=(\alph*)]
    \item Those quasars that radiate near the Eddington limit and, therefore, are very close to a physical limit for the luminosity. This can form a relation between this limit and the black hole mass. 
    \item The black hole mass can be obtained through the virialized relation which depends on the observed FWHM of H$\beta$ in the spectra. 
    \item Using a determined ionization parameter \cite{Marziani:2014mnras, Dultzin2020-xA} we can reach an expression for the luminosity, and the distance modulus will depend solely on the FWHM and the continuum value $f_\lambda \lambda$.  
\end{enumerate}

For a complete description of the used method for this sample, in \cite{Dultzin2020-xA} is presented a complete review, and the references therein show different possible approaches. We start by writing the virial luminosity equation \cite{Marziani:2014mnras} as  
\begin{equation}
    L(\rm FWHM) = 7.88 \times 10^{44} (\rm FWHM)_{1000}^4,
\end{equation}
where the FWHM is expressed in units of 1000 km/s. Using this expression and the H$\beta$ line we can calculate the distance modulus as\cite{negrete2018highly}: 
\begin{equation}
    \mu = 2.5[\log L - \log(f_\lambda \lambda)] - 100.19 + 5\log(1+z),
\end{equation}
where $f_\lambda \lambda$ is the rest-frame flux measured in 5100 \AA\ \cite{Negrete:2017front}. In this work, we combined both samples \cite{Negrete:2017front,Marziani:2018front,negrete2018highly} to build a model-independent baseline with $\sim 250$ objects containing a measure for $\mu$ and $\delta \mu$. The total quasar sample can be denoted using a $\chi_{\text{xA}}^2$ given by 
\begin{equation}
    \chi_{\text{xA}}^2 = -\frac12 \sum_i \left[\frac{(\mu_i - \mu(z_i,\Theta))^2}{\delta \mu_i^2} + \ln(\delta \mu_i^2)\right],
\end{equation}
where $\mu$ and $\delta \mu_i$ are the measurements and their uncertainty, respectively, for every obtained distance modulus $\mu_i(z_i, \Theta)$. In Figure ~\ref{fig:samples} we show the two quasar samples that will complement our Hubble diagram.  
\end{itemize}


\begin{table}[!ht]
    \centering
    \begin{tabular}{|l|l|l|l|l|l|l|l|l|l|}
    \hline
        \textbf{Measurement} & \textbf{$H_0$ [km/s/Mpc]} & \textbf{Reference} \\ \hline
        SH0ES (R21) & $73.3 \pm 1.04$  & \cite{Riess:2021jrx} \\ \hline
        GAIA  & $74.03 \pm 1.42$  & \cite{Abdalla:2022yfr} \\ \hline
        F20 & $69.8 \pm 0.8$  & \cite{Freedman:2019jwv} \\ \hline
        Planck 2018 (P18) & $67.36 \pm 0.54$  & \cite{Planck:2018vyg} \\ \hline
        ACT & $67.9 \pm 1.5$  & \cite{ACT:2020gnv} \\ \hline
    \end{tabular}
    \caption{Priors used to calibrate baseline and QSO samples. The first column denotes the measurements. The second column indicates the $H_0$ values in km/s/Mpc. References for each data are indicated in the last column.}
    \label{tab:priors}
\end{table}

\section{Model constraints using the baselines: SNIa+$H(z)$+BAO \& QSO}
\label{sec:results-baseline}

This section presents the results for the different considerations of the baselines described and the constraints derived for each of the four $f(T)$ models. 
Also, we divided the analyses using first the xA quasar sample, and afterward the quasar sample nUVX. 

As mentioned, the BAO measurements depend strongly on the physics assumed in the early Universe. Due to this fact, there can be a bias in the estimation for the late-time parameters, such as $H_0$. For this reason, we describe separate the results for the samples with and without BAO measurements, i.e. one group of data will use only $H(z)$ and SNIa measurements, and the other will include $H(z)$, SNIa, and BAO. In the case of SN catalog, we leave also the absolute magnitude $M$ as a free value in our analyzes. 

Furthermore, we present the results and their discussions on the cases with a $H(z)$+SNIa+BAO baseline and the two described QSO samples.
We divided the $H(z)$ +SNIa +BAO +QSO analysis into two parts in order not to produce a bias due to the physical assumptions on the calibration performed in the QSO samples. 

In Table \ref{tab:priors} we described the $H_0$ priors required to calibrated our QSO samples. For each case we consider a free parameter vector related with the model at hand, e.g. for $f_1 (T)$ model: $\{\Omega_m, b_1\}$,  for $f_2 (T)$ model: $\{\Omega_m, 1/b_2\}$,  for $f_3 (T)$ model: $\{\Omega_m, b_3\}$, and for $\Lambda$CDM and $f_1 (T)$ models we have one free parameter, $\Omega_m$. Furthermore, we include the calculation of the absolute magnitude $M$ for each baseline in order to compare the possible degeneracy with the priors described in Table \ref{tab:priors}. 

Additionally, the QSO-nUVX sample has an extra nuisance parameter, $\beta'$,  that will be reported in Tables for each case. Along the analysis, we will notice that the QSO-nUVX measurements can reduce significantly the $H_0$ tension at 2-$\sigma$ in all the models, including the standard $\Lambda$CDM model.


\subsection{\texorpdfstring{$\Lambda$}{}CDM model}

\begin{figure}[H]
  \centering
  \includegraphics[width=.49\linewidth]{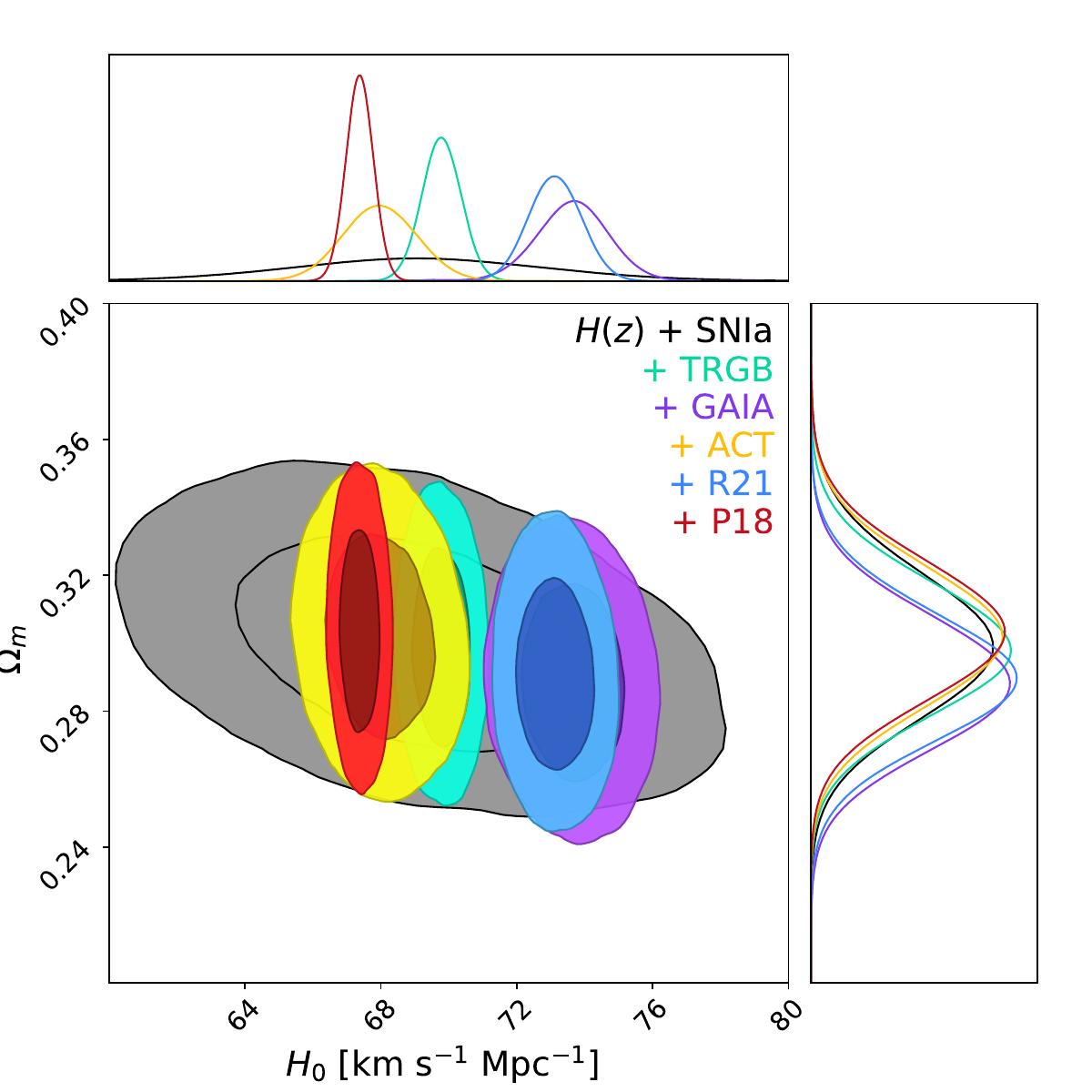}
  \includegraphics[width=.49\linewidth]{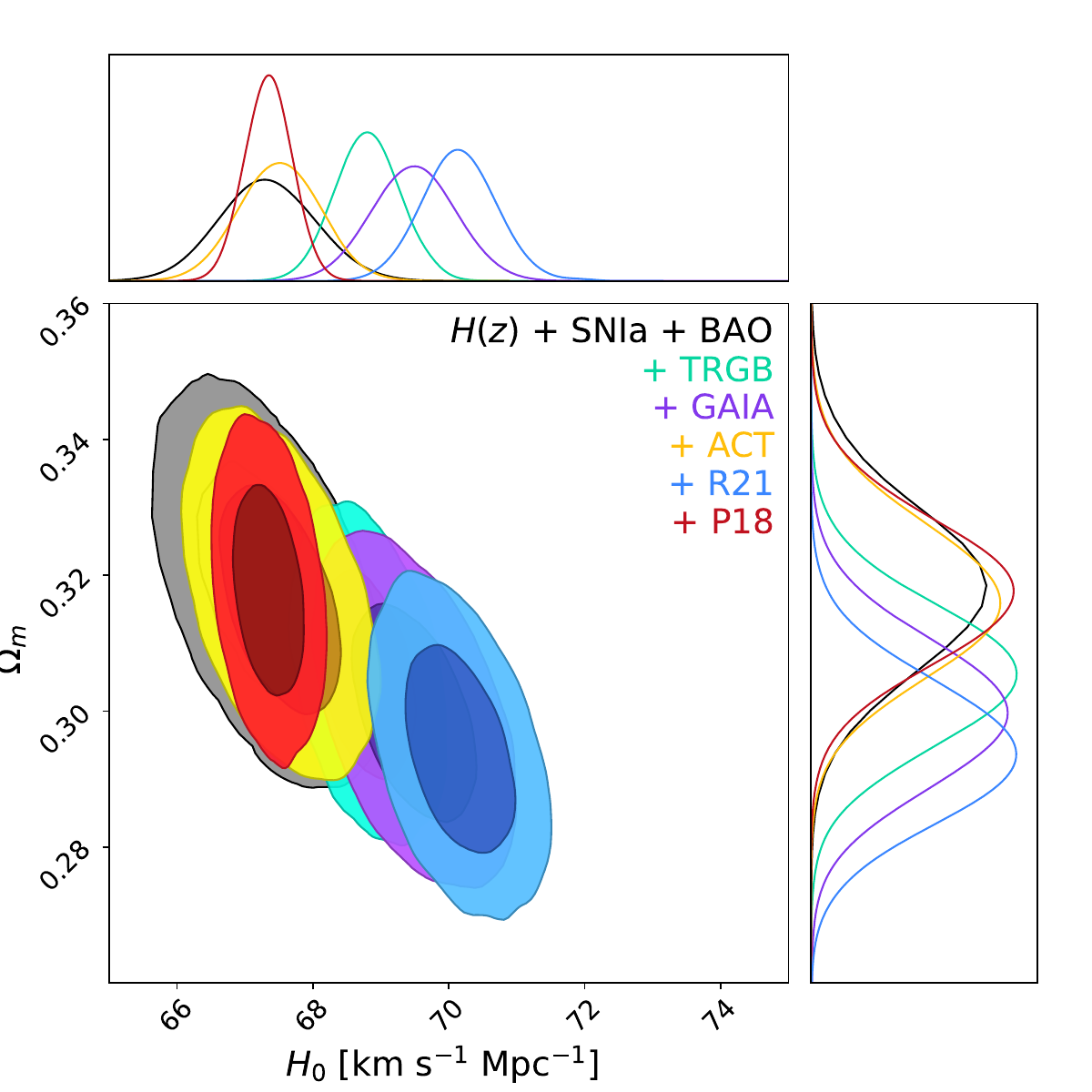}
\caption{1-2$\sigma$ C.L results for the $\Lambda$CDM model using: \textit{Left:} $H(z)$ and Pantheon data sets. \textit{Right:} including BAO. Blue color denotes R21, purple color for GAIA, red color for P18, green color for F20
, and yellow color for ACT priors. Additionally, the model was constrained with the sample baselines without prior, here denoted in black color.}
\label{fig:lcdm}
\end{figure}

The 1-2$\sigma$ C.L. constraints for this model are given in Figure \ref{fig:lcdm}. The results for each of the constrained parameters are given in Table \ref{tab:cc+pn_lcdm} for the baseline proposed $H(z)$+SNIa with and without BAO measurements. We also include the computing absolute magnitude $M$ to compare the possible degeneracy between this parameter and the $H_0$ for each prior consideration.

\begin{figure}[H]
  \centering
   \includegraphics[width=.49\linewidth]{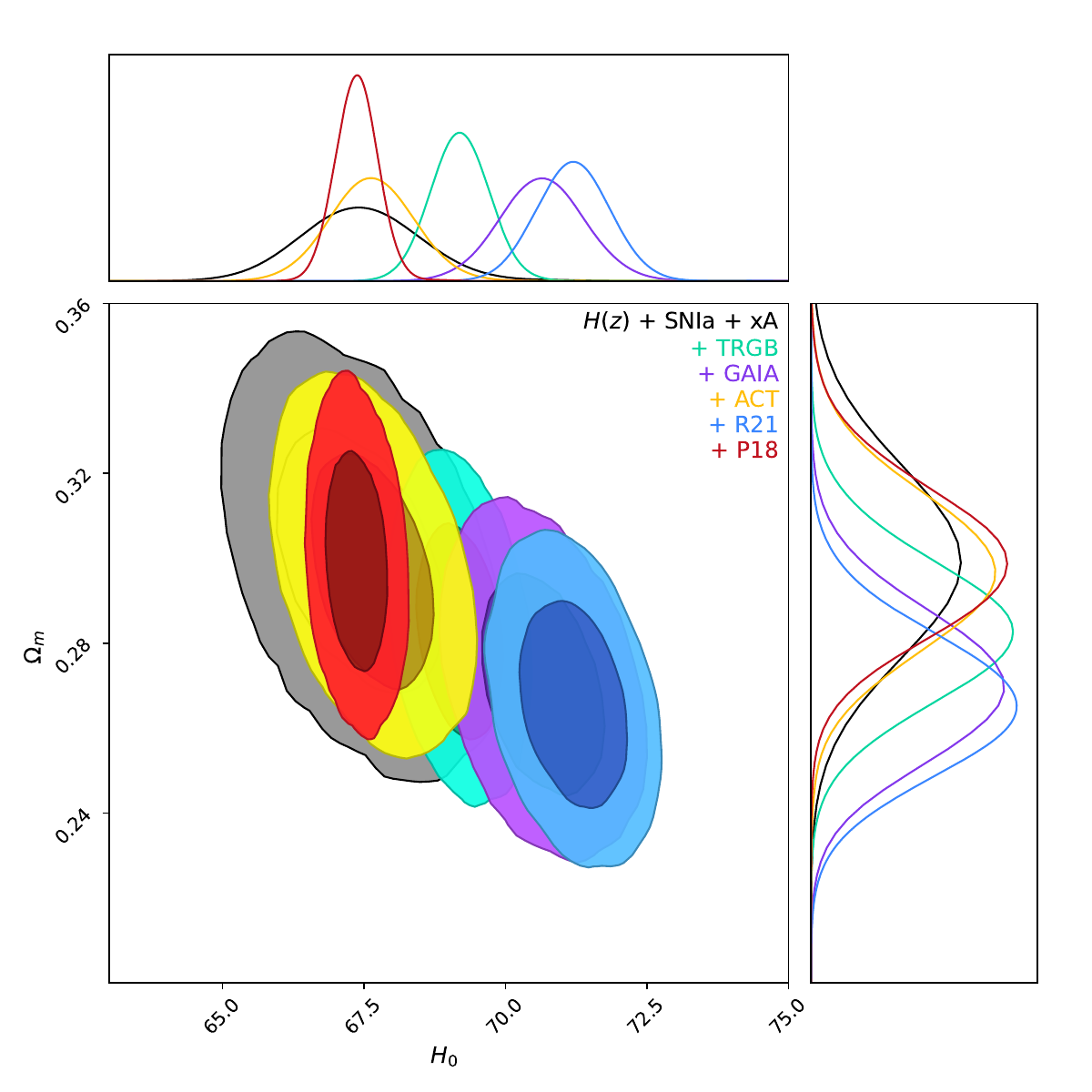}
 \includegraphics[width=.49\linewidth]{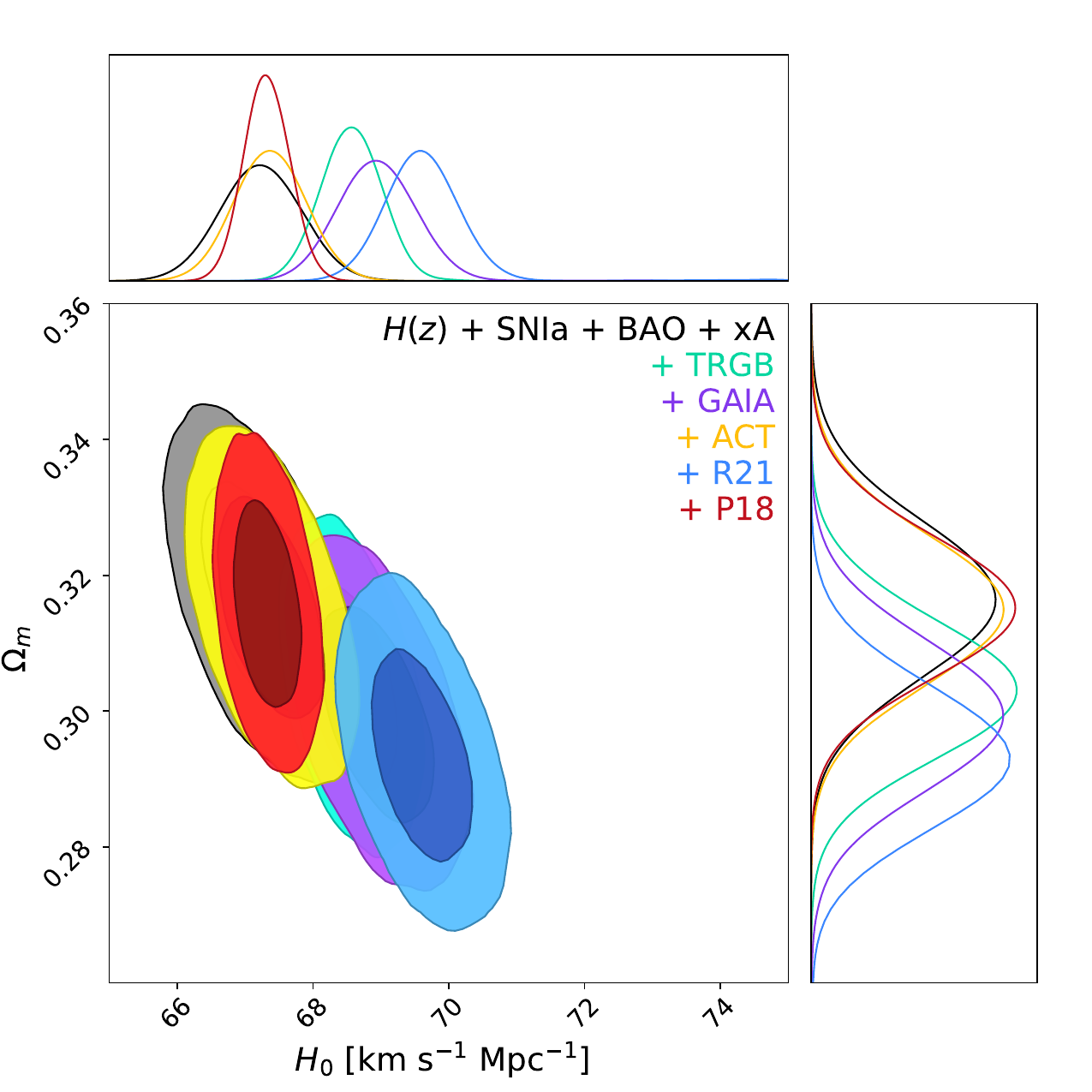}
  \includegraphics[width=.49\linewidth]{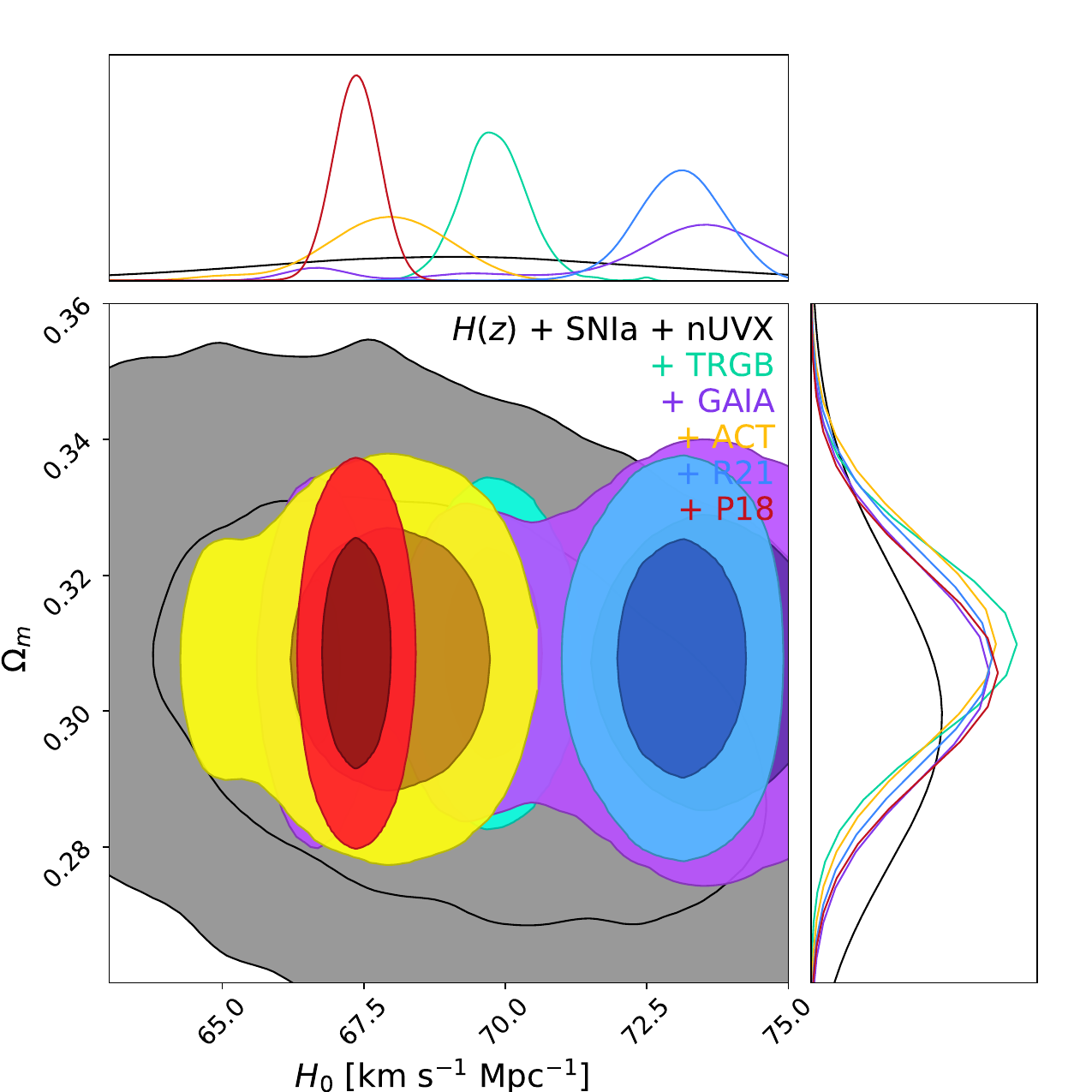}
  \includegraphics[width=.49\linewidth]{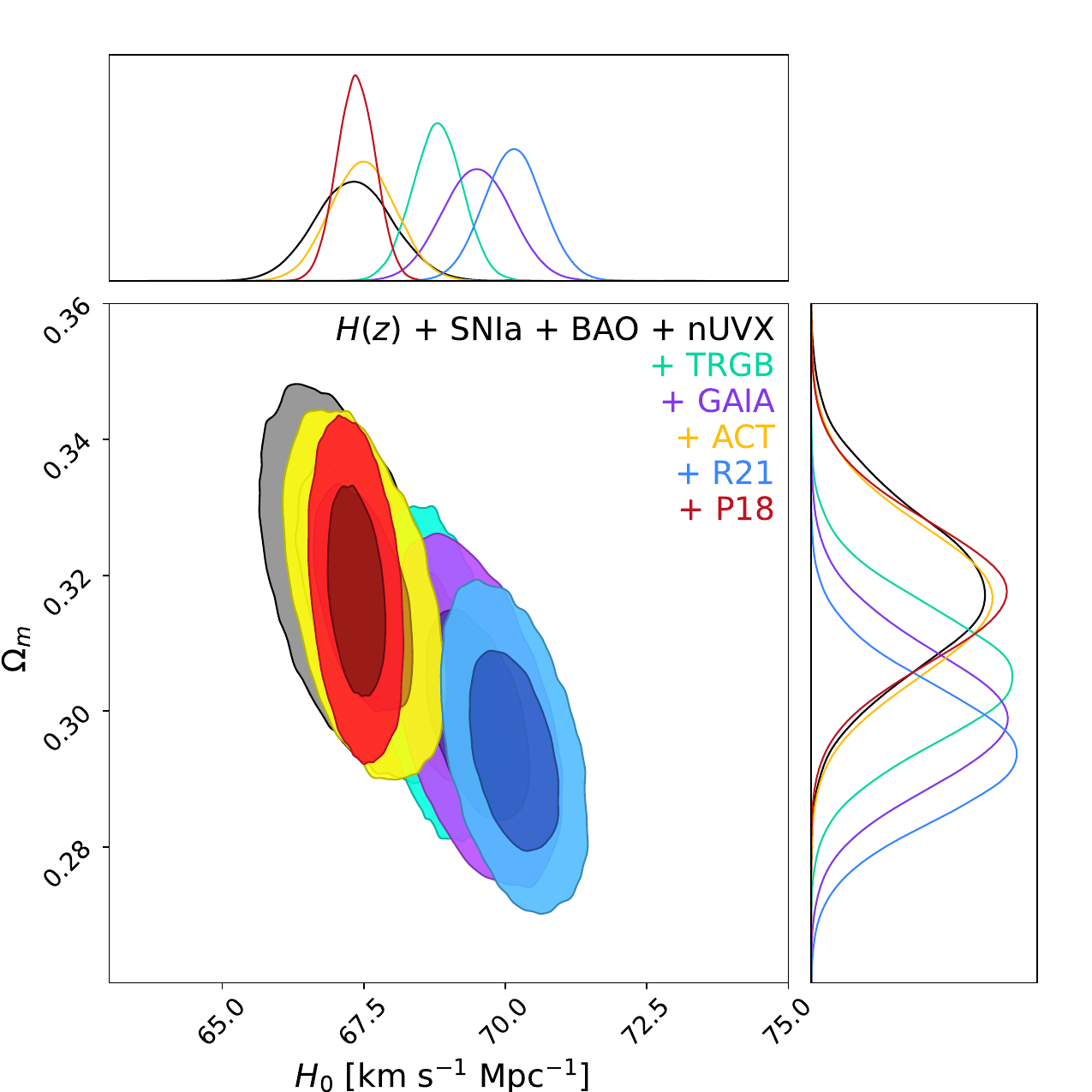}
\caption{1-2$\sigma$ C.L results for the $\Lambda$CDM model using: \textit{Top left:} $H(z)$+SNIa and including the xA sample. \textit{Top right:} $H(z)$+SNIa+BAO and including the xA sample. 
\textit{Bottom left:} $H(z)$+SNIa and including the nUVX sample. \textit{Bottom right:} \textit{Top right:} $H(z)$+SNIa+BAO and including the nUVX sample.
Blue color denotes R21, purple color for GAIA, red color for P18, green color for F20, and yellow color for ACT priors. Additionally, the model was constrained with the sample baselines without prior, here denoted in black color.}
\label{fig:lcdm_qso1}
\end{figure}

Also, the 1-2-$\sigma$ constraints of this model are reported in Figure  \ref{fig:lcdm_qso1}, with their constraints reported in Table \ref{tab:cc+pn+xA_lcdm}  for the QSO-xA sample and in Table \ref{tab:cc+pn+nUVX_lcdm} for the QSO-nUVX sample.

Notice that using BAO data the estimation for $H_0$ has a lower value in comparison with the estimations obtained for the $H(z)$+SNIa measurements. As seen in Table \ref{tab:cc+pn_lcdm}, the estimation without prior is $H_0 = 69.6^{+3.0}_{-4.1}$ km s$^{-1}$ Mpc$^{-1}$, with a significant uncertainty value. 

When using the R21 prior we recover an increase in the value $H_0 = 73.1^{+0.7}_{-0.8}$ km s$^{-1}$ Mpc$^{-1}$, which is consistent with the expected results using high Hubble priors \cite{Riess:2021jrx}. According to this analysis, $\Lambda$CDM model seems to not relax the $H_0$ tension -- as it is expected using the baseline-only data sets -- although the use of BAO sample has the effect of lowering this value to $H_0 = 70.2 \pm 0.5$ km s$^{-1}$ Mpc$^{-1}$ using the same R21 prior. For every combination of the different baselines, the minimum estimated uncertainty is the case obtained using the prior P18, $H_0 = 67.4 \pm 0.4$ km s$^{-1}$ Mpc$^{-1}$ for both data sets with and without BAO. This is another indicator that this measurement relies strongly on a lower $H_0$ value. Regarding fractional matter density $\Omega_m$, the highest value is obtained using the P18 and ACT priors.


\begin{table}[H]
\resizebox{\textwidth}{!}{%
    \centering
    \begin{tabular}{c|cccc}
    \hline 
    \hline & \\[-1.9ex]
    Dataset & $H_0$ [km s$^{-1}$ Mpc$^{-1}$] & $\Omega_m$ & $M$ & $\mathbf{\chi^2_\mathrm{min}}$ \\[0.5ex]
    \hline & \\[-1.8ex]
    $H(z)$ + SNIa & $69.6^{+3.0}_{-4.1}$ & $0.299\pm 0.021$ & $-19.37^{+0.10}_{-0.12}$ & 949.41 \\[0.9ex]
    $H(z)$ + SNIa + R21 & $73.1^{+0.7}_{-0.8}$ & $0.289^{+0.020}_{-0.017}$ & $-19.26\pm 0.02$ & 950.72 \\[0.5ex] 
    $H(z)$ + SNIa + P18 & $67.4 \pm 0.4$ & $0.305^{+0.018}_{-0.020}$ & $-19.43^{+0.02}_{-0.01}$ & 949.65 \\[0.5ex]
     $H(z)$ + SNIa + F20 & $69.8^{+0.6}_{-0.5}$ & $0.298^{+0.020}_{-0.018}$ & $-19.36\pm 0.02$ & 949.45\\[0.5ex]
    $H(z)$ + SNIa + GAIA & $73.6^{+1.0}_{-0.9}$ & $0.289^{+0.018}_{-0.02}$ & $-19.24\pm 0.03$ & 951.15 \\[0.5ex]
    $H(z)$ + SNIa + ACT & $68.0^{+1.0}_{-1.1}$ & $0.301^{+0.021}_{-0.019}$ & $-19.42^{+0.04}_{-0.03}$ & 949.52\\[0.5ex]
   \hline\hline
      $H(z)$ + SNIa + BAO & $67.3^{+0.7}_{-0.6}$ & $0.318\pm 0.013$ & $-19.43\pm 0.02$  & 960.30 \\[0.9ex]
    $H(z)$ + SNIa + BAO + R21 & $70.2\pm 0.5$ & $\left( 293.5^{+10.4}_{-9.6} \right) \times 10^{-3}$ & $-19.35\pm 0.02$  & 995.28 \\[0.5ex] 
    $H(z)$ + SNIa + BAO + P18 & $67.4^{+0.3}_{-0.4}$ & $0.317\pm 0.010$ & $-19.43\pm 0.01$  & 960.29 \\[0.5ex]
    $H(z)$ + SNIa + BAO + F20 & $68.8^{+0.4}_{-0.5}$ & $\left( 305.7^{+9.6}_{-10.3} \right) \times 10^{-3}$ & $-19.38^{+0.01}_{-0.02}$  & 968.01 \\[0.5ex]
    $H(z)$ + SNIa + BAO + GAIA & $69.5\pm 0.6$ & $\left( 299.2^{+10.9}_{-10.0} \right) \times 10^{-3}$ & $-19.37\pm 0.02$  & 998.59 \\[0.5ex]
    $H(z)$ + SNIa + BAO + ACT & $67.5\pm 0.6$ & $\left( 315.3^{+11.9}_{-9.8} \right) \times 10^{-3}$ & $-19.42\pm 0.02$ & 960.49 \\[0.5ex]
    \hline    
    \end{tabular}
    }
    \caption{\textit{Top line:} $\Lambda$CDM model results using $H(z)$ and SNIa datasets. \textit{Below line}: $\Lambda$CDM model results using $H(z)$, SNIa and BAO datasets.
    We include the analysis with the priors described in Table \ref{tab:priors}.}
    \label{tab:cc+pn_lcdm}
\end{table}


\begin{table}[H]
\resizebox{\textwidth}{!}{%
    \centering
    \begin{tabular}{c|cccc}
    \hline 
    \hline & \\[-1.9ex]
    Dataset & $H_0$ [km s$^{-1}$ Mpc$^{-1}$] & $\Omega_m$ & $M$ & $\chi^2_\mathrm{min}$ \\[0.5ex]
    \hline & \\[-1.8ex]
    $H(z)$ + SNIa + xA & $67.4 \pm 1.0$ & $0.299\pm 0.021$ & $-19.43\pm 0.03$ & 2616.59 \\[0.9ex]
    $H(z)$ + SNIa + xA + R21 & $71.2\pm 0.6$ & $0.265^{+0.016}_{-0.015}$ & $-19.33\pm 0.02$  & 2639.99 \\[0.5ex]
    $H(z)$ + SNIa + xA + P18 & $67.4^{+0.3}_{-0.4}$ & $0.299^{+0.018}_{-0.017}$ & $-19.43\pm 0.01$  & 2617.00\\[0.5ex]
    $H(z)$ + SNIa + xA + F20 & $69.2\pm 0.5 $ & $0.283^{+0.016}_{-0.017}$ & $-19.38\pm 0.02$  & 2621.45\\[0.5ex]
    $H(z)$ + SNIa + xA + GAIA & $70.6\pm 0.7$ & $0.269^{+0.019}_{-0.017}$ & $-19.34\pm 0.02$  & 2639.22 \\[0.5ex]
    $H(z)$ + SNIa + xA + ACT & $67.6\pm 0.7$ & $0.297^{+0.018}_{-0.019}$ & $-19.42\pm 0.02$  & 2617.09\\[0.5ex]
    \hline & \\[-1.9ex]
    $H(z)$ + SNIa + BAO + xA & $67.2^{+0.6}_{-0.5}$ & $0.316^{+0.012}_{-0.010}$ & $-19.43\pm 0.02$  & 2628.24\\[0.9ex]
    $H(z)$ + SNIa + BAO + xA + R21 & $69.6\pm 0.5$ & $\left( 292.5^{+11.2}_{-9.7} \right) \times 10^{-3}$ & $-19.37^{+0.02}_{-0.01}$  & 2670.17 \\[0.5ex]
    $H(z)$ + SNIa + BAO + xA + P18 & $67.3^{+0.4}_{-0.3}$ & $\left( 315.2^{+10.0}_{-9.8} \right) \times 10^{-3}$ & $-19.43\pm 0.01$  & 2628.27 \\[0.5ex]
    $H(z)$ + SNIa + BAO + xA + F20 & $68.6^{+0.4}_{-0.5}$ & $\left( 303.7^{+9.2}_{-10.3} \right) \times 10^{-3}$ & $-19.39^{+0.01}_{-0.02}$  & 2638.32 \\[0.5ex]
    $H(z)$ + SNIa + BAO + xA + GAIA & $69.0^{+0.5}_{-0.6}$ & $\left( 298.1^{+11.4}_{-9.0} \right) \times 10^{-3}$ & $-19.38\pm 0.02$  & 2662.63 \\[0.5ex]
    $H(z)$ + SNIa + BAO + xA + ACT & $67.4 \pm 0.5$ & $0.314^{+0.011}_{-0.010}$ & $-19.43\pm 0.02$  & 2628.53\\[0.5ex]
    \hline 
    \end{tabular}
    }
    \caption{$\Lambda$CDM model constraints using the: \textit{Top line:} $H(z)$+SNIa sample (in the first block), \textit{Below line:} and with BAO sample, both using QSO-xA sample.}
    \label{tab:cc+pn+xA_lcdm}
\end{table}

\begin{table}[H]
\resizebox{\textwidth}{!}{%
    \centering
    \begin{tabular}{c|ccccc}
    \hline 
    \hline & \\[-1.9ex]
    Dataset & $H_0$ [km s$^{-1}$ Mpc$^{-1}$] & $\Omega_m$ & $M$ & $\beta'$ & $\chi^2_\mathrm{min}$ \\[0.5ex]
    \hline & \\[-1.8ex]
    $H(z)$ + SNIa + nUVX & $69.0^{+3.6}_{-3.5}$ & $0.300^{+0.020}_{-0.022}$ & $-19.37\pm 0.11$ & $-11.42^{+6.6}_{-7.7}$ & 2993.82 \\[0.9ex]
    $H(z)$ + SNIa + nUVX + R21 & $73.2^{+0.7}_{-0.8}$ & $0.308\pm 0.011$ & $-19.25\pm 0.02$ & $-11.429^{+0.095}_{-0.094}$ & 3044.75 \\[0.5ex]
    $H(z)$ + SNIa + nUVX + P18 & $67.4\pm 0.4 $ & $0.309\pm 0.011$ & $-19.43\pm 0.01$ & $-11.39^{+0.15}_{-0.16}$ & 3068.76 \\[0.5ex]
    $H(z)$ + SNIa + nUVX + F20 & $69.7^{+0.7}_{-0.5}$ & $\left( 308.5^{+10.0}_{-9.9} \right) \times 10^{-3}$ & $-19.35\pm0.02$ & $-11.40\pm 0.14$ &  3067.9 \\[0.5ex]
    $H(z)$ + SNIa + nUVX + GAIA & $73.5^{+2.2}_{-2.0}$ & $0.307\pm 0.012$ & $-19.24\pm 0.06$ & $-11.42\pm0.12$ & 3047.23 \\[0.5ex]
    $H(z)$ + SNIa + nUVX + ACT & $67.9^{+1.6}_{-1.4}$ & $0.308\pm 0.011$ & $-19.41 \pm 0.05$ & $-11.40\pm 0.12$ & 3068.20 \\[0.5ex]
    \hline & \\[-1.9ex]
    $H(z)$ + SNIa + BAO + nUVX  & $67.3 \pm 0.7$ & $0.317^{+0.012}_{-0.011}$ & $-19.43 \pm 0.02$ & $-11.03^{+1.69}_{-1.75}$  & 3117.92  \\[0.9ex]
    $H(z)$ + SNIa + BAO + nUVX + R21 & $70.2\pm 0.5$ & $\left( 293.8^{+9.9}_{-9.8} \right) \times 10^{-3}$ & $-19.35\pm 0.02$ & $-11.46 \pm 1.78$  & 3152.89 \\[0.5ex]
    $H(z)$ + SNIa + BAO + nUVX + P18 & $67.3\pm 0.3$ & $\left( 318.1^{+9.7}_{-10.9} \right) \times 10^{-3}$ & $-19.43\pm 0.01$ & $-11.12^{+1.54}_{-1.91}$ & 3117.91 \\[0.5ex]
    $H(z)$ + SNIa + BAO + nUVX + F20 & $68.77^{+0.48}_{-0.41}$ & $\left( 305.4^{+9.7}_{-10.3} \right) \times 10^{-3}$ & $-19.384^{+0.013}_{-0.014}$ & $-11.46^{+1.62}_{-1.77}$ & 3125.63 \\[0.5ex]
    $H(z)$ + SNIa + BAO + nUVX + GAIA & $69.5\pm 0.6$ & $0.299\pm 0.010$ & $-19.37 \pm 0.02$ & $-11.49^{+1.77}_{-1.65}$ & 3148.20\\[0.5ex]
    $H(z)$ + SNIa + BAO + nUVX + ACT &  $67.5\pm 0.6$ & $0.317^{+0.010}_{-0.012}$ & $-19.42 \pm 0.02$ & $-11.47^{+1.64}_{-1.86}$ & 3118.18\\[0.5ex]
    \hline
    \end{tabular}
    }
    \caption{$\Lambda$CDM model constraints using the: \textit{Top line:} $H(z)$+SNIa sample (in the first block), \textit{Below line:} and with BAO sample, both using QSO-nUVX sample.}
    \label{tab:cc+pn+nUVX_lcdm}
\end{table}


\subsection{Power Law Model -- \texorpdfstring{$f_1(T)$}{} model}

\begin{figure}[H]
  \centering
  \includegraphics[width=.47\linewidth]{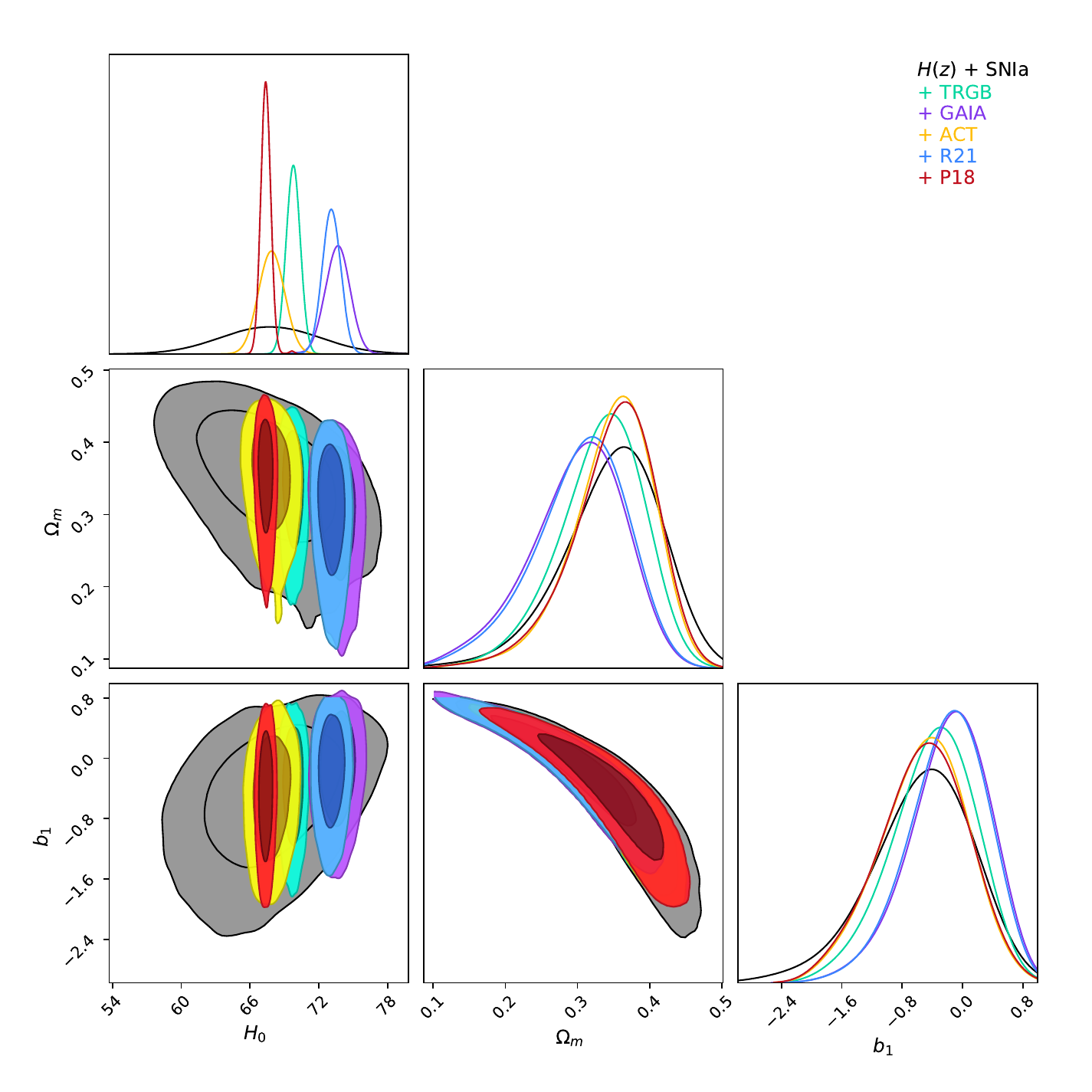}
  \includegraphics[width=.47\linewidth]{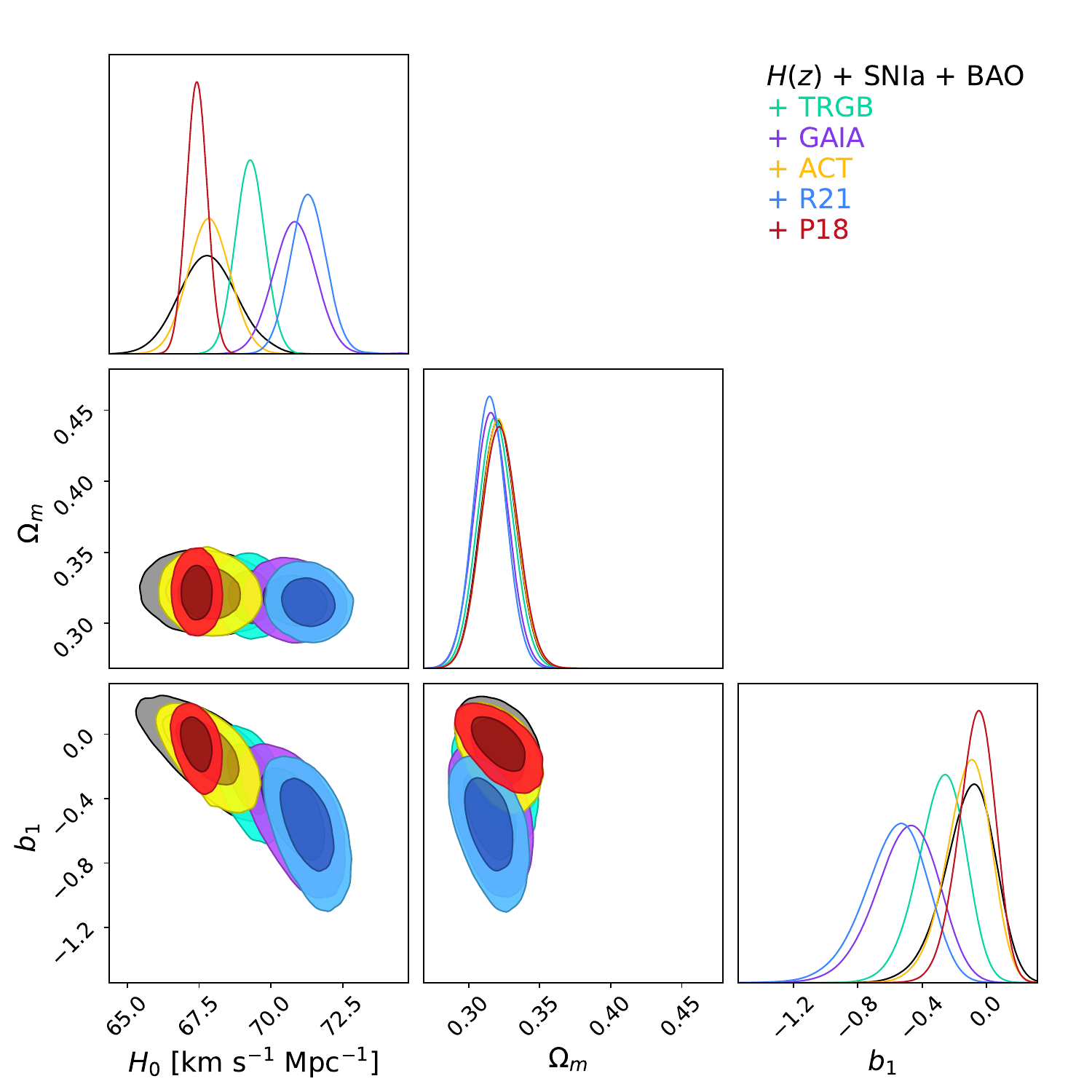}
\caption{1-2$\sigma$ C.L results for the $f_1 (T)$ model using: \textit{Left:} $H(z)$ and Pantheon data sets. \textit{Right:} including BAO. Blue color denotes R21, purple color for GAIA, red color for P18, green color for F20, and yellow color for ACT priors. Additionally, the model was constrained with the sample baselines without prior, here denoted in black color.}
\label{fig:f1}
\end{figure}

The 1-2$\sigma$ C.L. constraints for this model are given in Figure \ref{fig:f1}. We show the results for each of the constrained cosmological parameters in Table \ref{tab:cc+pn_f1}, where the nuisance parameter $M$ is also given for each case.

\begin{figure}[H]
  \includegraphics[width=.49\linewidth]{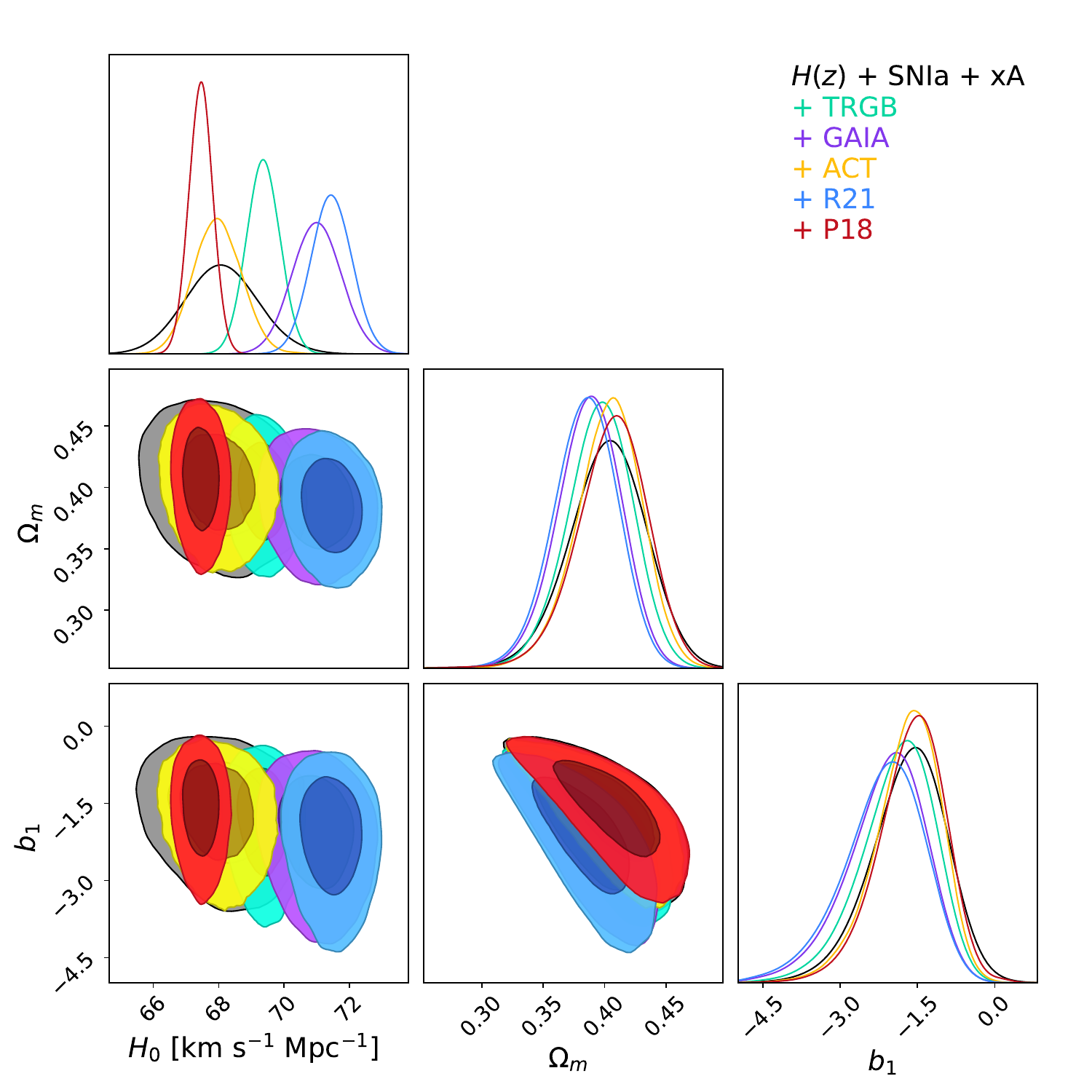}
   \includegraphics[width=.49\linewidth]{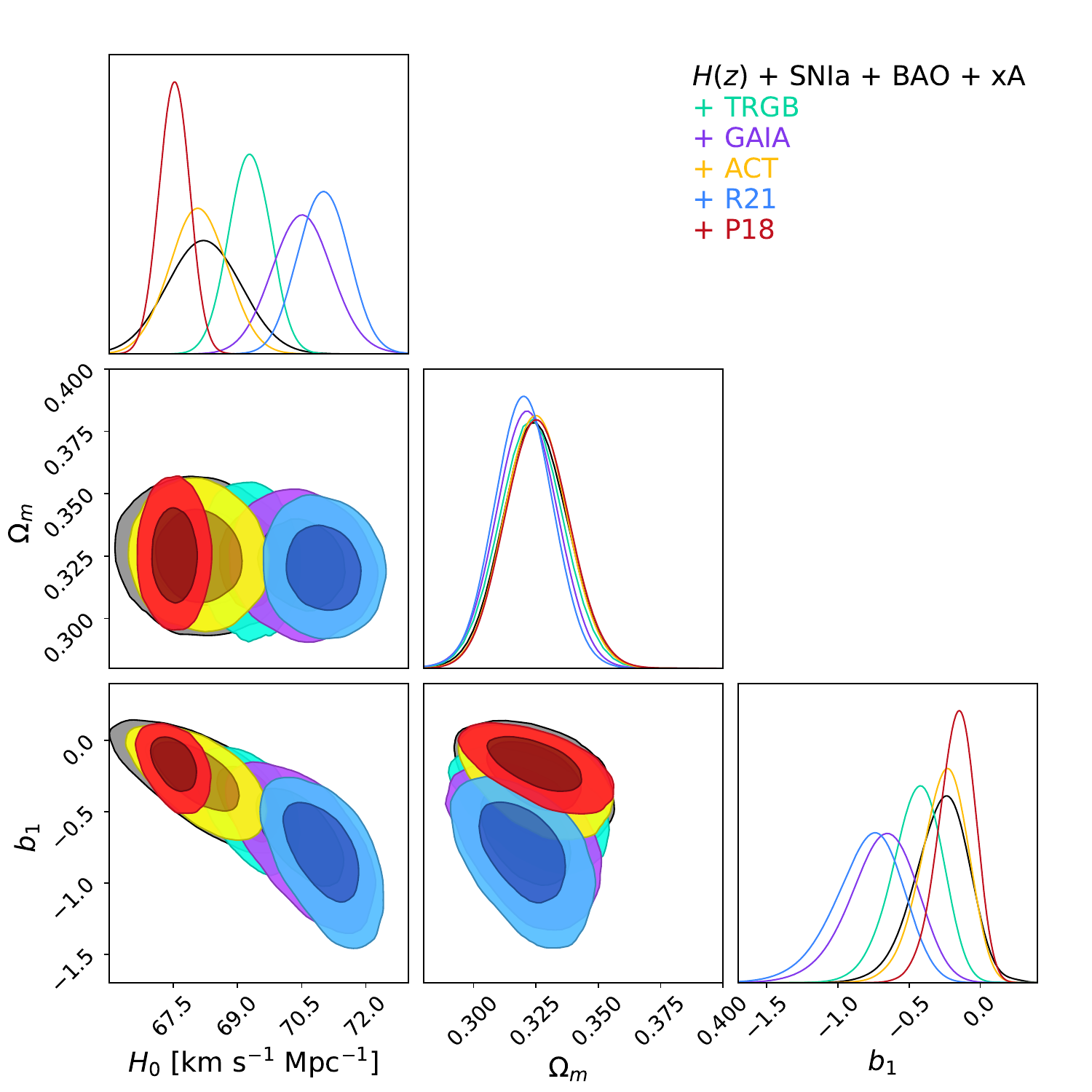}
  \includegraphics[width=.49\linewidth]{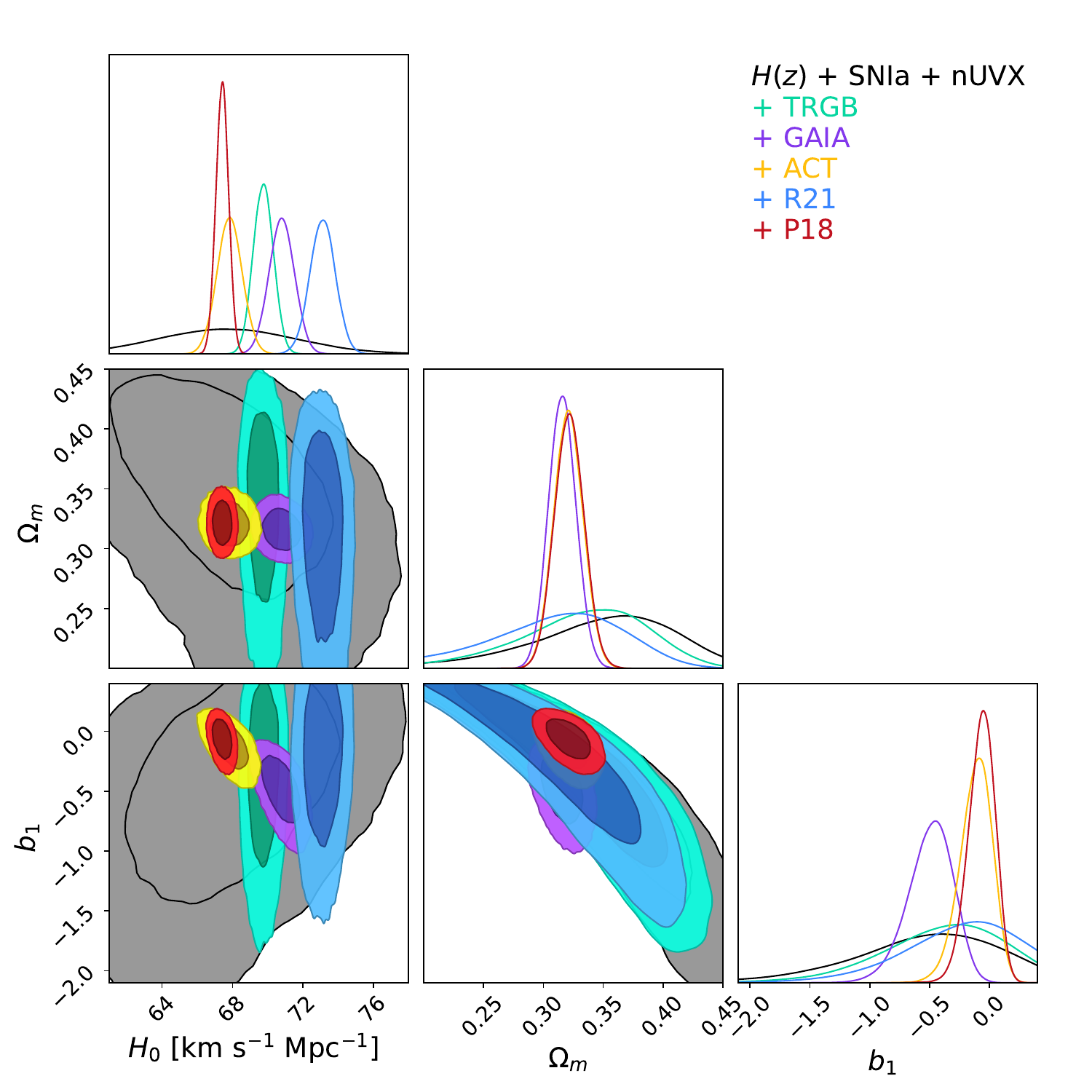}
    \includegraphics[width=.49\linewidth]{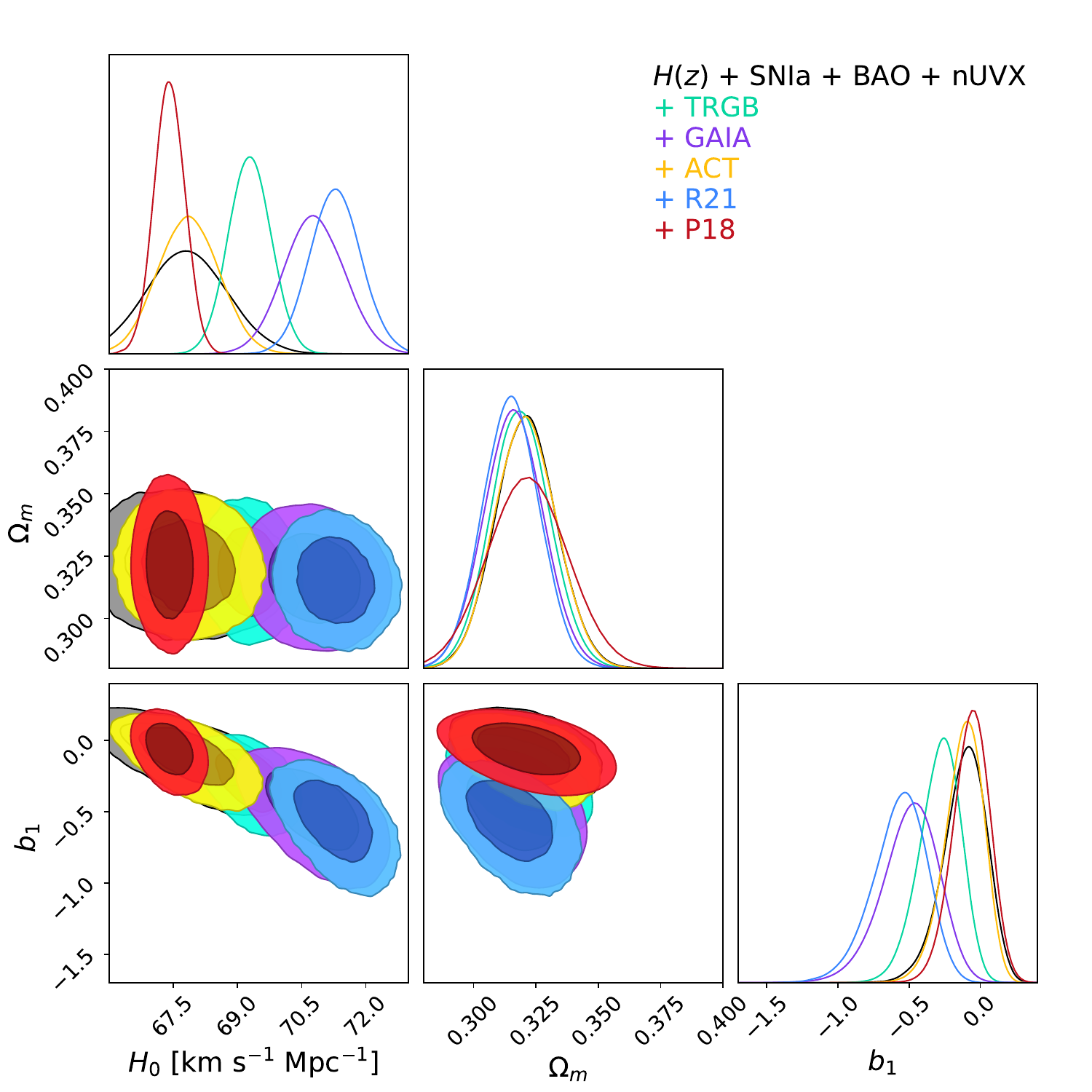}
\caption{1-2$\sigma$ C.L results for the $f_1(T)$ model using: \textit{Top left:} $H(z)$+SNIa and including the xA sample. \textit{Top right:} $H(z)$+SNIa+BAO and including the xA sample. 
\textit{Bottom left:} $H(z)$+SNIa and including the nUVX sample. \textit{Bottom right:} \textit{Top right:} $H(z)$+SNIa+BAO and including the nUVX sample.
Blue color denotes R21, purple color for GAIA, red color for P18, green color for F20, and yellow color for ACT priors. Additionally, the model was constrained with the sample baselines without prior, here denoted in black color.}
\label{fig:f1_qso1}
\end{figure}

Also, it is reported the 1-2-$\sigma$ constraints of this model in Figure \ref{fig:f1_qso1}, with their constraints reported in Table \ref{tab:cc+pn+qso_f1}  for the QSO-xA sample and in Table \ref{tab:cc+pn+qso2_f1} for the QSO-nUVX sample.

We notice that the highest value for the Hubble constant in this analysis was obtained using only $H(z)$ and SNIa measurements with the GAIA prior $H_0 = 73.8^{+0.9}_{-1.1}$ km s$^{-1}$ Mpc$^{-1}$. In comparison to the $\Lambda$CDM model, when we consider the BAO sample, the priors associated with early Universe physics, such as ACT and P18, prefer a value of $b_1 \to 0$. The latter case gives $b_1 = -0.05^{+0.10}_{-0.11}$, which recover the case of $\Lambda$CDM as it is predicted from the model Eq.(\ref{eq:f1}).

Furthermore, using the R21 prior we obtain $H_0 = 71.3 \pm 0.6$ km s$^{-1}$ Mpc$^{-1}$, and $b_1 = -0.52^{+0.16}_{-0.21}$, denoting a deviation from $\Lambda$CDM of more than 2$\sigma$. 


\begin{table}[H]
\resizebox{\textwidth}{!}{%
    \centering
    \begin{tabular}{c|ccccc}
    \hline 
    \hline & \\[-1.9ex]
    Dataset & $H_0$ [km s$^{-1}$ Mpc$^{-1}$] & $\Omega_m$ & $b_1$ & $M$ & $\chi^2_\mathrm{min}$ \\[0.5ex]
    \hline & \\[-1.8ex]
    $H(z)$ + SNIa & $68.1^{+3.6}_{-4.1}$ & $0.370^{+0.051}_{-0.068}$ & $-0.41^{+0.59}_{-0.61}$ & $-19.41^{+0.11}_{-0.13}$  & 949.01\\[0.9ex]
    $H(z)$ + SNIa + R21 & $73.0^{+0.8}_{-0.7}$ & $0.322^{+0.054}_{-0.063}$ & $-0.08^{+0.45}_{-0.52}$ & $-19.26^{+0.02}_{-0.03}$  & 950.70 \\[0.5ex] 
    $H(z)$ + SNIa + P18 & $67.4 \pm 0.4$ & $0.371^{+0.042}_{-0.059}$ & $-0.42^{+0.52}_{-0.59}$ & $-19.44\pm 0.02$  & 949.05 \\[0.5ex]
    $H(z)$ + SNIa + F20 & $69.8^{+0.5}_{-0.6}$ & $0.348^{+0.048}_{-0.056}$ & $-0.27^{+0.47}_{-0.54}$ & $-19.36\pm 0.02$  & 949.19 \\[0.5ex]
    $H(z)$ + SNIa + GAIA & $73.8^{+0.9}_{-1.1}$ & $0.327^{+0.046}_{-0.075}$ & $-0.07^{+0.47}_{-0.49}$ & $-19.25\pm 0.03$  & 951.15 \\[0.5ex]
    $H(z)$ + SNIa + ACT & $67.9^{+1.0}_{-1.1}$ & $0.365^{+0.045}_{-0.055}$ & $-0.32^{+0.41}_{-0.68}$ & $-19.42\pm 0.04$  & 949.01 \\[0.5ex]
    \hline\hline
     $H(z)$ + SNIa + BAO & $67.8^{+0.9}_{-1.0}$ & $0.321^{+0.011}_{-0.013}$ & $-0.06^{+0.12}_{-0.17}$ & $-19.42 \pm 0.02$  & 959.98 \\[0.9ex]
    $H(z)$ + SNIa +BAO + R21 & $71.3\pm 0.6$ & $0.315\pm 0.011$ & $-0.52^{+0.16}_{-0.21}$ & $-19.34 \pm 0.02$  & 981.26 \\[0.5ex] 
    $H(z)$ + SNIa + BAO + P18 & $67.4\pm 0.4$ & $0.322^{+0.012}_{-0.013}$ & $-0.05^{+0.10}_{-0.11}$ & $-19.43 \pm 0.01$  & 960.09 \\[0.5ex]
    $H(z)$ + SNIa + BAO + F20 & $69.8^{+0.5}_{-0.6}$ & $0.343^{+0.030}_{-0.033}$ & $-0.21^{+0.24}_{-0.34}$ & $-19.36\pm 0.02$  & 963.50 \\[0.5ex]
    $H(z)$ + SNIa + BAO + GAIA & $70.8\pm 0.7$ & $0.315^{+0.012}_{-0.011}$ & $-0.46^{+0.17}_{-0.20}$ & $-19.35\pm 0.02$  & 980.64 \\[0.5ex]
    $H(z)$ + SNIa + BAO + ACT & $67.9^{+0.6}_{-0.8}$ & $0.321^{+0.013}_{-0.012}$ & $-0.09^{+0.12}_{-0.15}$ & $-19.42\pm 0.02$  & 959.99 \\[0.5ex]
    \hline    
    \end{tabular}
    }
    \caption{\textit{Top line:} $f_1 (T)$ model results using $H(z)$ and SNIa datasets. \textit{Below line}: $f_1 (T)$ model results using $H(z)$, SNIa and BAO datasets.
    We include the analysis with the priors described in Table \ref{tab:priors}.}
    \label{tab:cc+pn_f1}
\end{table}


\begin{table}[H]
\resizebox{\textwidth}{!}{%
    \centering
    \begin{tabular}{c|ccccc}
    \hline 
    \hline & \\[-1.9ex]
    Dataset & $H_0$ [km s$^{-1}$ Mpc$^{-1}$] & $\Omega_m$ & $b_1$ & $M$ & $\chi^2_\mathrm{min}$ \\[0.5ex]
    \hline
    $H(z)$ + SNIa + xA & $68.1 \pm 1.0$ & $0.405^{+0.027}_{-0.028}$ & $-1.43^{+0.49}_{-0.77}$ & $-19.44 \pm 0.03$ &  2606.16\\[0.9ex]
    $H(z)$ + SNIa + xA + R21 &  $71.4 \pm 0.6$ & $0.387^{+0.024}_{-0.026}$ & $-1.97^{+0.69}_{-0.79}$ & $-19.35 \pm 0.02$ & 2624.62 \\[0.5ex]
    $H(z)$ + SNIa + xA + P18 & $67.5^{+0.3}_{-0.4}$ & $0.411^{+0.024}_{-0.030}$ & $-1.45^{+0.54}_{-0.69}$ & $-19.453^{+0.02}_{-0.01}$ & 2606.55 \\[0.5ex]
    $H(z)$ + SNIa + xA + F20 & $69.4\pm 0.5$ & $0.399^{+0.025}_{-0.027}$ & $-1.73^{+0.64}_{-0.71}$ & $-19.40\pm 0.02$ & 2608.69 \\[0.5ex]
    $H(z)$ + SNIa + xA + GAIA &  $71.0\pm 0.7$ & $0.389\pm 0.025$ & $-1.88^{+0.62}_{-0.82}$ & $-19.36 \pm 0.02$ & 2624.50 \\[0.5ex]
    $H(z)$ + SNIa + xA + ACT & $67.96^{+0.73}_{-0.74}$ & $0.407^{+0.026}_{-0.029}$ & $-1.58^{+0.62}_{-0.65}$ & $-19.440\pm 0.023$ & 2606.18 \\[0.5ex]
    \hline & \\[-1.8ex]
    $H(z)$ + SNIa + BAO + xA & $68.3^{+0.8}_{-0.9}$ & $0.324^{+0.014}_{-0.013}$ & $-0.23^{+0.15}_{-0.21}$ & $-19.41 \pm 0.02$ & 2625.79 \\[0.9ex]
    $H(z)$ + SNIa + BAO + xA + R21 & $71.0 \pm 0.6$ & $0.320\pm 0.011$ & $-0.72^{+0.19}_{-0.26}$ & $-19.35 \pm 0.01$ & 2648.17 \\[0.5ex]
    $H(z)$ + SNIa + BAO + xA + P18 & $67.5^{+0.3}_{-0.4}$ & $0.325^{+0.013}_{-0.012}$ & $-0.15^{+0.12}_{-0.13}$ & $-19.43\pm 0.01$ & 2626.52 \\[0.5ex]
    $H(z)$ + SNIa + BAO + xA + F20 & $69.3\pm 0.5$ & $0.322^{+0.016}_{-0.014}$ & $-0.41^{+0.16}_{-0.19}$ & $-19.39\pm 0.01$ & 2628.70 \\[0.5ex]
    $H(z)$ + SNIa + BAO + xA + GAIA & $70.5^{+0.6}_{-0.7}$ & $0.322^{+0.011}_{-0.012}$ & $-0.65^{+0.22}_{-0.23}$ & $-19.36\pm 0.02$ & 2646.78 \\[0.5ex]
    $H(z)$ + SNIa + BAO + xA + ACT & $68.1\pm 0.7$ & $0.326^{+0.011}_{-0.013}$ & $-0.23^{+0.15}_{-0.17}$ & $-19.412\pm 0.02$ & 2625.81 \\[0.5ex]
    \hline    
    \end{tabular}
    }
    \caption{$f_1(T)$ model constraints using the: \textit{Top line:} $H(z)$+SNIa sample (in the first block), \textit{Below line:} and with BAO sample, both using QSO-xA sample.}
    \label{tab:cc+pn+qso_f1}
\end{table}

\begin{table}[H]
\resizebox{\textwidth}{!}{%
    \centering
    \begin{tabular}{c|cccccc}
    \hline 
    \hline & \\[-1.9ex]
    Data Set & $H_0$ [km s$^{-1}$ Mpc$^{-1}$] & $\Omega_m$ & $b_1$ & $M$ & $\beta'$ & $\chi^2_\mathrm{min}$ \\[0.5ex]
    \hline
    $H(z)$ + SNIa + xA & $67.2^{+4.3}_{-3.5}$ & $0.369^{+0.052}_{-0.069}$ & $-0.41^{+0.63}_{-0.64}$ & $-19.42^{+0.12}_{-0.13}$ & $-11.74^{+1.31}_{-1.82}$ & 3106.06 \\[0.9ex]
    $H(z)$ + SNIa + nUVX + R21 &  $73.2^{+0.6}_{-0.8}$ & $0.327^{+0.051}_{-0.067}$ & $-0.11^{+0.47}_{-0.55}$ & $-19.26^{+0.03}_{-0.02}$ & $-10.57^{+1.13}_{-1.94}$ & 3108.32 \\[0.5ex]
    $H(z)$ + SNIa + nUVX + P18 & $67.5^{+0.3}_{-0.4}$ & $0.322^{+0.012}_{-0.013}$ & $-0.05\pm 0.11$ & $-19.43\pm 0.01$ & $-11.54^{+1.28}_{-1.84}$ & 3117.72 \\[0.5ex]
    $H(z)$ + SNIa + nUVX + F20 & $69.8^{+0.5}_{-0.6}$ & $0.351^{+0.045}_{-0.062}$ & $-0.25^{+0.47}_{-0.57}$ & $-19.36\pm 0.02$ & $-10.9^{+1.06}_{-1.99}$&  3106.82 \\[0.5ex]
    $H(z)$ + SNIa + nUVX + GAIA & $70.8\pm 0.7$ & $0.317^{+0.010}_{-0.012}$ & $-0.45^{+0.16}_{-0.21}$ & $-19.35\pm 0.02$ & $-11.09^{+1.42}_{-1.69}$  & 3138.27 \\[0.5ex]
    $H(z)$ + SNIa + nUVX + ACT & $67.9\pm 0.7$ & $0.321^{+0.013}_{-0.012}$ & $-0.10\pm 0.13$ & $-19.42\pm 0.02$ & $-10.12^{+1.14}_{-1.97}$ & 3117.61 \\[0.5ex]
    \hline & \\[-1.8ex]
    $H(z)$ + SNIa + BAO + nUVX & $67.8^{+1.0}_{-0.9}$ & $0.321\pm 0.012$ & $-0.08^{+0.14}_{-0.15}$ & $-19.42^{+0.03}_{-0.02}$ & $-11.82^{+1.31}_{-1.80}$ & 3117.59 \\[0.9ex]
    $H(z)$ + SNIa + BAO + nUVX + R21 & $71.3\pm 0.6$ & $0.315^{+0.011}_{-0.012}$ & $-0.53^{+0.17}_{-0.20}$ & $-19.34\pm 0.02$ & $-11.36^{+1.27}_{-1.86}$ & 3138.83 \\[0.5ex]
    $H(z)$ + SNIa + BAO + nUVX + P18 & $67.4^{+0.4}_{-0.3}$ & $0.322^{+0.014}_{-0.015}$ & $-0.05^{+0.11}_{-0.13}$ & $-19.43 \pm 0.01$ & $-10.78^{+1.17}_{-1.92}$ & 3117.71 \\[0.5ex]
    $H(z)$ + SNIa + BAO + nUVX + F20 & $69.3 \pm 0.5$ & $0.318^{+0.013}_{-0.011}$ & $-0.26^{+0.13}_{-0.16}$ & $-19.38\pm 0.01$ & $-10.73^{+1.25}_{-1.92}$ &  3121.12 \\[0.5ex]
    $H(z)$ + SNIa + BAO + nUVX + GAIA & $70.8 \pm 0.7$ & $0.316\pm 0.012$ & $-0.46^{+0.18}_{-0.20}$ & $-19.35\pm 0.02$ & $-11.01^{+1.12}_{-1.98}$ & 3138.27 \\[0.5ex]
    $H(z)$ + SNIa + BAO + nUVX + ACT & $67.8 \pm 0.7$ & $0.321^{+0.012}_{-0.013}$ & $-0.10^{+0.13}_{-0.14}$ & $-19.42\pm 0.02$ & $-10.89^{+1.15}_{-1.83}$ & 3117.67 \\[0.5ex]
    \hline    
    \end{tabular}
    }
    \caption{$f_1(T)$ model constraints using the: \textit{Top line:} $H(z)$+SNIa sample (in the first block), \textit{Below line:} and with BAO sample, both using QSO-nUVX sample.}
    \label{tab:cc+pn+qso2_f1}
\end{table}


\subsection{Linder Model -- \texorpdfstring{$f_2(T)$}{} model}

\begin{figure}[H]
  \centering
  \includegraphics[width=.49\linewidth]{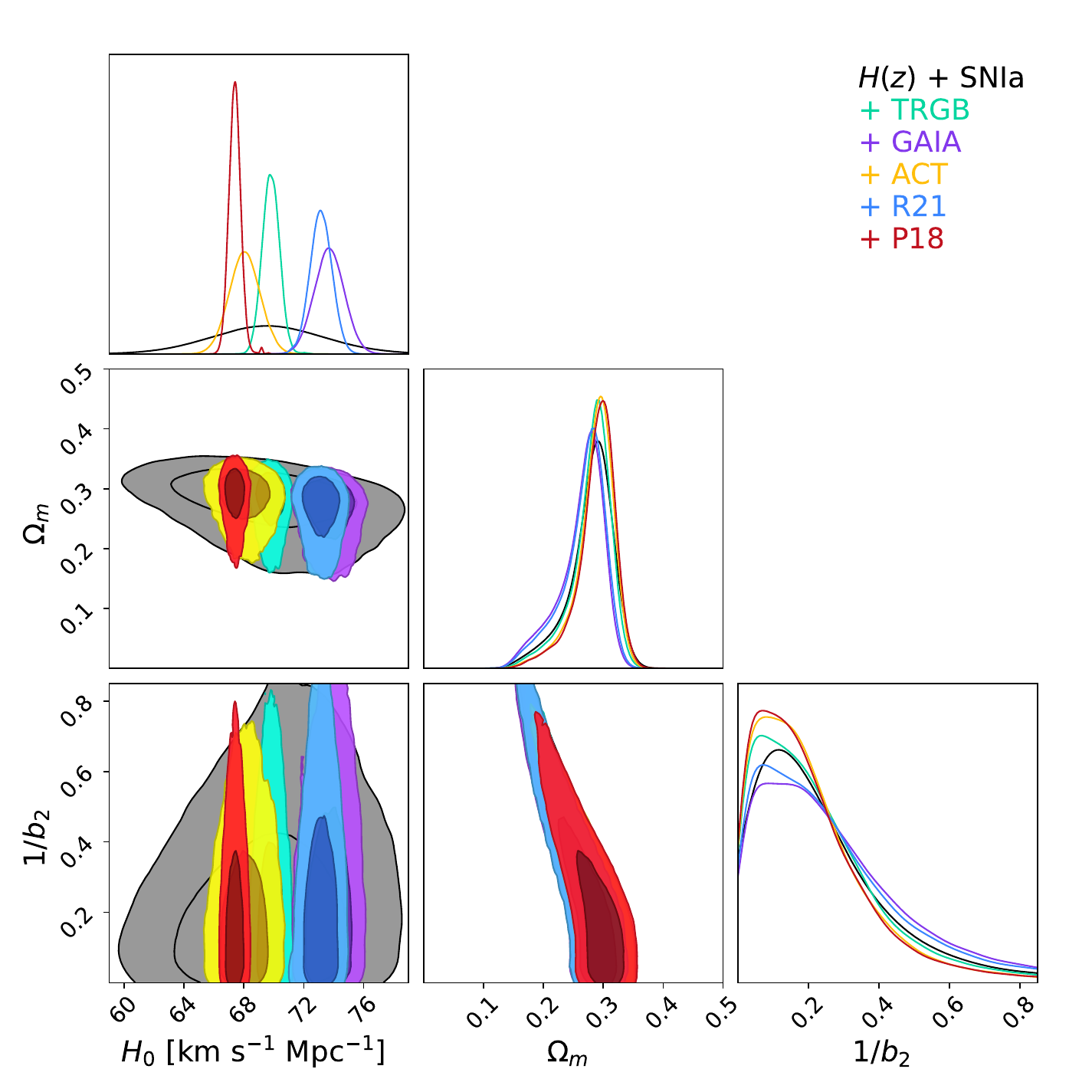}
  \includegraphics[width=.49\linewidth]{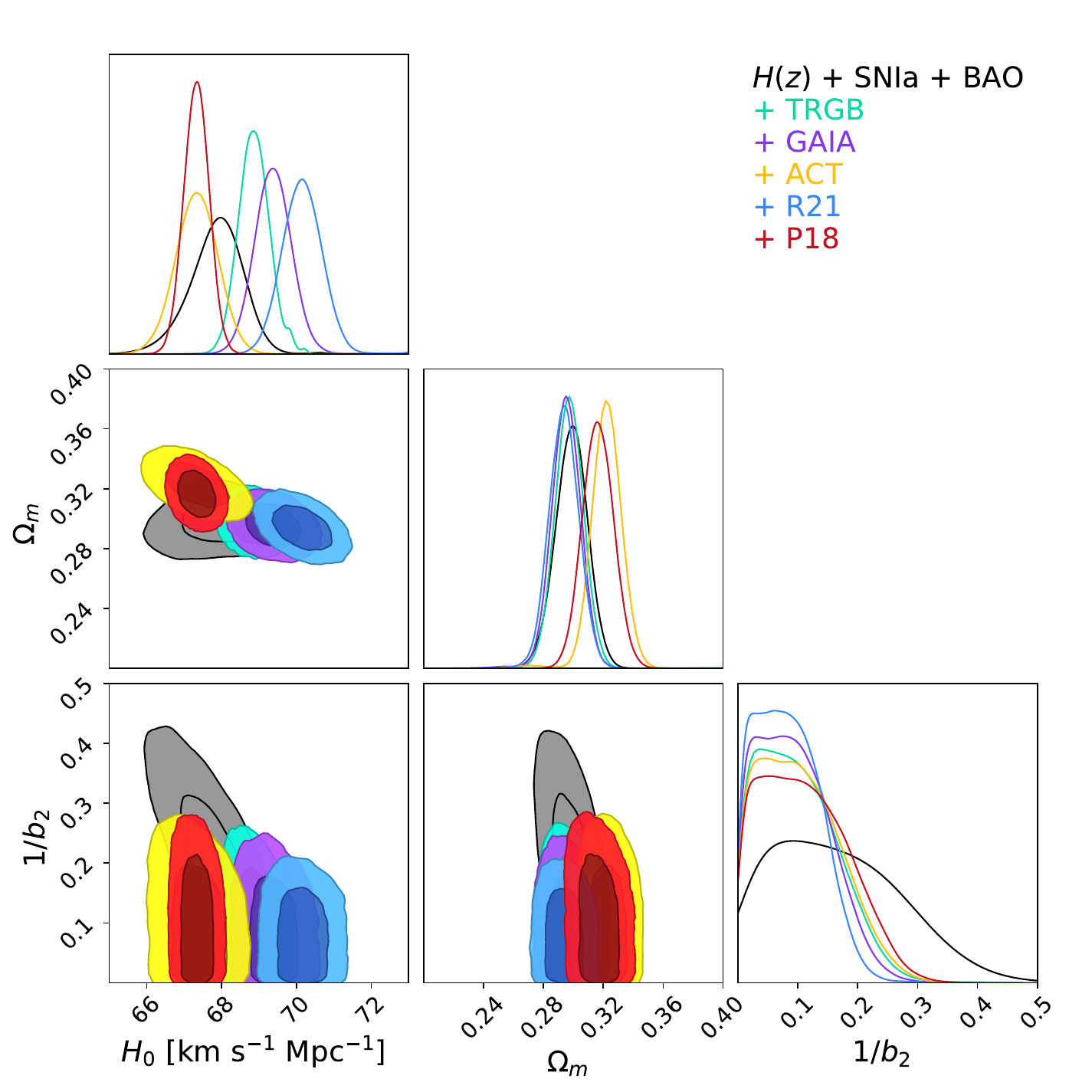}
\caption{1-2$\sigma$ C.L results for the $f_2 (T)$ model using: \textit{Left:} $H(z)$ and Pantheon data sets. \textit{Right:} including BAO. Blue color denotes R21, purple color for GAIA, red color for P18, green color for F20, and yellow color for ACT priors. Additionally, the model was constrained with the sample baselines without prior, here denoted in black color.
}
\label{fig:f2}
\end{figure}

\begin{figure}[H]
  \centering
  \includegraphics[width=.49\linewidth]{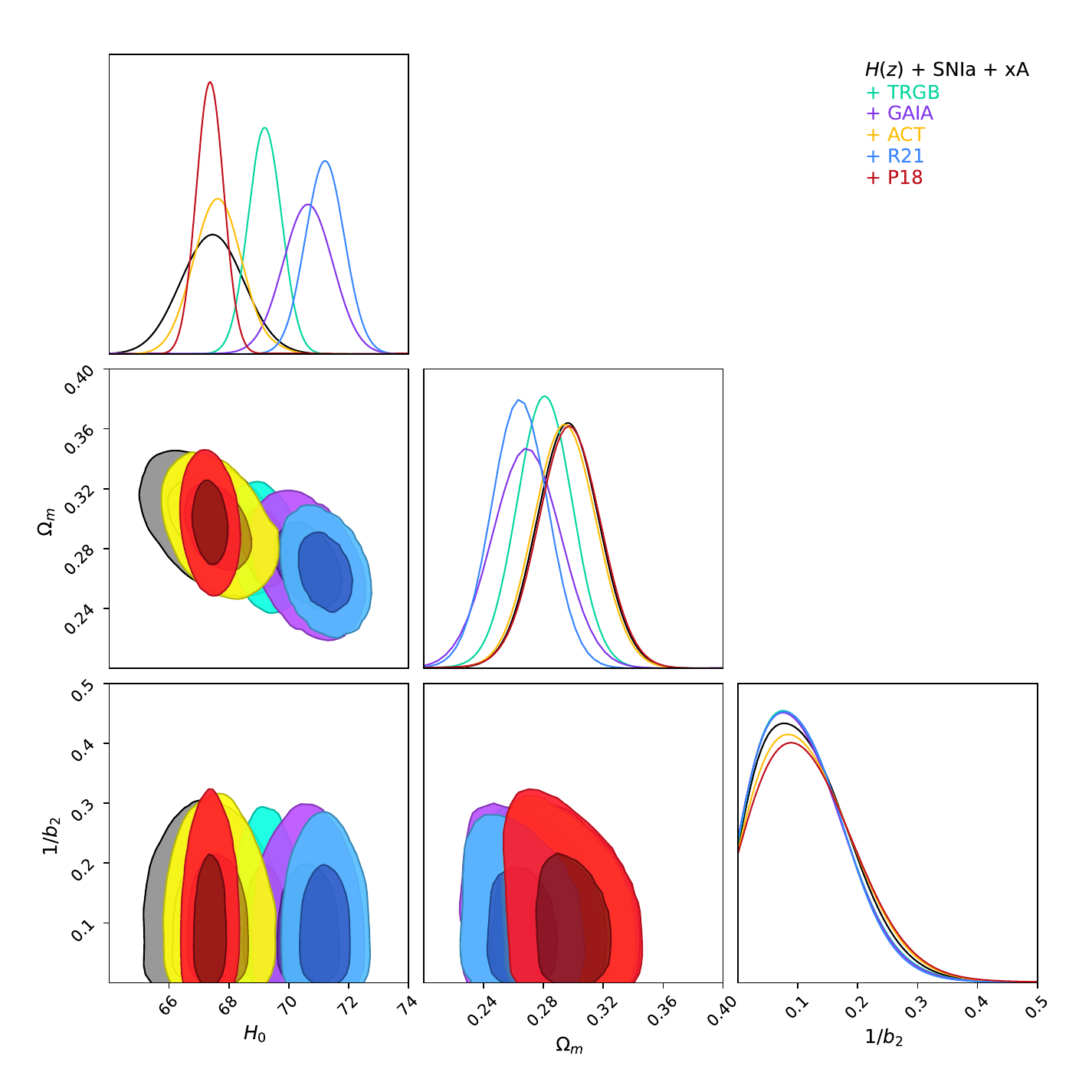}
  \includegraphics[width=.49\linewidth]{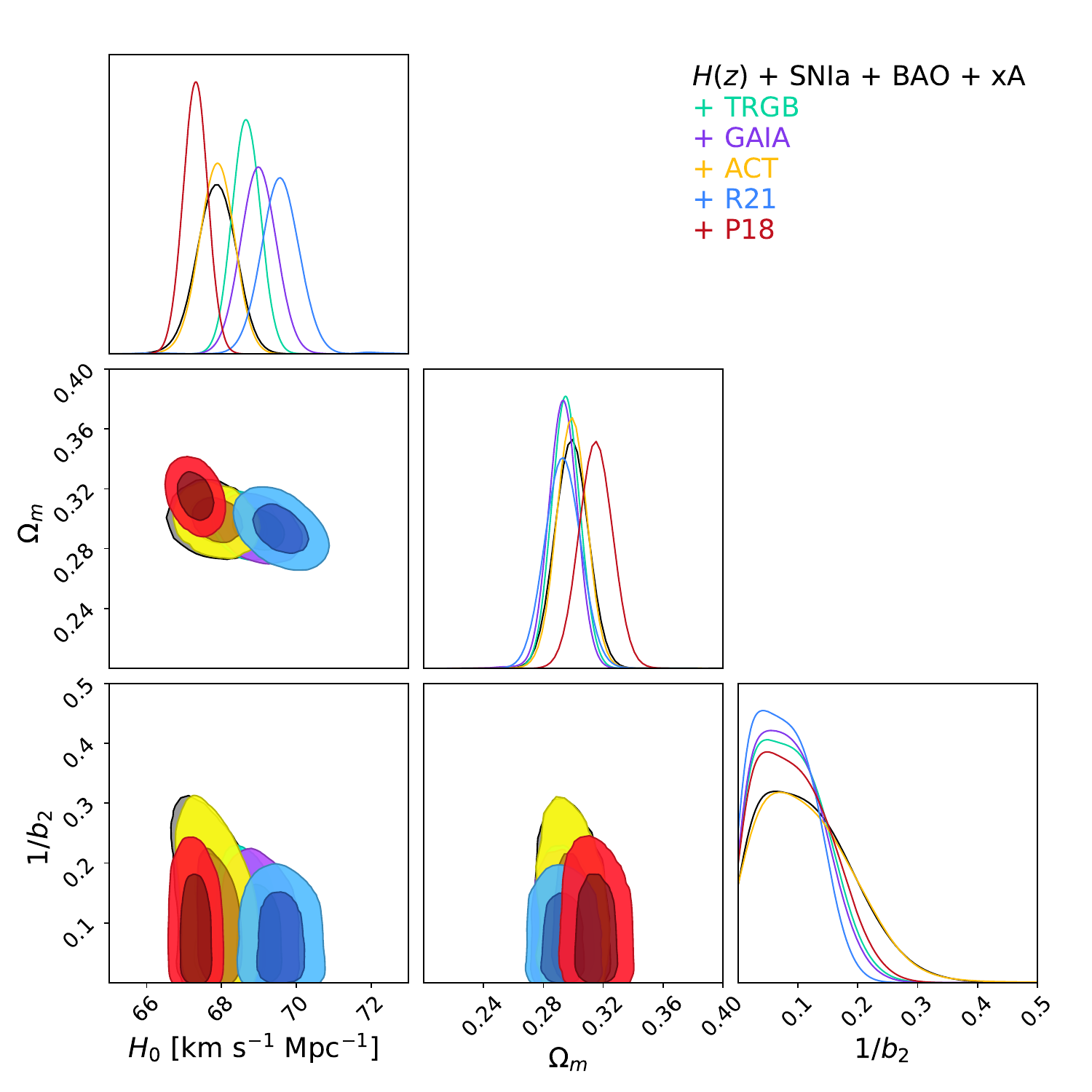}
    \includegraphics[width=.49\linewidth]{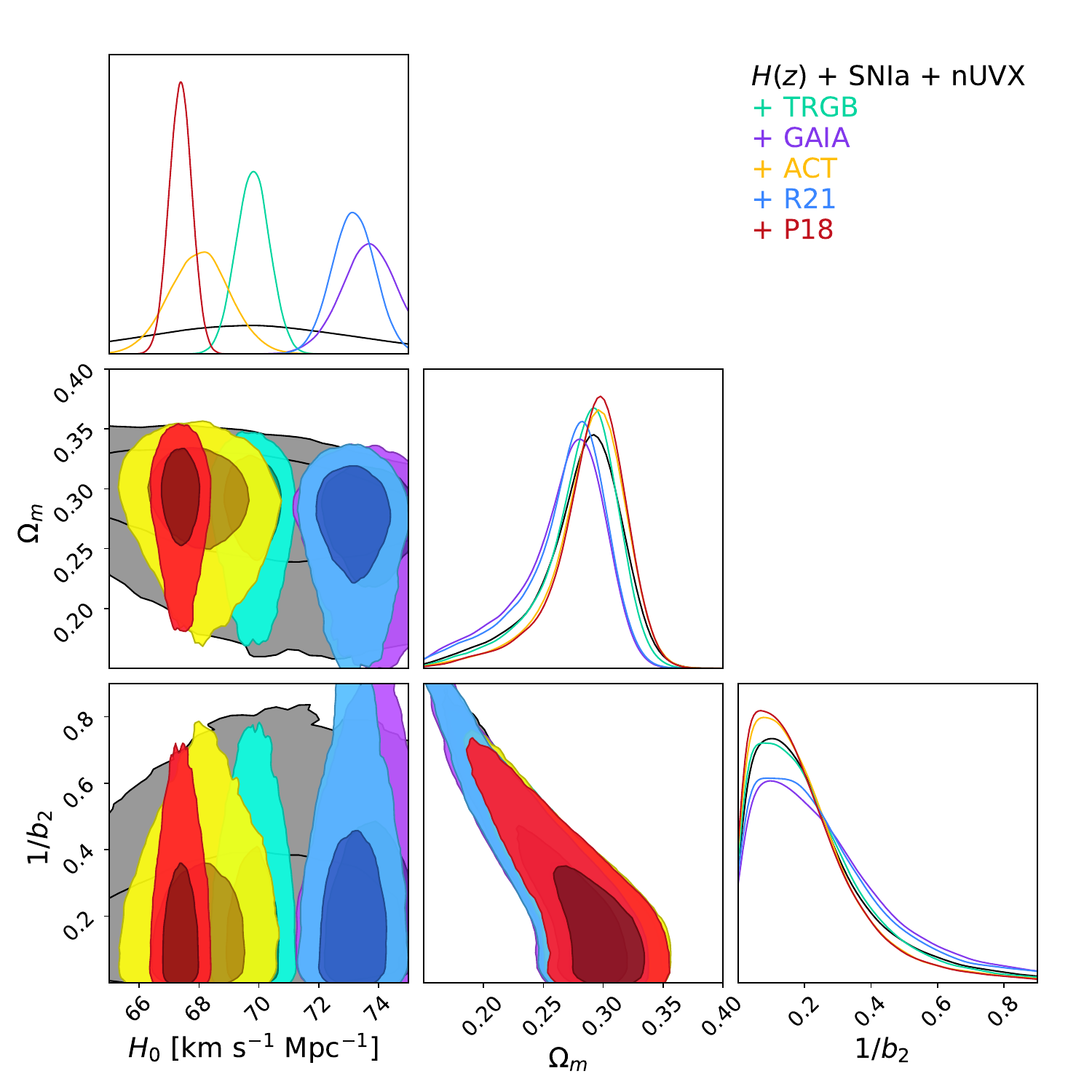}
      \includegraphics[width=.49\linewidth]{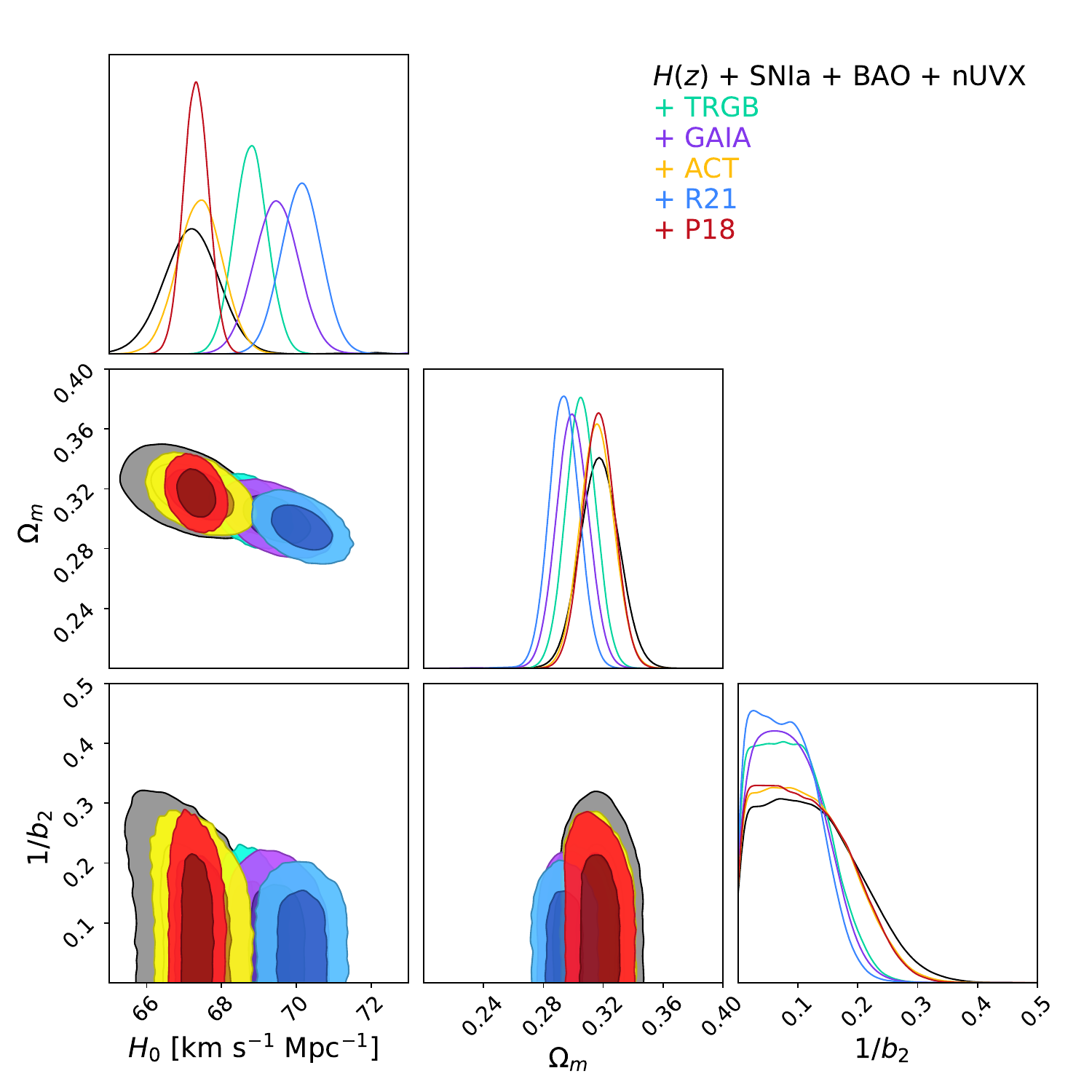}
\caption{1-2$\sigma$ C.L results for the $f_2(T)$ model using: \textit{Top left:} $H(z)$+SNIa and including the xA sample. \textit{Top right:} $H(z)$+SNIa+BAO and including the xA sample. 
\textit{Bottom left:} $H(z)$+SNIa and including the nUVX sample. \textit{Bottom right:} \textit{Top right:} $H(z)$+SNIa+BAO and including the nUVX sample.
Blue color denotes R21, purple color for GAIA, red color for P18, green color for F20, and yellow color for ACT priors. Additionally, the model was constrained with the sample baselines without prior, here denoted in black color.}
\label{fig:f2_qso1}
\end{figure}

The 1-2$\sigma$ C.L. constraints for this model are given in Figures \ref{fig:f2} and \ref{fig:f2_qso1}. The cosmological constraints for this model are given in Table \ref{tab:cc+pn_f2}.
In this case,  we wrote the free parameter for the model as $1/b_2$, to avoid the divergence problem previously mentioned in recovering the LCDM limit (i.e. $b_2 \to \infty$) in Eq.(\ref{eq:f2}), i.e. 
$1/b_2 \to 0$ recover the $\Lambda$CDM case.  Also, it is reported the 1-2-$\sigma$ constraints of this model in Figure \ref{fig:f2}, with their constraints reported in Table \ref{tab:cc+pn+qso_f2}  for the QSO-xA sample and in Table \ref{tab:cc+pn+qso2_f2} for the QSO-nUVX sample.

As it is expected, the constraints for this model using the baseline described prefer values of $1/b_2$ near $0$, recovering the $\Lambda$CDM case. Regarding $H_0$, its low value using a BAO sample is also obtained. On one hand, the highest $H_0$ estimation for this model is the one using the GAIA prior with $H(z)$ and SNIa combined baseline with $H_0 = 73.7 \pm 1.0$ km s$^{-1}$ Mpc$^{-1}$, with the lowest matter density $\Omega_m = 0.281^{+0.028}_{-0.037}$, and $1/b_2 \sim 0$. 
On the other hand, the lowest $H_0$ estimations are the ones obtained with ACT and P18 priors using the BAO sample.


\begin{table}[H]
\resizebox{\textwidth}{!}{%
    \centering
    \begin{tabular}{c|ccccc}
    \hline 
    \hline & \\[-1.9ex]
    Dataset & $H_0$ [km s$^{-1}$ Mpc$^{-1}$] & $\Omega_m$ & $1/b_2$ & $M$ & $\chi^2_\mathrm{min}$ \\[0.5ex]
    \hline & \\[-1.8ex]
    $H(z)$ + SNIa &  $69.8^{+3.4}_{-3.8}$ & $0.292^{+0.029}_{-0.034}$ & $0.00^{+0.27}_{-0.00}$ & $-19.36\pm 0.11$  & 949.41 \\[0.9ex]
    $H(z)$ + SNIa + R21 & $73.1^{+0.8}_{-0.7}$ & $0.283^{+0.030}_{-0.039}$ & $0.00^{+0.31}_{-0.00}$ & $-19.25^{+0.03}_{-0.02}$  & 950.71  \\[0.5ex] 
    $H(z)$ + SNIa + P18 & $67.4\pm 0.4$ & $0.300^{+0.024}_{-0.031}$ & $0.027^{+0.224}_{-0.024}$ & $-19.43\pm0.02$  & 949.66 \\[0.5ex]
    $H(z)$ + SNIa + F20 &  $69.7^{+0.6}_{-0.5}$ & $0.290^{+0.029}_{-0.031}$ & $0.00^{+0.27}_{-0.00}$ & $-19.35\pm 0.02$  & 949.45 \\[0.5ex]
    $H(z)$ + SNIa + GAIA & $73.7 \pm 1.0$ & $0.281^{+0.028}_{-0.037}$ & $0.00^{+0.34}_{-0.00}$ & $-19.24\pm 0.03$  & 951.15 \\[0.5ex]
    $H(z)$ + SNIa + ACT & $68.0^{+1.1}_{-1.0}$ & $0.295^{+0.026}_{-0.029}$ & $0.00^{+0.25}_{-0.00}$ & $-19.41^{+0.03}_{-0.04}$   & 949.52 \\[0.5ex]
    \hline\hline
      $H(z)$ + SNIa + BAO & $68.0\pm 0.7$ & $\left( 300.6^{+8.6}_{-11.6} \right) \times 10^{-3}$ & $0.048^{+0.178}_{-0.046}$ & $-19.41\pm 0.02$  & 961.25 \\[0.9ex]
    $H(z)$ + SNIa + BAO + R21 & $70.2 \pm 0.5$ & $\left( 294.0^{+9.6}_{-10.3} \right) \times 10^{-3}$ & $0.059^{+0.059}_{-0.058}$ & $-19.35\pm 0.2$  & 995.28 \\[0.5ex] 
    $H(z)$ + SNIa + BAO + P18 & $67.4^{+0.3}_{-0.4}$ & $0.316^{+0.011}_{-0.010}$ & $0.047^{+0.107}_{-0.046}$ & $-19.42 \pm 0.01$  & 960.30 \\[0.5ex]
    $H(z)$ + SNIa + BAO + F20 & $68.9\pm 0.4$ & $\left( 296.7^{+11.0}_{-9.6} \right) \times 10^{-3}$ & $0.00^{+0.14}_{-0.00}$ & $-19.38^{+0.01}_{-0.02}$  & 964.89 \\[0.5ex]
    $H(z)$ + SNIa + BAO + GAIA & $69.4 \pm 0.5$ & $\left( 294.9^{+10.3}_{-9.5} \right) \times 10^{-3}$ & $0.083^{+0.047}_{-0.081}$ & $-19.37\pm 0.02$  & 987.31 \\[0.5ex]
    $H(z)$ + SNIa + BAO + ACT & $67.4\pm 0.6$ & $0.323^{+0.010}_{-0.012}$ & $0.045^{+0.106}_{-0.042}$ & $-19.42 \pm 0.02$  & 961.25 \\[0.5ex]
    \hline    
    \end{tabular}
    }
    \caption{$f_2 (T)$ model results using $H(z)$ and SNIa datasets. \textit{Below line}: $f_2 (T)$ model results using $H(z)$, SNIa and BAO datasets.
    We include the analysis with the priors described in Table \ref{tab:priors}.
    }
    \label{tab:cc+pn_f2}
\end{table}


\begin{table}[H]
\resizebox{\textwidth}{!}{%
    \centering
    \begin{tabular}{c|ccccc}
    \hline 
    \hline & \\[-1.9ex]
    Data Set & $H_0$ [km s$^{-1}$ Mpc$^{-1}$] & $\Omega_m$ & $1/b_2$ & $M$ & $\chi^2_\mathrm{min}$ \\[0.5ex]
    \hline & \\[-1.8ex]
    $H(z)$ + SNIa + xA & $67.4^{+1.0}_{-0.9}$ & $0.296\pm 0.019$ & $0.00^{+0.15}_{-0.00}$ & $-19.43 \pm 0.03$ & 2616.99 \\[0.9ex]
    $H(z)$ + SNIa + xA + R21 & $71.2\pm 0.6 $ & $0.264\pm 0.019$ & $0.00^{+0.14}_{-0.00}$ & $-19.33 \pm 0.02$ & 2639.99 \\[0.5ex]
    $H(z)$ + SNIa + xA + P18 & $67.4 \pm 0.4$ & $0.297\pm 0.019$ & $0.094^{+0.066}_{-0.091}$ & $-19.43^{+0.01}_{-0.02}$ & 2617.00 \\[0.5ex]
    $H(z)$ + SNIa + xA + F20 & $69.2^{+0.6}_{-0.5}$ & $0.281\pm 0.017$ & $0.053^{+0.093}_{-0.051}$ & $-19.4 \pm 0.02$ & 2621.45 \\[0.5ex]
    $H(z)$ + SNIa + xA + GAIA &  $70.7^{+0.7}_{-0.8}$ & $0.269\pm 0.020$ & $0.00^{+0.14}_{-0.00}$ & $-19.34\pm 0.02$ & 2639.23 \\[0.5ex]
    $H(z)$ + SNIa + xA + ACT & $67.7^{+0.7}_{-0.8}$ & $0.294^{+0.020}_{-0.018}$ & $0.018^{+0.141}_{-0.014}$ & $-19.42 \pm 0.02$  & 2617.09 \\[0.5ex]
    \hline & \\[-1.8ex]
    $H(z)$ + SNIa + BAO + xA & $67.9 \pm 0.5$ & $0.298^{+0.012}_{-0.011}$ & $0.025^{+0.134}_{-0.023}$ & $-19.41\pm 0.02$ & 2628.34 \\[0.9ex]
    $H(z)$ + SNIa + BAO + xA + R21 & $69.55\pm 0.5$ & $0.293\pm 0.011$ & $0.00^{+0.10}_{-0.00}$ & $-19.37^{+0.01}_{-0.02}$ & 2670.16 \\[0.5ex]
    $H(z)$ + SNIa + BAO + xA + P18 & $67.3\pm 0.3$ & $0.314^{+0.012}_{-0.011}$ & $0.022^{+0.107}_{-0.021}$ & $-19.43\pm 0.01$ & 2628.27 \\[0.5ex]
    $H(z)$ + SNIa + BAO + xA + F20 & $68.6^{+0.4}_{-0.3}$ & $\left( 295.0^{+9.0}_{-9.3} \right) \times 10^{-3}$ & $0.038^{+0.087}_{-0.036}$ & $-19.39\pm 0.01$ & 2634.83\\[0.5ex]
    $H(z)$ + SNIa + BAO + xA + GAIA & $69.0^{+0.5}_{-0.4}$ & $\left( 293.2^{+9.1}_{-9.4} \right) \times 10^{-3}$ & $0.051^{+0.072}_{-0.050}$ & $-19.38^{+0.02}_{-0.01}$& 2658.71 \\[0.5ex]
    $H(z)$ + SNIa + BAO + xA + ACT & $67.9^{+0.4}_{-0.5}$ & $0.300^{+0.010}_{-0.011}$ & $0.056^{+0.107}_{-0.052}$ & $-19.42 \pm 0.02$ & 2628.34\\[0.5ex]
    \hline    
    \end{tabular}
    }
    \caption{$f_2(T)$ model constraints using the: \textit{Top line:} $H(z)$+SNIa sample (in the first block), \textit{Below line:} and with BAO sample, both using QSO-xA sample.
    }
    \label{tab:cc+pn+qso_f2}
\end{table}

\begin{table}[H]
\resizebox{\textwidth}{!}{%
    \centering
    \begin{tabular}{c|cccccc}
    \hline 
    \hline & \\[-1.9ex]
    Data Set & $H_0$ [km s$^{-1}$ Mpc$^{-1}$] & $\Omega_m$ & $1/b_2$ & $M$ & $\beta'$ & $\chi^2_\mathrm{min}$ \\[0.5ex]
    \hline
    $H(z)$ + SNIa + nUVX & $69.8^{+3.0}_{-4.2}$ & $0.296^{+0.022}_{-0.023}$ & $0.00^{+0.14}_{-0.00}$ & $-19.36^{+0.10}_{-0.12}$ & $-12.64^{+0.14}_{-0.73}$ & 3107.04 \\[0.9ex]
    $H(z)$ + SNIa + nUVX + R21 & $73.2^{+0.7}_{-0.8}$ & $0.288^{+0.019}_{-0.021}$ & $0.059^{+0.090}_{-0.058}$ & $-19.27\pm 0.02$ & $-12.21^{+0.13}_{-0.70}$ & 3108.34 \\[0.5ex]
    $H(z)$ + SNIa + nUVX + P18 & $67.4\pm 0.4$ & $0.301^{+0.021}_{-0.022}$ & $0.021^{+0.124}_{-0.020}$ & $-19.43\pm 0.02$ & $-12.62^{+0.61}_{-0.53}$ & 3107.27\\[0.5ex]
    $H(z)$ + SNIa + nUVX + F20 & $69.8\pm 0.6$ & $0.295\pm 0.021$ & $0.077^{+0.065}_{-0.076}$ & $-19.36\pm 0.02$ & $-12.51^{+0.66}_{-0.49}$ & 3107.07 \\[0.5ex]
    $H(z)$ + SNIa + nUVX + GAIA & $73.8^{+0.9}_{-1.1}$ & $0.285\pm 0.021$ & $0.018^{+0.133}_{-0.017}$ & $-19.24\pm 0.03$ & $-12.63^{+0.48}_{-0.63}$ & 3107.76\\[0.5ex]
    $H(z)$ + SNIa + nUVX + ACT & $68.0\pm 1.0$ & $0.299^{+0.021}_{-0.022}$ & $0.00^{+0.14}_{-0.00}$ & $-19.41\pm 0.03$ & $-12.52^{+0.68}_{-0.47}$ & 3107.14 \\[0.5ex]
    \hline & \\[-1.8ex]
    $H(z)$ + SNIa + BAO + nUVX & $67.2^{+0.8}_{-0.7}$ & $0.317^{+0.013}_{-0.012}$ & $0.071^{+0.097}_{-0.070}$ & $-19.43\pm 0.02$ & $-10.83^{+1.59}_{-1.51}$ & 3117.91 \\[0.9ex]
    $H(z)$ + SNIa + BAO + nUVX + R21 & $70.2^{+0.5}_{-0.6}$ & $\left( 294.8^{+8.9}_{-10.9} \right) \times 10^{-3}$ & $0.025^{+0.092}_{-0.024}$ & $-19.35 \pm 0.02$ & $-10.85^{+1.02}_{-1.26}$ & 3152.89 \\[0.5ex]
    $H(z)$ + SNIa + BAO + nUVX + P18 & $67.3\pm 0.3$ & $0.317^{+0.011}_{-0.010}$ & $0.030^{+0.126}_{-0.029}$ & $-19.43 \pm 0.01$ & $-10.52^{+1.15}_{-1.01}$ & 3117.91 \\[0.5ex]
    $H(z)$ + SNIa + BAO + nUVX + F20 & $68.8^{+0.4}_{-0.5}$ & $0.305\pm 0.010$ & $0.075^{+0.052}_{-0.074}$ & $-19.39\pm 0.01$ & $-10.13^{+1.11}_{-1.98}$ & 3125.64 \\[0.5ex]
    $H(z)$ + SNIa + BAO + nUVX + GAIA & $69.4 \pm 0.6$ & $0.300^{+0.010}_{-0.011}$ & $0.054^{+0.075}_{-0.052}$ & $-19.37 \pm 0.02$ & $-10.34^{+1.14}_{-1.96}$ & 3148.19\\[0.5ex]
    $H(z)$ + SNIa + BAO + nUVX + ACT & $67.5 \pm 0.6$ & $0.316\pm 0.011$ & $0.055^{+0.102}_{-0.054}$ & $-19.42 \pm 0.02$ & $-10.34^{+1.17}_{-1.96}$ & 3118.12\\[0.5ex]
    \hline    
    \end{tabular}
    }
    \caption{$f_2(T)$ model constraints using the: \textit{Top line:} $H(z)$+SNIa sample (in the first block), \textit{Below line:} and with BAO sample, both using QSO-nUVX sample.}
        \label{tab:cc+pn+qso2_f2}
\end{table}


\subsection{Variant Linder Model -- \texorpdfstring{$f_3(T)$}{} model}

\begin{figure}[H]
  \centering
  \includegraphics[width=.49\linewidth]{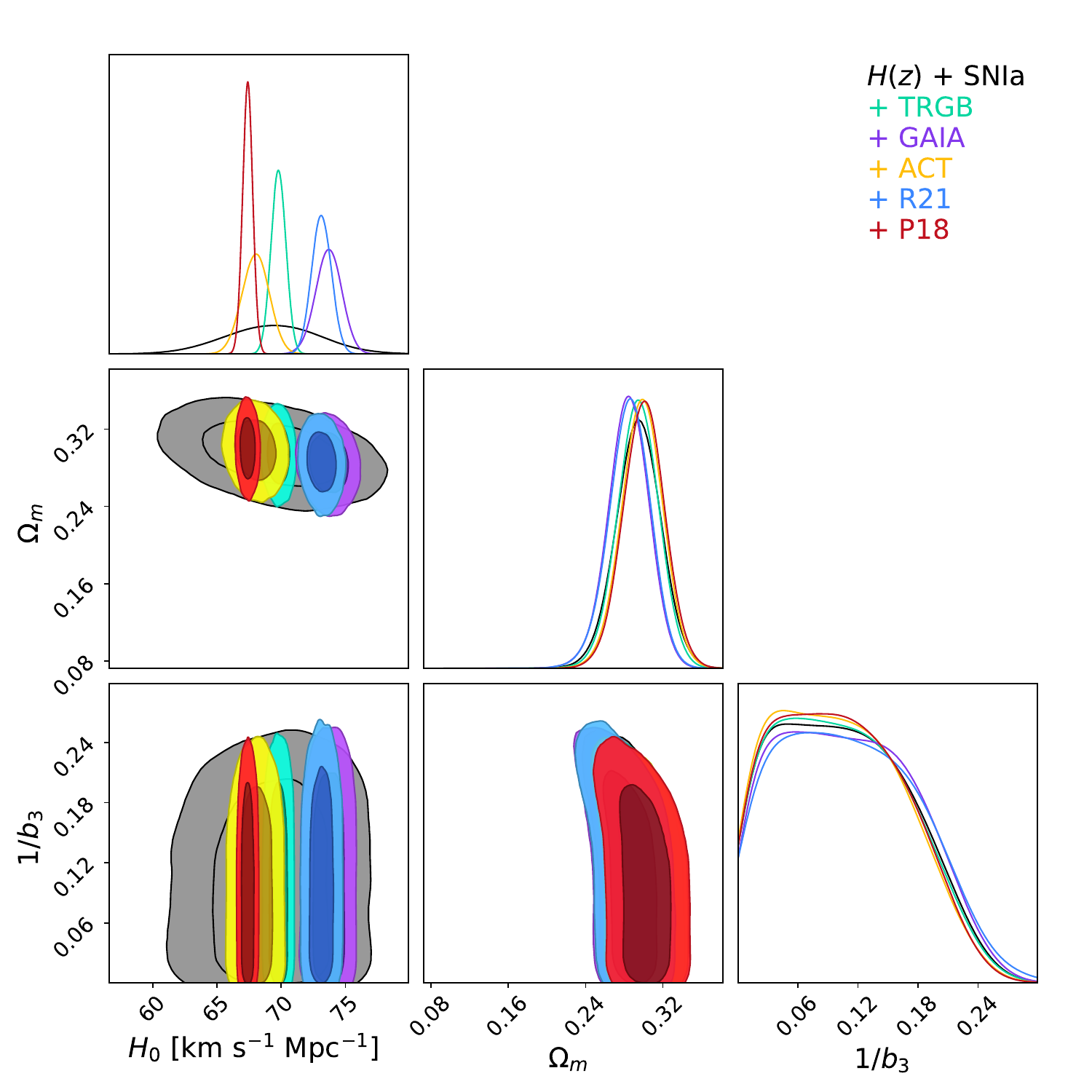}
  \includegraphics[width=.49\linewidth]{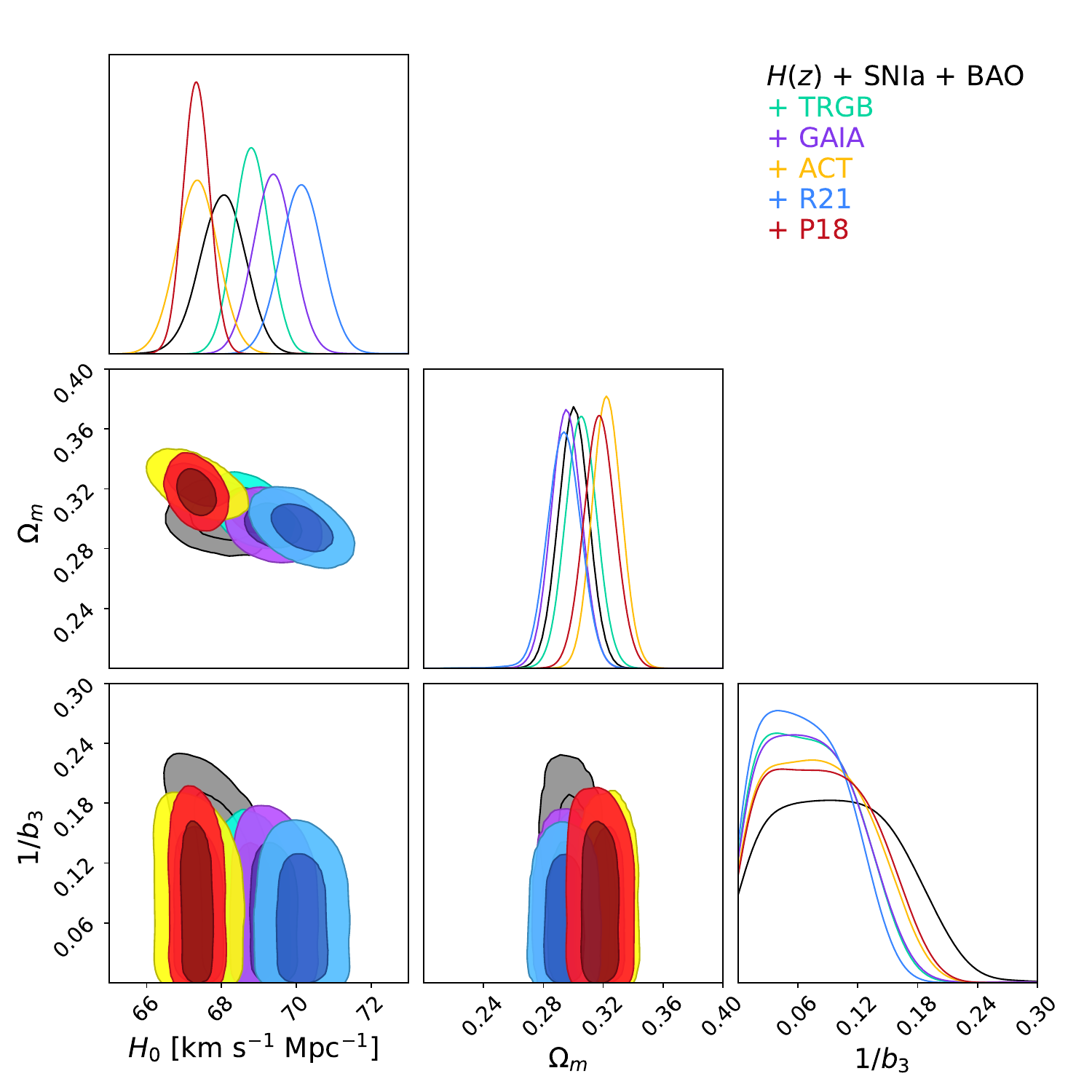}
\caption{1-2$\sigma$ C.L results for the $f_3 (T)$ model using: \textit{Left:} $H(z)$ and Pantheon data sets. \textit{Right:} including BAO. Blue color denotes R21, purple color for GAIA, red color for P18, green color for F20, and yellow color for ACT priors. Additionally, the model was constrained with the sample baselines without prior, here denoted in black color.}
\label{fig:f3}
\end{figure}

\begin{figure}[H]
  \centering
  \includegraphics[width=.49\linewidth]{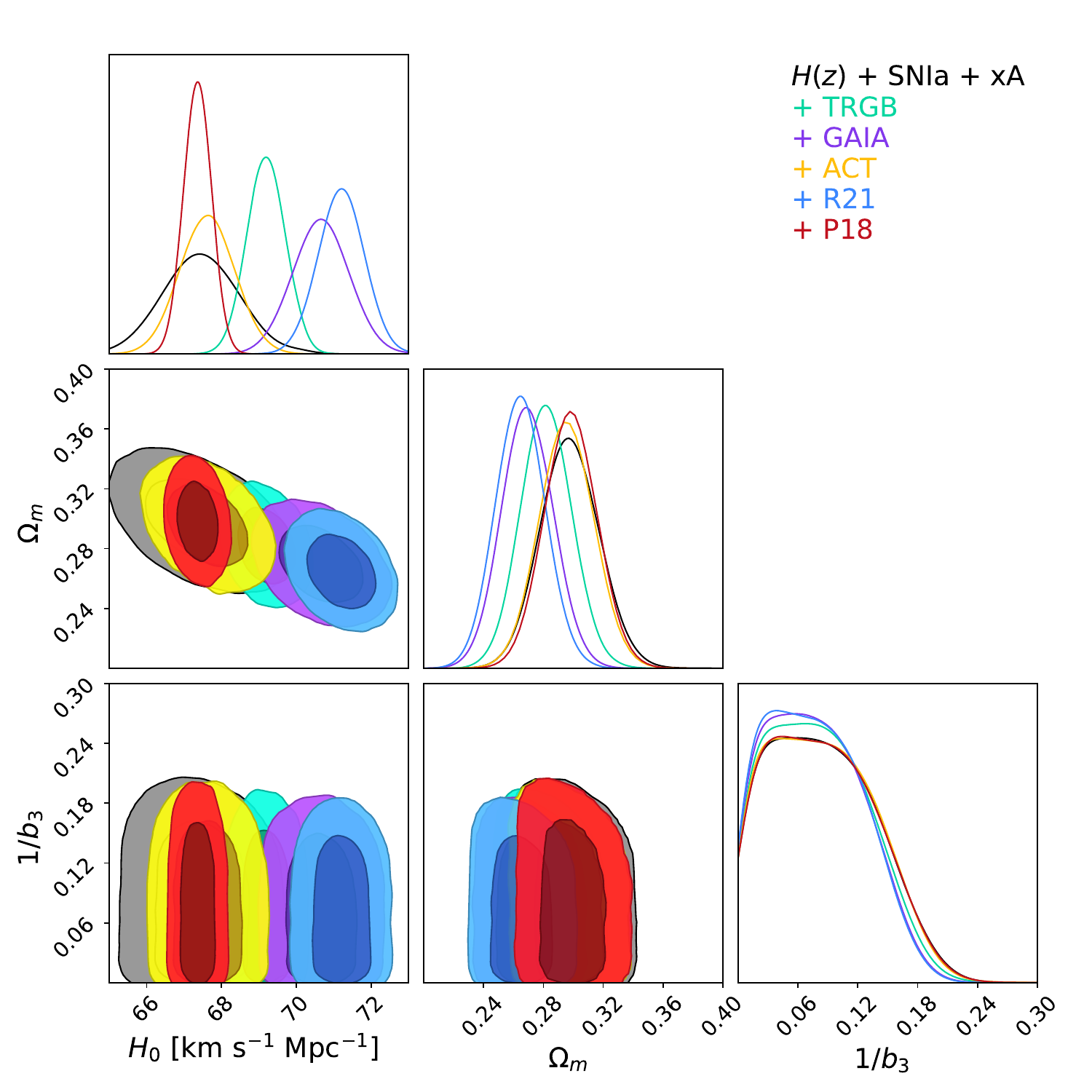}
  \includegraphics[width=.49\linewidth]{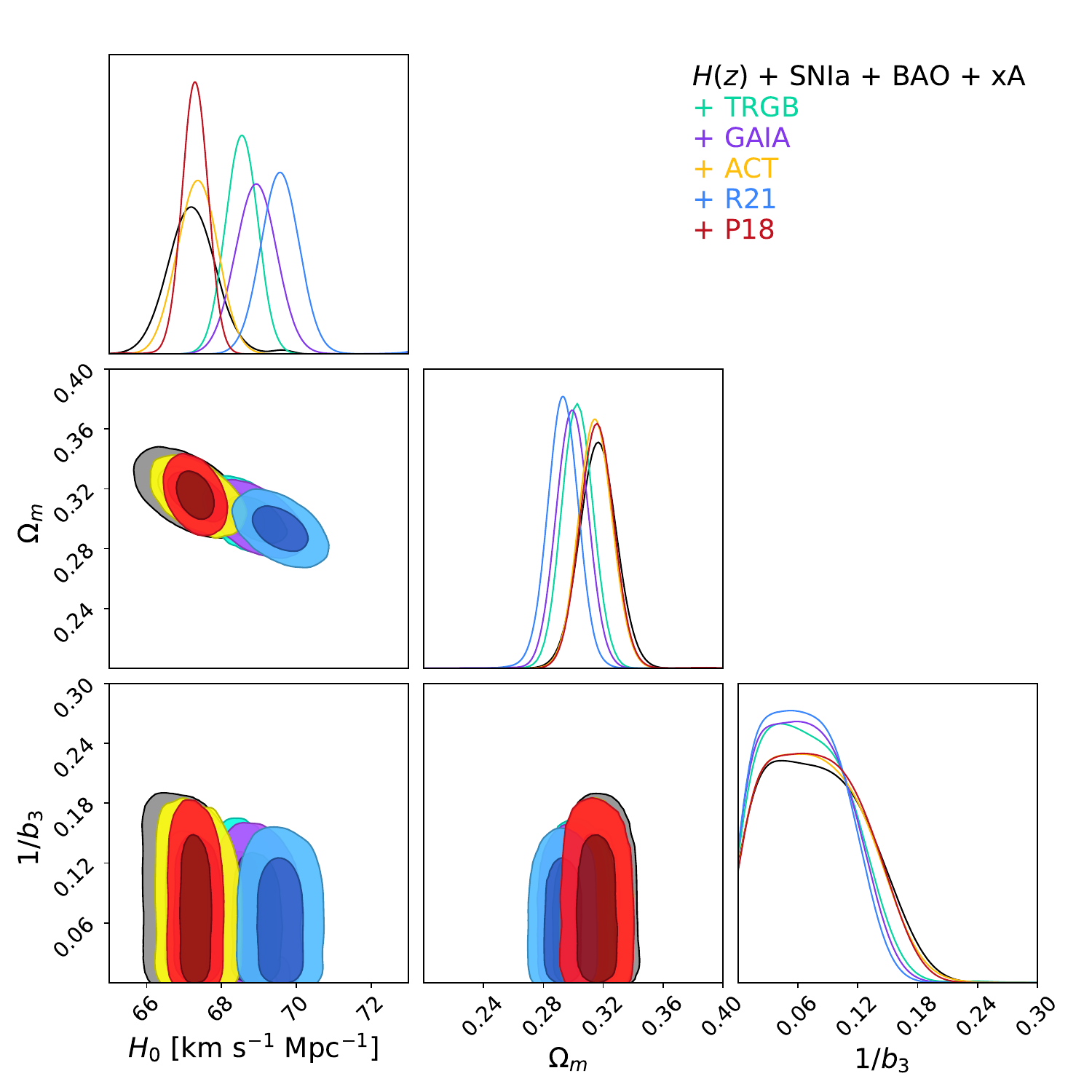}
    \includegraphics[width=.49\linewidth]{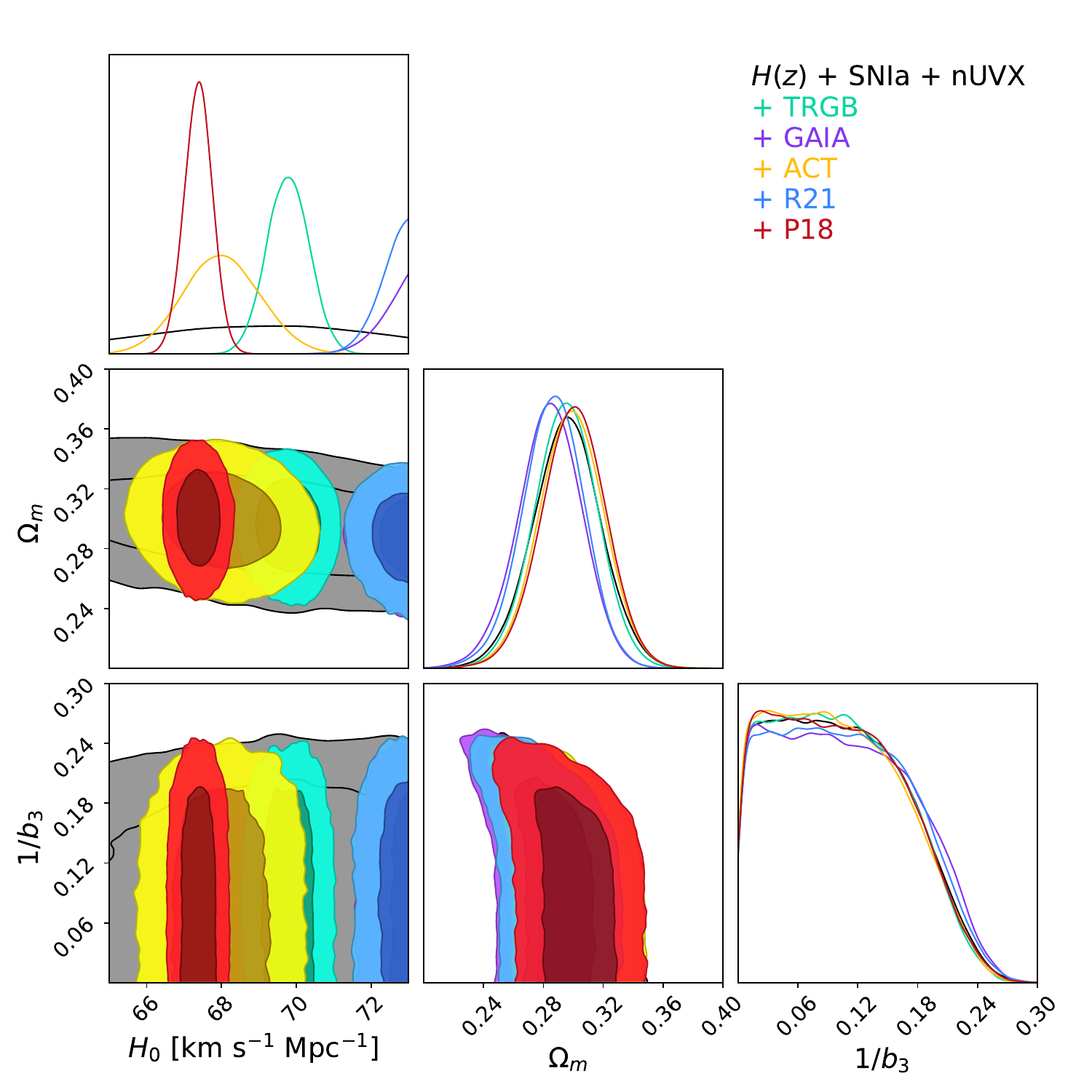}
  \includegraphics[width=.49\linewidth]{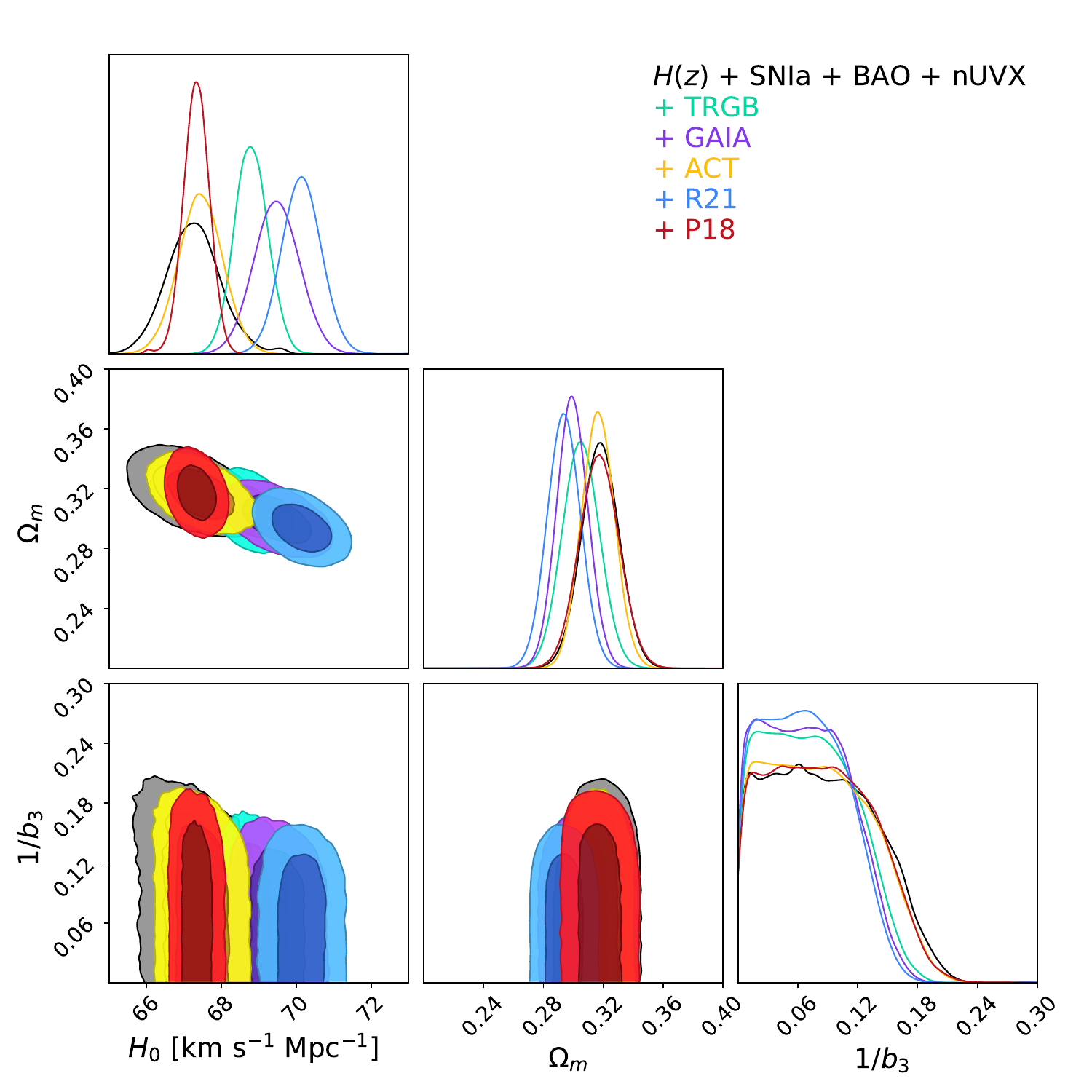}
\caption{1-2$\sigma$ C.L results for the $f_3(T)$ model using: \textit{Top left:} $H(z)$+SNIa and including the xA sample. \textit{Top right:} $H(z)$+SNIa+BAO and including the xA sample. 
\textit{Bottom left:} $H(z)$+SNIa and including the nUVX sample. \textit{Bottom right:} \textit{Top right:} $H(z)$+SNIa+BAO and including the nUVX sample.
Blue color denotes R21, purple color for GAIA, red color for P18, green color for F20, and yellow color for ACT priors. Additionally, the model was constrained with the sample baselines without prior, here denoted in black color.}
\label{fig:f3_qso1}
\end{figure}

The 1-2$\sigma$ C.L. constraints for this model are given in Figure \ref{fig:f3}. The cosmological constraints for this model are given in Table \ref{tab:cc+pn_f3}. Also, it is reported the 1-2-$\sigma$ constraints of this model in Figure \ref{fig:f3_qso1}, with their constraints reported in Table \ref{tab:cc+pn+qso_f3}  for the QSO-xA sample and in Table \ref{tab:cc+pn+qso2_f3} for the QSO-nUVX sample.

We notice that the model goes to $\Lambda$CDM as its constraint in $b_3 \rightarrow +\infty$, which is analogous to the $f_2$CDM model, see Eq.(\ref{eq:f3}). We reported this quantity as the inverse to avoid divergencies in the model. The result obtained using the $H(z)$+SNIa sample seems to raise the $H_0$ value in comparison to the other cases. However, as is expected, the introduction of the BAO sample tends to lower the $H_0$ values for the priors considered.


\begin{table}[H]
\resizebox{\textwidth}{!}{%
    \centering
    \begin{tabular}{c|ccccc}
    \hline 
    \hline & \\[-1.9ex]
    Dataset & $H_0$ [km s$^{-1}$ Mpc$^{-1}$] & $\Omega_m$ & $1/b_3$ & $M$ & $\chi^2_\mathrm{min}$ \\[0.5ex]
    \hline & \\[-1.8ex]
    $H(z)$ + SNIa &  $69.5^{+3.5}_{-3.7}$ & $0.294^{+0.024}_{-0.022}$ & $0.015^{+0.130}_{-0.014}$ & $-19.37\pm 0.11$  & 949.41 \\[0.9ex]
    $H(z)$ + SNIa + R21 & $73.1\pm 0.8$ & $0.286^{+0.024}_{-0.023}$ & $0.00^{+0.12}_{-0.00}$ & $-19.26\pm0.03$  & 950.71 \\[0.5ex] 
    $H(z)$ + SNIa + P18 & $67.4 \pm 0.4$ & $0.302^{+0.019}_{-0.022}$ & $0.00^{+0.14}_{-0.00}$ & $-19.43\pm 0.02$  & 949.65 \\[0.5ex]
    $H(z)$ + SNIa + F20 &  $69.8^{+0.6}_{-0.5}$ & $0.294^{+0.021}_{-0.020}$ & $0.00^{+0.14}_{-0.00}$ & $-19.36\pm0.02$  & 949.45 \\[0.5ex]
    $H(z)$ + SNIa + GAIA & $73.7\pm 1.0$ & $0.285^{+0.022}_{-0.020}$ & $0.050^{+0.106}_{-0.049}$ & $-19.235\pm 0.03$  & 951.15 \\[0.5ex]
    $H(z)$ + SNIa + ACT & $67.7^{+1.2}_{-0.7}$ & $0.299^{+0.023}_{-0.024}$ &  & $-19.41\pm 0.04$  & 949.51\\[0.5ex]
    \hline\hline
     $H(z)$ + SNIa + BAO & $68.1^{+0.5}_{-0.6}$ & $0.300^{+0.012}_{-0.011}$ & $0.093^{+0.053}_{-0.091}$ & $-19.41\pm 0.02$  & 960.24\\[0.9ex]
    $H(z)$ + SNIa + BAO + R21 & $70.2^{+0.5}_{-0.6}$ & $0.294^{+0.011}_{-0.010}$ & $0.001^{+0.091}_{-0.000}$ & $-19.35\pm 0.02$  & 995.28 \\[0.5ex] 
    $H(z)$ + SNIa + BAO + P18 & $67.3^{+0.3}_{-0.4}$ & $0.318\pm 0.010$ & $0.030^{+0.087}_{-0.029}$ & $-19.42\pm 0.01$  & 960.30 \\[0.5ex]
    $H(z)$ + SNIa + BAO + F20 & $68.8^{+0.5}_{-0.4}$ & $\left( 305.5^{+9.8}_{-10.6} \right) \times 10^{-3}$ & $\left( 7.3^{+95.6}_{-6.2} \right) \times 10^{-3}$ & $-19.38^{+0.01}_{-0.02}$  & 968.02 \\[0.5ex]
    $H(z)$ + SNIa + BAO + GAIA & $69.4\pm 0.5$ & $\left( 294.8^{+10.9}_{-9.7} \right) \times 10^{-3}$ & $0.062^{+0.042}_{-0.061}$ & $-19.37\pm 0.02$  & 987.32 \\[0.5ex]
    $H(z)$ + SNIa + BAO + ACT & $67.4^{+0.5}_{-0.6}$ & $\left( 322.3^{+9.9}_{-9.8} \right) \times 10^{-3}$ & $0.077^{+0.037}_{-0.076}$ & $-19.42\pm 0.02$  & 961.25 \\[0.5ex]
    \hline    
    \end{tabular}
    }
    \caption{$f_3 (T)$ model results using $H(z)$ and SNIa datasets. \textit{Below line}: $f_3 (T)$ model results using $H(z)$, SNIa and BAO datasets.
    We include the analysis with the priors described in Table \ref{tab:priors}.}
    \label{tab:cc+pn_f3}
\end{table}


\begin{table}[H]
\resizebox{\textwidth}{!}{%
    \centering
    \begin{tabular}{c|ccccc}
    \hline 
    \hline & \\[-1.9ex]
    Dataset & $H_0$ [km s$^{-1}$ Mpc$^{-1}$] & $\Omega_m$ & $1/b_3$ & $M$ & $\chi^2_\mathrm{min}$ \\[0.5ex]
    \hline & \\[-1.8ex]
    $H(z)$ + SNIa + xA & $67.4 \pm 1.0$ & $0.297^{+0.020}_{-0.018}$ & $\left( 6.9^{+108.4}_{-6.1} \right) \times 10^{-3}$ & $-19.43 \pm 0.03$ & 2616.99 \\[0.9ex]
    $H(z)$ + SNIa + xA + R21 & $71.2 \pm 0.6$ & $0.265^{+0.015}_{-0.017}$ & $0.028^{+0.077}_{-0.028}$ & $-19.33 \pm 0.02$ & 2639.99 \\[0.5ex]
    $H(z)$ + SNIa + xA + P18 & $67.4^{+0.3}_{-0.4}$ & $0.299^{+0.017}_{-0.020}$ & $0.033^{+0.088}_{-0.032}$ & $-19.43 \pm 0.02$ & 2617.00 \\[0.5ex]
    $H(z)$ + SNIa + xA + F20 & $69.2\pm 0.5$ & $0.281^{+0.017}_{-0.016}$ & $0.083^{+0.030}_{-0.082}$ & $-19.38\pm 0.02$ & 2621.45 \\[0.5ex]
    $H(z)$ + SNIa + xA + GAIA & $70.7 \pm 0.7$ & $0.269^{+0.018}_{-0.016}$ & $0.055^{+0.051}_{-0.055}$ & $-19.34 \pm 0.02$ & 2639.22 \\[0.5ex]
    $H(z)$ + SNIa + xA + ACT & $67.7^{+0.6}_{-0.8}$ & $0.294^{+0.024}_{-0.021}$ & $0.032^{+0.079}_{-0.031}$ & $-19.42^{+0.02}_{-0.03}$ & 2617.09 \\[0.5ex]
    \hline & \\[-1.8ex]
    $H(z)$ + SNIa + BAO + xA & $67.2\pm 0.6$ & $0.317^{+0.011}_{-0.012}$ & $0.000^{+0.099}_{-0.000}$ & $-19.429^{+0.017}_{-0.019}$ & 2628.24 \\[0.9ex]
    $H(z)$ + SNIa + BAO + xA + R21 & $69.6\pm 0.5$ & $\left( 293.0^{+10.1}_{-9.7} \right) \times 10^{-3}$ & $0.052^{+0.043}_{-0.051}$ & $-19.37\pm 0.02$ & 2670.17 \\[0.5ex]
    $H(z)$ + SNIa + BAO + xA + P18 & $67.3\pm 0.3$ & $0.315^{+0.011}_{-0.010}$ & $0.064^{+0.044}_{-0.063}$ & $-19.43\pm 0.01$ & 2628.27 \\[0.5ex]
    $H(z)$ + SNIa + BAO + xA + F20 & $68.6^{+0.4}_{-0.5}$ & $0.302^{+0.012}_{-0.011}$ & $0.038^{+0.062}_{-0.037}$ & $-19.39\pm 0.01$ & 2638.32 \\[0.5ex]
    $H(z)$ + SNIa + BAO + xA + GAIA & $68.9^{+0.5}_{-0.6}$ & $\left( 298.9^{+10.7}_{-10.0} \right) \times 10^{-3}$ & $0.008^{+0.076}_{-0.007}$ & $-19.38\pm 0.02$ & 2662.63 \\[0.5ex]
    $H(z)$ + SNIa + BAO + xA + ACT & $67.4\pm 0.5$ & $\left( 313.9^{+11.3}_{-9.9} \right) \times 10^{-3}$ & $0.078^{+0.031}_{-0.077}$ & $-19.43 \pm 0.02$ & 2628.54 \\[0.5ex]
    \hline    
    \end{tabular}
    }
    \caption{$f_3(T)$ model constraints using the: \textit{Top line:} $H(z)$+SNIa sample (in the first block), \textit{Below line:} and with BAO sample, both using QSO-xA sample.}
      \label{tab:cc+pn+qso_f3}
\end{table}

\begin{table}[H]
\resizebox{\textwidth}{!}{%
    \centering
    \begin{tabular}{c|cccccc}
    \hline 
    \hline & \\[-1.9ex]
    Dataset & $H_0$ [km s$^{-1}$ Mpc$^{-1}$] & $\Omega_m$ & $1/b_3$ & $M$ & $\beta'$ & $\chi^2_\mathrm{min}$ \\[0.5ex]
    \hline & \\[-1.8ex]
    $H(z)$ + SNIa + nUVX & $69.8^{+3.0}_{-4.2}$ & $0.296^{+0.022}_{-0.023}$ & $0.00^{+0.14}_{-0.00}$ & $-19.36^{+0.10}_{-0.12}$ & $-12.64^{+0.14}_{-0.73}$ & 3107.02 \\[0.9ex]
    $H(z)$ + SNIa + nUVX + R21 & $73.2^{+0.7}_{-0.8}$ & $0.288^{+0.019}_{-0.021}$ & $0.059^{+0.090}_{-0.058}$ & $-19.27\pm 0.02$ & $-12.21^{+0.13}_{-0.70}$ & 3108.34 \\[0.5ex]
    $H(z)$ + SNIa + nUVX + P18 & $67.4\pm 0.4$ & $0.301^{+0.021}_{-0.022}$ & $0.021^{+0.124}_{-0.020}$ & $-19.43\pm 0.02$ & $-12.62^{+0.61}_{-0.53}$ & 3107.26 \\[0.5ex]
    $H(z)$ + SNIa + nUVX + F20 & $69.8\pm 0.6$ & $0.295\pm 0.021$ & $0.077^{+0.065}_{-0.076}$ & $-19.36\pm 0.02$ & $-12.51^{+0.66}_{-0.49}$ & 3107.06 \\[0.5ex]
    $H(z)$ + SNIa + nUVX + GAIA & $73.8^{+0.9}_{-1.1}$ & $0.285\pm 0.021$ & $0.018^{+0.133}_{-0.017}$ & $-19.24\pm 0.03$ & $-12.63^{+0.48}_{-0.63}$ & 3108.77 \\[0.5ex]
    $H(z)$ + SNIa + nUVX + ACT & $68.0\pm 1.0$ & $0.299^{+0.021}_{-0.022}$ & $0.00^{+0.14}_{-0.00}$ & $-19.41\pm 0.03$ & $-12.52^{+0.68}_{-0.47}$ & 3107.14 \\[0.5ex]
    \hline & \\[-1.8ex]
    $H(z)$ + SNIa + BAO + nUVX & $67.3^{+0.6}_{-0.8}$ & $0.318\pm 0.012$ & $0.061^{+0.058}_{-0.060}$ & $-19.43\pm 0.02$ & $-11.48^{+0.19}_{-0.11}$ & 3117.91 \\[0.9ex]
    $H(z)$ + SNIa + BAO + nUVX + R21 & $70.1\pm 0.2$ & $0.294\pm 0.010$ & $0.067^{+0.032}_{-0.066}$ & $-19.39\pm 0.02$ & $-12.61^{+0.45}_{-0.73}$ & 3152.90 \\[0.5ex]
    $H(z)$ + SNIa + BAO + nUVX + P18 & $67.3^{+0.4}_{-0.3}$ & $0.318\pm 0.012$ & $0.00^{+0.12}_{-0.00}$ & $-19.43\pm 0.01$ & $-11.39^{+0.77}_{-0.36}$ & 3117.92 \\[0.5ex]
    $H(z)$ + SNIa + BAO + nUVX + F20 & $68.8\pm 0.4$ & $0.304^{+0.012}_{-0.011}$ & $0.020^{+0.085}_{-0.018}$ & $-19.39^{+0.02}_{-0.01}$ & $-11.77^{+0.77}_{-0.32}$ & 3125.63 \\[0.5ex]
    $H(z)$ + SNIa + BAO + nUVX + GAIA & $69.5\pm 0.6$ & $0.299^{+0.011}_{-0.010}$ & $0.018^{+0.079}_{-0.018}$ & $-19.37\pm 0.02$ & $-11.37^{+0.88}_{-0.62}$ & 3148.20 \\[0.5ex]
    $H(z)$ + SNIa + BAO + nUVX + ACT & $67.4^{+0.6}_{-0.5}$ & $0.316\pm 0.011$ & $0.00^{+0.12}_{-0.00}$ & $-19.42\pm 0.02$ & $-11.45^{+0.11}_{-0.10}$ & 3118.11\\[0.5ex]
    \hline    
    \end{tabular}
    }
    \caption{$f_3(T)$ model constraints using the: \textit{Top line:} $H(z)$+SNIa sample (in the first block), \textit{Below line:} and with BAO sample, both using QSO-nUVX sample.}
     \label{tab:cc+pn+qso2_f3}
\end{table}


\subsection{Logarithmic Model -- \texorpdfstring{$f_4(T)$}{} model}

\begin{figure}[H]
  \includegraphics[width=.49\linewidth]{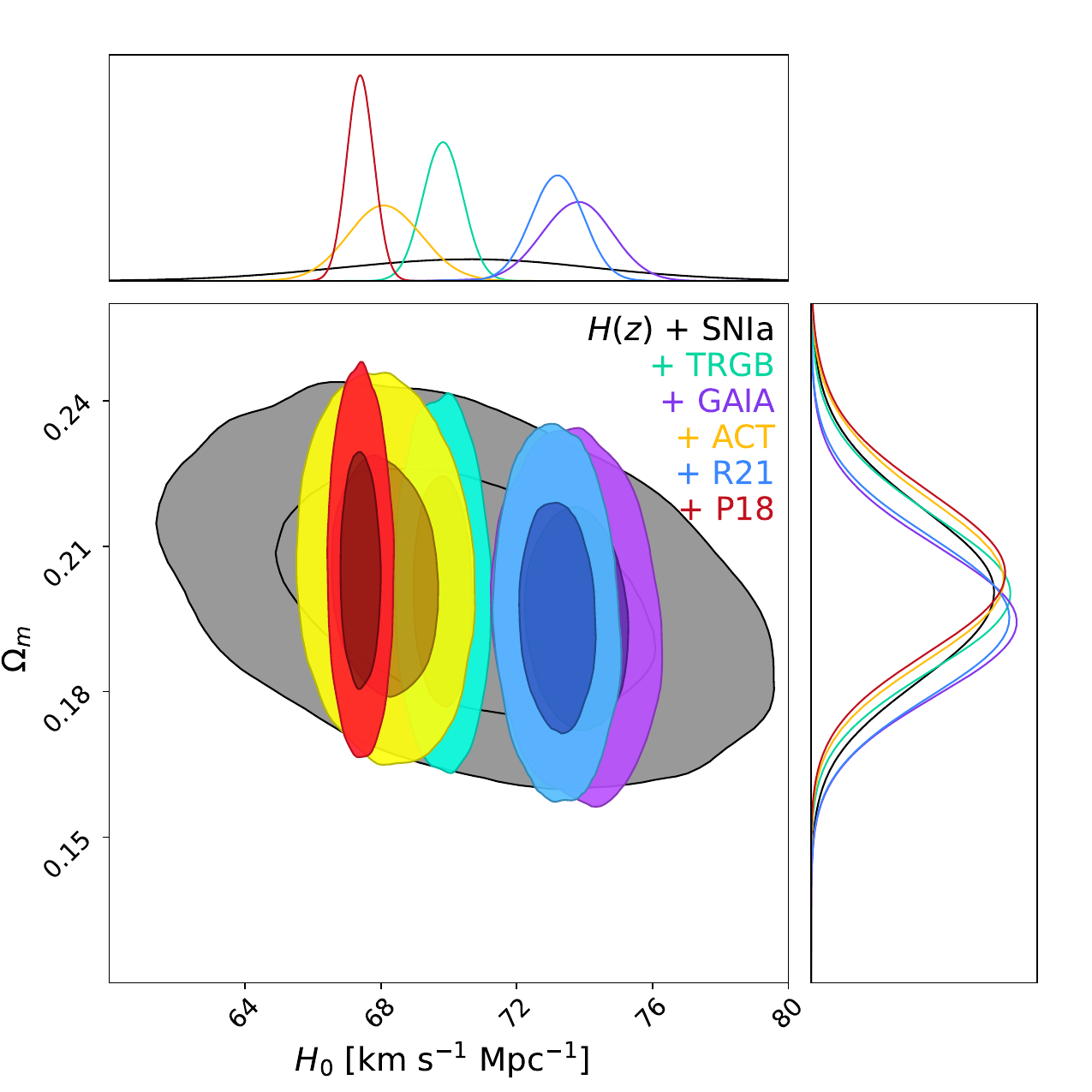}
  \includegraphics[width=.49\linewidth]{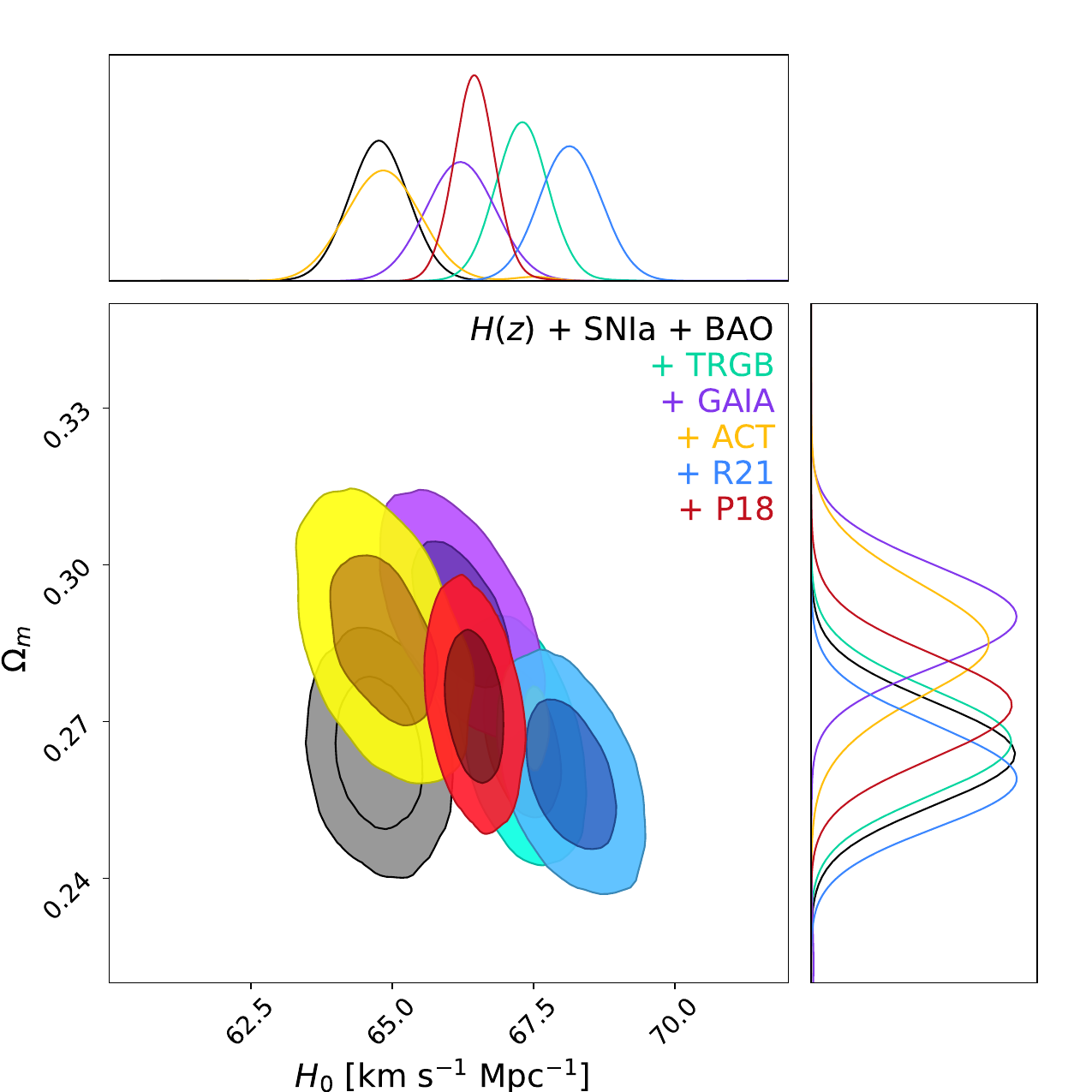}
\caption{1-2$\sigma$ C.L results for the $f_4 (T)$ model using: \textit{Left:} $H(z)$ and Pantheon data sets. \textit{Right:} including BAO. Blue color denotes R21, purple color for GAIA, red color for P18, green color for F20, and yellow color for ACT priors. Additionally, the model was constrained with the sample baselines without prior, here denoted in black color.}
\label{fig:f4}
\end{figure}

\begin{figure}[H]
  \centering
  \includegraphics[width=.49\linewidth]{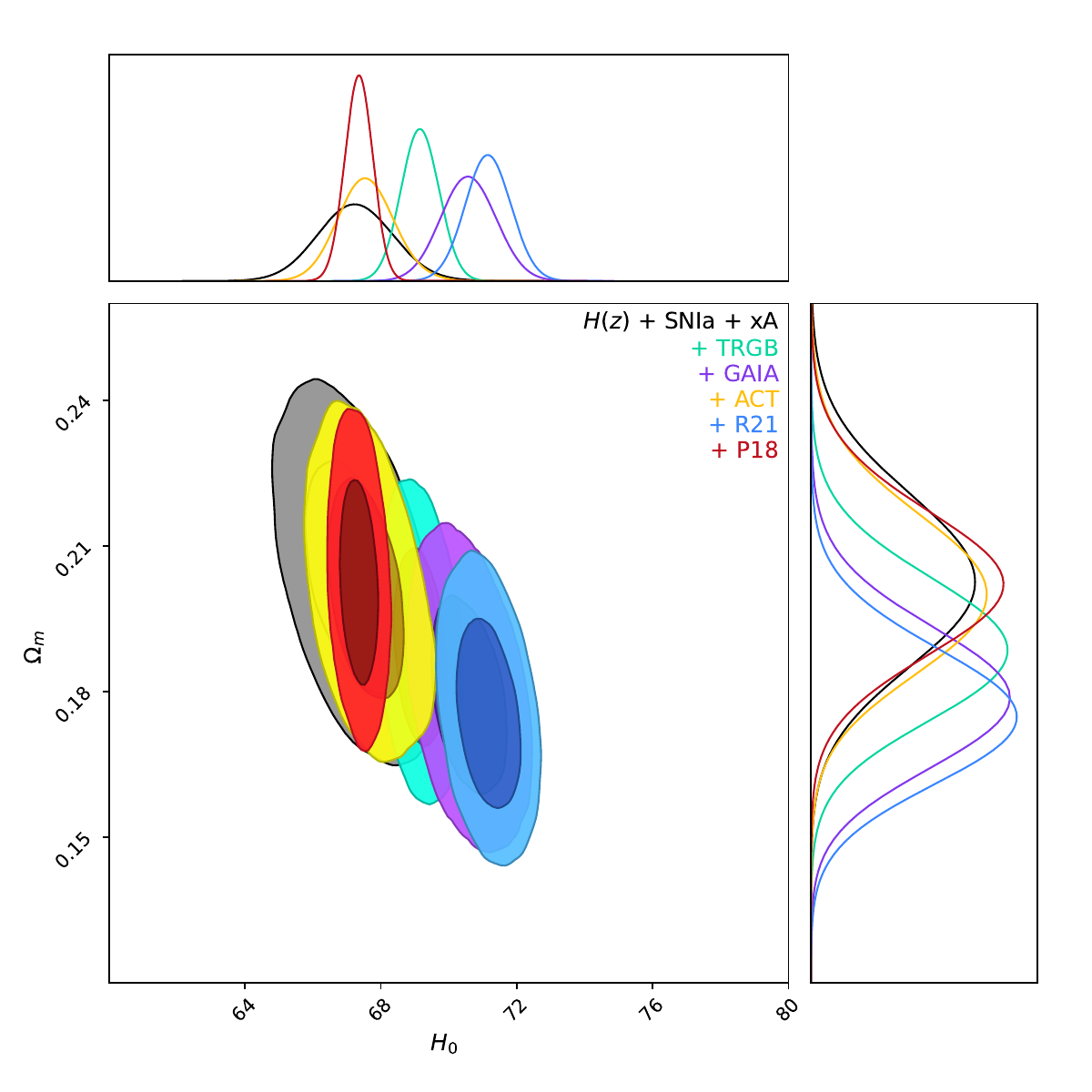}
  \includegraphics[width=.49\linewidth]{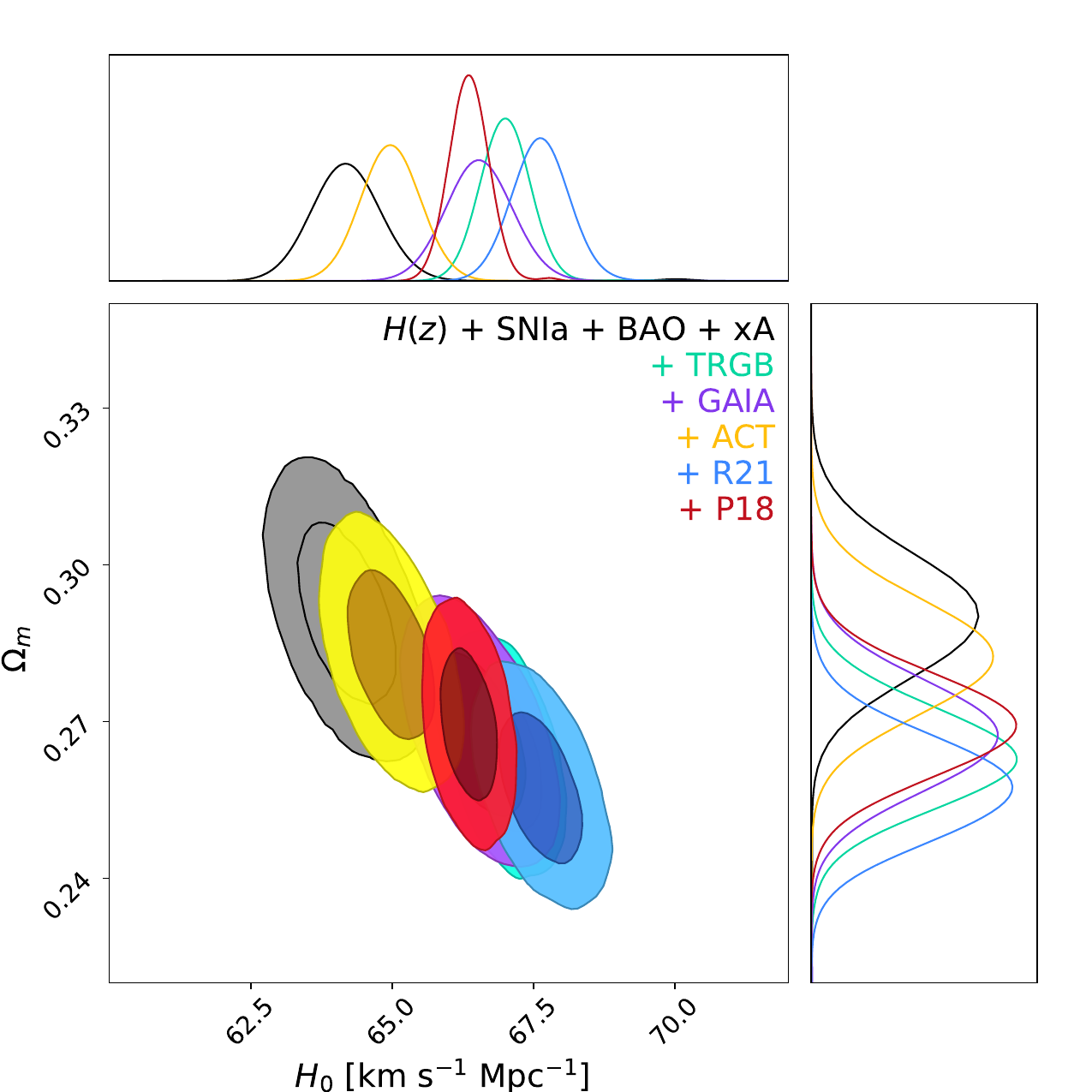}
    \includegraphics[width=.49\linewidth]{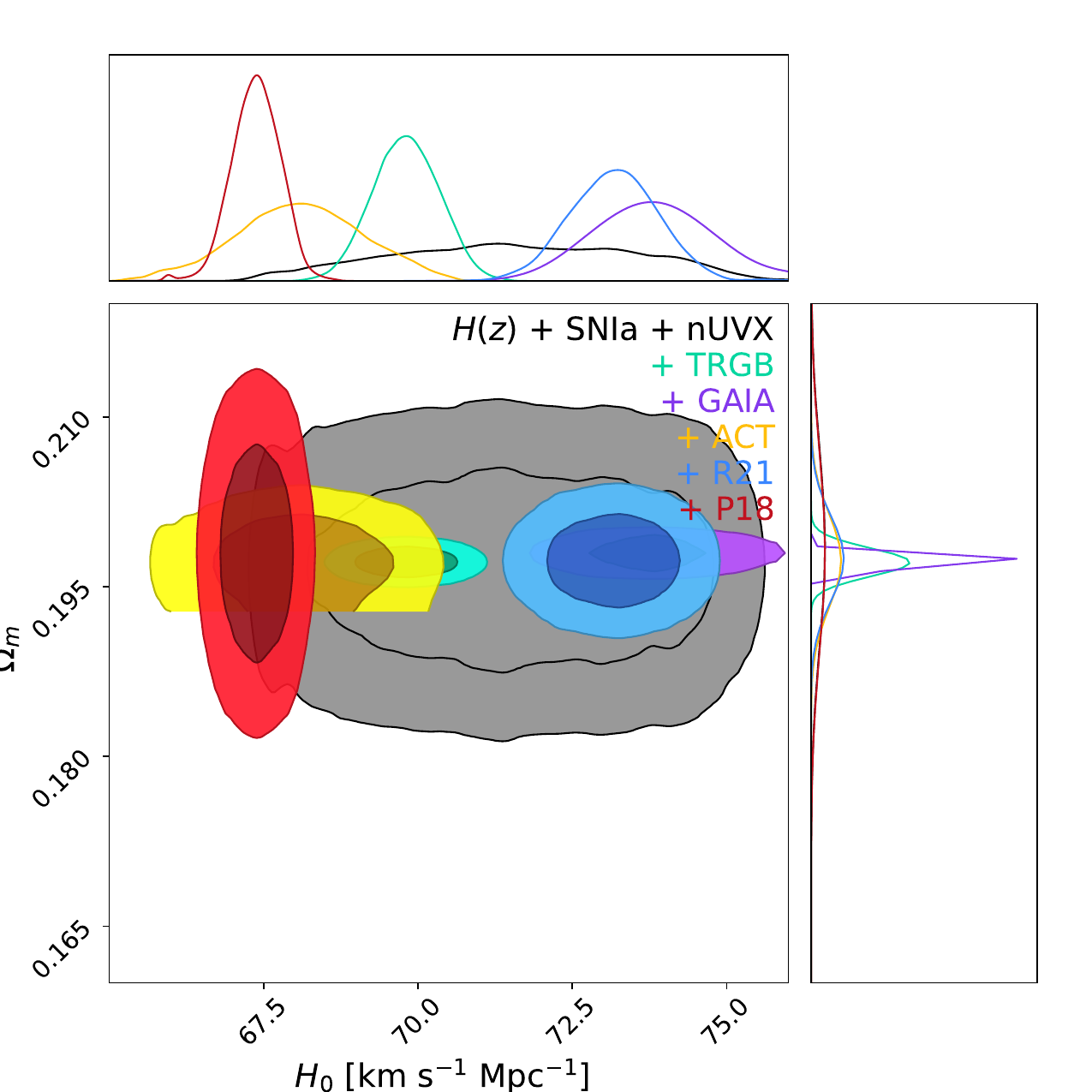}
  \includegraphics[width=.49\linewidth]{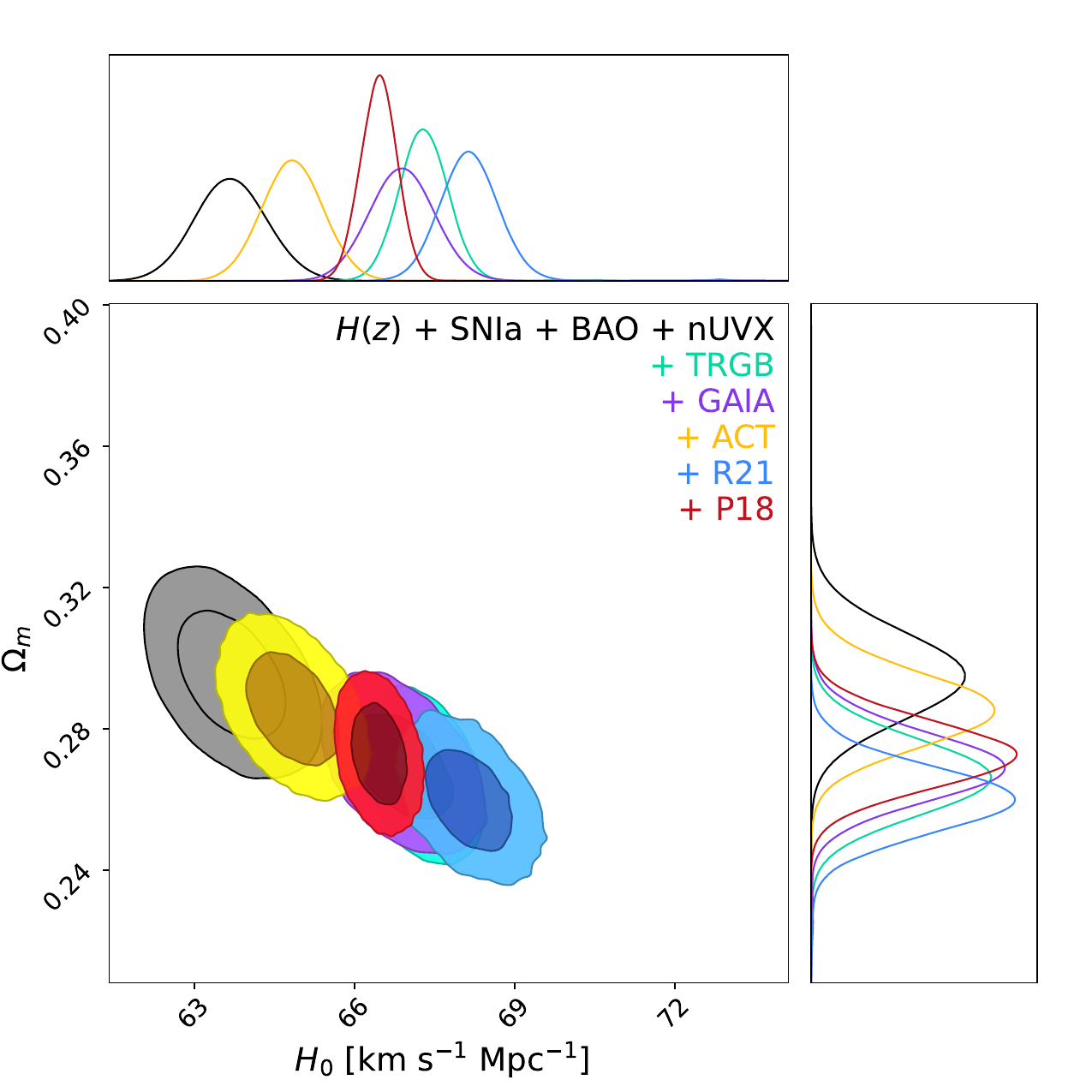}
\caption{1-2$\sigma$ C.L results for the $4_1(T)$ model using: \textit{Top left:} $H(z)$+SNIa and including the xA sample. \textit{Top right:} $H(z)$+SNIa+BAO and including the xA sample. 
\textit{Bottom left:} $H(z)$+SNIa and including the nUVX sample. \textit{Bottom right:} \textit{Top right:} $H(z)$+SNIa+BAO and including the nUVX sample.
Blue color denotes R21, purple color for GAIA, red color for P18, green color for F20, and yellow color for ACT priors. Additionally, the model was constrained with the sample baselines without prior, here denoted in black color.
}
\label{fig:f4_qso1}
\end{figure}

The 1-2$\sigma$ C.L. constraints for this model are given in Figure \ref{fig:f4}. The cosmological constraints for this model are given in Table \ref{tab:cc+pn_f4}. Also, it is reported the 1-2-$\sigma$ constraints of this model in Figure \ref{fig:f4_qso1}, with their constraints reported in Table \ref{tab:cc+pn+xA_f4}  for the QSO-xA sample and in Table \ref{tab:cc+pn+xUV_f4} for the QSO-nUVX sample.

Notice that due to the nature of this model, from Eq.(\ref{eq:f4}), we do not treat with free parameters related to $b_4$. As a consequence, the background behaviour of this model cannot feature per se any bias compared with the $\Lambda$CDM model. The case with the $H(z)$+SNIa+GAIA prior is the only one that shows a high $H_0$ value.


\begin{table}[H]
    \centering
    \begin{tabular}{c|cccc}
    \hline 
    \hline & \\[-1.9ex]
    Dataset & $H_0$ [km s$^{-1}$ Mpc$^{-1}$] & $\Omega_m$ & $M$ & $\chi^2_\mathrm{min}$ \\[0.5ex]
    \hline & \\[-1.8ex]
    $H(z)$ + SNIa &$70.1^{+3.9}_{-3.3}$ & $0.202^{+0.014}_{-0.019}$ & $-19.30^{+0.08}_{-0.13}$  & 951.61 \\[0.9ex]
    $H(z)$ + SNIa + R21  & $73.2^{+0.7}_{-0.8}$ & $0.195^{+0.016}_{-0.015}$ & $-19.25^{+0.03}_{-0.02}$  & 952.17\\[0.5ex] 
    $H(z)$ + SNIa + P18 & $67.4\pm 0.04$ & $0.205\pm 0.016$ & $-19.42^{+0.01}_{-0.02}$  & 952.41 \\[0.5ex]
    $H(z)$ + SNIa + F20  & $69.8^{+0.6}_{-0.5}$ & $0.200^{+0.016}_{-0.015}$ & $-19.35\pm 0.02$  & 951.66 \\[0.5ex]
    $H(z)$ + SNIa + GAIA &  $73.9^{+0.9}_{-1.0}$ & $0.194^{+0.016}_{-0.014}$ & $-19.23\pm 0.03$  & 952.47\\[0.5ex]
    $H(z)$ + SNIa + ACT & $68.0^{+1.1}_{-1.0}$ & $0.203^{+0.017}_{-0.016}$ & $-19.40^{+0.04}_{-0.03}$  & 952.12\\[0.5ex]
    \hline\hline
     $H(z)$ + SNIa + BAO & $64.8\pm 0.5$ & $\left( 264.0^{+9.7}_{-9.9} \right) \times 10^{-3}$ & $-19.48 \pm 0.02$  & 974.34 \\[0.9ex]
    $H(z)$ + SNIa + BAO + R21 & $68.2 \pm 0.6$ & $\left( 258.6^{+10.3}_{-8.6} \right) \times 10^{-3}$ & $-19.37\pm 0.02$  & 1088.61 \\[0.5ex] 
    $H(z)$ + SNIa + BAO + P18 & $66.4^{+0.4}_{-0.3}$ & $\left( 273.8^{+8.8}_{-10.6} \right) \times 10^{-3}$ & $-19.42\pm 0.01$  & 1018.55 \\[0.5ex]
    $H(z)$ + SNIa + BAO + F20 &$67.4^{+0.4}_{-0.5}$ & $\left( 265.3^{+10.2}_{-8.8} \right) \times 10^{-3}$ & $-19.39\pm 0.01$  & 1044.32 \\[0.5ex]
    $H(z)$ + SNIa + BAO + GAIA & $66.2^{+0.6}_{-0.5}$ & $\left( 289.9^{+9.3}_{-8.8} \right) \times 10^{-3}$ & $-19.42 \pm 0.02$  & 1087.61 \\[0.5ex]
    $H(z)$ + SNIa + BAO + ACT & $64.9^{+0.5}_{-0.7}$ & $\left( 284.4^{+12.0}_{-9.1} \right) \times 10^{-3}$ & $-19.47 \pm 0.02$  & 1006.92\\[0.5ex]
    \hline    
    \end{tabular}
    \caption{$f_4 (T)$ model results using $H(z)$ and SNIa datasets. \textit{Below line}: $f_4 (T)$ model results using $H(z)$, SNIa and BAO datasets.
    We include the analysis with the priors described in Table \ref{tab:priors}.}
    \label{tab:cc+pn_f4}
\end{table}


\begin{table}[H]
\resizebox{\textwidth}{!}{%
    \centering
    \begin{tabular}{c|cccc}
    \hline 
    \hline & \\[-1.9ex]
    Dataset & $H_0$ [km s$^{-1}$ Mpc$^{-1}$] & $\Omega_m$ & $M$ & $\chi^2_\mathrm{min}$\\[0.5ex]
    \hline & \\[-1.8ex]
    $H(z)$ + SNIa + xA & $67.3\pm 1.0$ & $0.201^{+0.017}_{-0.014}$ & $-19.43\pm 0.03$ & 2627.82 \\[0.9ex]
    $H(z)$ + SNIa + xA + R21 & $71.1^{+0.7}_{-0.5}$ & $0.175^{+0.012}_{-0.013}$ & $-19.32\pm 0.02$ & 2652.04 \\[0.5ex]
    $H(z)$ + SNIa + xA + P18 & $67.4\pm 0.4$ & $0.202^{+0.014}_{-0.013}$ & $-19.42\pm 0.01$ & 2627.82 \\[0.5ex]
    $H(z)$ + SNIa + xA + F20 & $69.1 \pm 0.5$ & $0.188^{+0.014}_{-0.012}$ & $-19.375\pm 0.02$ & 2632.91 \\[0.5ex]
    $H(z)$ + SNIa + xA + GAIA & $70.6^{+0.7}_{-0.8}$ & $0.178^{+0.014}_{-0.013}$ & $-19.34\pm 0.02$ & 2651.11\\[0.5ex]
    $H(z)$ + SNIa + xA + ACT & $67.5^{+0.8}_{-0.7}$ & $0.199^{+0.016}_{-0.013}$ & $-19.42\pm 0.02$ & 2628.01 \\[0.5ex]
    \hline& \\[-1.8ex]
     $H(z)$ + SNIa + BAO + xA & $64.2\pm 0.6$ & $0.291^{+0.011}_{-0.013}$ & $-19.49\pm 0.02$ & 2675.19 \\[0.9ex]
     $H(z)$ + SNIa + BAO + xA + R21 & $67.6\pm 0.5$ & $\left( 257.7^{+9.1}_{-10.0} \right) \times 10^{-3}$ & $-19.39\pm 0.02$ & 2772.13\\[0.5ex]
     $H(z)$ + SNIa + BAO + xA + P18 & $66.4^{+0.3}_{-0.4}$ & $\left( 269.0^{+9.7}_{-8.9} \right) \times 10^{-3}$ & $-19.42\pm 0.01$ & 2697.35 \\[0.5ex]
     $H(z)$ + SNIa + BAO + xA + F20 & $67.0\pm 0.4$ & $\left( 262.7^{+9.6}_{-9.0} \right) \times 10^{-3}$ & $-19.41\pm 0.01$& 2725.01 \\[0.5ex]
     $H(z)$ + SNIa + BAO + xA + GAIA & $66.5^{+0.6}_{-0.5}$ & $0.267\pm 0.010$ & $-19.42\pm 0.02$ & 2748.71 \\[0.5ex]
     $H(z)$ + SNIa + BAO + xA + ACT & $65.0\pm 0.5$ & $0.282^{+0.011}_{-0.010}$ & $-19.46\pm 0.02$ & 2684.88 \\[0.5ex]  
    \hline     
    \end{tabular}
    }
    \caption{$f_4(T)$ model constraints using the: \textit{Top line:} $H(z)$+SNIa sample (in the first block), \textit{Below line:} and with BAO sample, both using QSO-xA sample.
    }
    \label{tab:cc+pn+xA_f4}
\end{table}

\begin{table}[H]
\resizebox{\textwidth}{!}{%
    \centering
    \begin{tabular}{c|ccccc}
    \hline 
    \hline & \\[-1.9ex]
    Dataset & $H_0$ [km s$^{-1}$ Mpc$^{-1}$] & $\Omega_m$ & $M$ & $\beta'$ & $\chi^2_\mathrm{min}$ \\[0.5ex]
    \hline & \\[-1.8ex]
    $H(z)$ + SNIa + nUVX & $71.4^{+2.4}_{-2.0}$ & $\left( 196.3^{+6.5}_{-6.1} \right) \times 10^{-3}$ & $-19.31\pm 0.03$ & $-10.7^{+8.5}_{-7.2}$ & 3404.16\\[0.9ex]
    $H(z)$ + SNIa + nUVX + R21 & $73.3^{+0.7}_{-0.8}$ & $\left( 197.3\pm 2.7 \right) \times 10^{-3}$ & $-19.25\pm 0.02$ & $-10.7^{+0.7}_{-0.9}$ & 3405.16 \\[0.5ex]
    $H(z)$ + SNIa + nUVX + P18 & $67.4\pm 0.4$ & $\left( 198.0^{+6.3}_{-6.4} \right) \times 10^{-3}$ & $-19.43\pm 0.01$ & $-11.3^{+0.9}_{-0.6}$ & 3405.14 \\[0.5ex]
    $H(z)$ + SNIa + nUVX + F20 & $69.8^{+0.5}_{-0.6}$ & $\left( 1972.1^{+8.7}_{-9.3} \right) \times 10^{-4}$ & $-19.35\pm 0.02$ & $-10.8^{+0.7}_{-0.8}$ & 3408.23 \\[0.5ex]
    $H(z)$ + SNIa + nUVX + GAIA & $73.8^{+0.9}_{-1.0}$ & $\left( 195.9^{+3.0}_{-0.0} \right) \times 10^{-3}$ & $-19.23\pm 0.03$ & $-10.4^{+0.7}_{-0.9}$ & 3405.06 \\[0.5ex]
    $H(z)$ + SNIa + nUVX + ACT & $68.2^{+0.9}_{-1.2}$ & $\left( 197.6^{+4.5}_{-5.3} \right) \times 10^{-3}$ & $-19.40^{+0.03}_{-0.04}$ & $-10.9\pm 0.8$ & 3407.80 \\[0.5ex]
    \hline & \\[-1.8ex]
    $H(z)$ + SNIa + BAO + nUVX & $63.7 \pm 0.7$ & $0.295\pm 0.012$ & $-19.50 \pm 0.02$ & $-10.79^{+1.61}_{-1.87}$ & 3153.12 \\[0.9ex]
    $H(z)$ + SNIa + BAO + nUVX + R21 & $68.2^{+0.5}_{-0.6}$ & $\left( 260.0\pm 9.5 \right) \times 10^{-3}$ & $-19.37\pm 0.02$ & $-10.7^{+1.25}_{-1.24}$ & 3246.22 \\[0.5ex]
    $H(z)$ + SNIa + BAO + nUVX + P18 & $66.5^{+0.3}_{-0.4}$ & $\left( 272.9^{+9.2}_{-9.5} \right) \times 10^{-3}$ & $-19.42\pm 0.01$ & $-10.65^{+1.52}_{-1.87}$ & 3176.16 \\[0.5ex]
    $H(z)$ + SNIa + BAO + nUVX + F20 & $67.3^{+0.5}_{-0.4}$ & $\left( 266.5^{+9.8}_{-10.3} \right) \times 10^{-3}$ & $-19.39\pm 0.02$ & $-11.12^{+1.67}_{-1.75}$ & 3201.93 \\[0.5ex]
    $H(z)$ + SNIa + BAO + nUVX + GAIA & $66.9\pm 0.6$ & $\left( 268.9^{+10.2}_{-9.6} \right) \times 10^{-3}$ & $-19.40\pm 0.02$ & $-10.24^{+1.62}_{-1.71}$ & 3226.86 \\[0.5ex]
    $H(z)$ + SNIa + BAO + nUVX + ACT & $64.8^{+0.6}_{-0.5}$ & $0.285^{+0.011}_{-0.010}$ & $-19.46\pm 0.02$ & $-12.11^{+1.86}_{-1.50}$ & 3164.54 \\[0.5ex]
    \hline 
    \end{tabular}
    }
    \caption{$f_4(T)$ model constraints using the: \textit{Top line:} $H(z)$+SNIa sample (in the first block), \textit{Below line:} and with BAO sample, both using QSO-nUVX sample.}
        \label{tab:cc+pn+xUV_f4}
\end{table}


\section{\label{sec:conc}Conclusions}


In this paper, we presented new constraints on $f(T)$ models
by adding to the current local observables two calibrated quasars datasets on the ultraviolet, x-ray, and optical plane. Our goal was to implement these high-redshift QSO samples based on fluxes distributions calibrated up to $z \sim 7 $, since due to the observed non-linear relation between the UV and the X-ray luminosity we can extend the distance ladder method to QSO higher redshift regime. The calibrations were performed using five $H_0$ prior scenarios, however, we added an analysis using a free prior.

It is important to notice that there is a variation in the trend on the correlation between the $\Omega_m-H_0$ parameters. This effect is probably a consequence of the use of the covariance matrix of the $H(z)$ measurements in which errors originating from different stellar modelings are included, and the error in the parameter determination is larger. Therefore, appears to have changed the trend in the $\Omega_m-H_0$ plane.

The methodology discussed presents new constraints on the $\Lambda$CDM model and four TEGR-inspired $f(T)$ models using a baseline constructed with SN + $H(z)$ \& BAO, and the calibrated QSO datasets. Both baselines (SN + $H(z)$ + BAO, SN + $H(z)$ + QSO) were employed to analyse the impact of considering objects as QSO as a fundamental study to relax the current statistical tension on $H_0$. In this matter, we found that our estimations provide the possibility to raise the $H_0$ value to solve the tension at 2$\sigma$ by using QSO ultraviolet measurements.


Furthermore, the models considered in this work offer important insights into the relative performance of each of them against the observational data set combinations. The $\Lambda$CDM case, Table~\ref{tab:priors} shows a $\mathbf{\chi^2_\mathrm{min}}$, that remains relatively flat around $\sim 950$ for the $H(z)$ + SNIa data set combination despite different priors being taken into consideration. When the same data sets are considered but with the addition of BAO data points, this becomes less consistent with a variance of $\sim 30$. This overall behaviour is retained when the QSO-xA and QSO-nUVX samples are added, as shown in Tables~\ref{tab:cc+pn+xA_lcdm},\ref{tab:cc+pn+nUVX_lcdm}. However, in all these cases, the constraints on the model parameters are stricter. This is a general observation but can be seen when considering the $H(z)$ + SNIa data set combination for the $f_1$CDM case.

The introduction of $f_i$CDM models nuances the $\mathbf{\chi^2_\mathrm{min}}$ tendencies we observe from the $\Lambda$CDM case. For instance, considering the $f_1$CDM model, in Tables~\ref{tab:cc+pn_f1},\ref{tab:cc+pn+qso_f1},\ref{tab:cc+pn+qso2_f1} one notices that the GAIA prior consistently has an effect on this $\mathbf{\chi^2_\mathrm{min}}$ value pointing to the need for further investigation, while the other priors have little to no impact against the baseline scenario in which no prior is considered. This tendency is also observed for the other models, namely $f_2$-$f_4$CDM, while the other priors seem to be very consistent with these models in terms of their compatibility against the baseline case. All the modified cosmology models similarly observe larger values of $\mathbf{\chi^2_\mathrm{min}}$ when BAO data is included in the analysis.

We show that the $f(T)$ models under investigation generically produce high values of Hubble constants which show stronger agreement with the more recent literature values reported in survey publications. Moreover, there is a better consistency for the matter density parameter in all cases as compared with the $\Lambda$CDM analog. This shows a stronger case for these models, and their potential ability to agree with observational data.


\section*{Acknowledgements}\label{sec:acknowledgements}
RS is supported by the CONACyT National Grant.
CE-R acknowledges the Royal Astronomical Society as FRAS 10147 and is supported by PAPIIT UNAM Project TA100122. 
The authors would like to acknowledge networking support by the COST Action CA18108 and funding support from Cosmology@MALTA which is supported by the University of Malta. This research has been carried out using computational facilities procured through the Cosmostatistics National Group project 
and the European Regional Development Fund, Project No. ERDF-080 ``A supercomputing laboratory for the University of Malta''. The authors also acknowledge funding from ``The Malta Council for Science and Technology'' in project IPAS-2020-007.
This article is based upon work from COST Action CA21136 Addressing observational tensions in cosmology with systematics and fundamental physics (CosmoVerse) supported by COST (European Cooperation in Science and Technology). JLS would also like to acknowledge funding from ``The Malta Council for Science and Technology'' as part of the ``FUSION R\&I: Research Excellence Programme'' REP-2023-019 (CosmoLearn) Project. JLS would also like to acknowledge funding from “The Malta Council for Science and Technology” in project IPAS-2023-010.


\bibliographystyle{JHEP}
\bibliography{references}

\end{document}